\documentclass[english,twocolumn,showpacs,aps,prd]{revtex4-1}

\usepackage{preprintcover}
\PreprintCoverPaperTitle{Search for new phenomena in events with three charged
  leptons at $\sqrt{s}=7\TeV$ with the ATLAS detector}
\PreprintIdNumber{CERN-PH-EP-2012-310}
\PreprintCoverAbstract{A generic search for anomalous production of events with at least three charged leptons is presented.  The search uses 
  a $pp$-collision data sample at a center-of-mass energy of $\sqrt{s} = 7\TeV$ corresponding to 4.6~\ifb\ of integrated luminosity 
  collected in 2011 by the ATLAS detector at the CERN Large Hadron Collider.  Events are required to contain at least two 
  electrons or muons, while the third lepton may either be an additional electron or muon, or a hadronically decaying tau lepton.  
  Events are categorized by the presence or absence of a reconstructed tau-lepton or $Z$-boson candidate 
  decaying to leptons.  No significant excess above backgrounds 
  expected from Standard Model processes is observed.  Results are presented as upper limits on event yields from non-Standard-Model 
  processes producing at least three prompt, isolated leptons, given as functions of lower bounds on several kinematic variables.
  Fiducial efficiencies for model testing are also provided.  The use of the results is illustrated by
  setting upper limits on the production of doubly-charged Higgs bosons decaying to same-sign lepton pairs.}
\PreprintJournalName{Phys. Rev. D}

\usepackage[hyperindex,breaklinks]{hyperref}
\usepackage{graphicx}
\usepackage{atlasphysics}
\usepackage{rotating} 
\usepackage{multirow}
\usepackage{subfigure}

\newcommand{\Ht}      {\ensuremath{H_{\mathrm{T}}^{\mathrm{jets}}}}
\newcommand{\Htjets}  {\Ht}
\newcommand{\St}      {\meff}
\newcommand{\st}      {\St}
\newcommand{\htlep}   {\ensuremath{H_{\mathrm{T}}^{\mathrm{leptons}}}}
\newcommand{\Htlep}   {\htlep}
\newcommand{\mt}      {\ensuremath{m_{\mathrm{T}}}}

\newcommand{\eg}      {\textit{e.g.}}

\newcommand{\ptmiss}  {\ensuremath{\mathbf{p}_{\mathrm{T}}^{\mathrm{miss}}}}

\newcommand{\ptvis}   {\ensuremath{p_{\mathrm{T}}^{\mathrm{vis}}}}
\newcommand{\etavis}  {\ensuremath{\eta^{\mathrm{vis}}}}

\newcommand{\Etisocal}   {\ensuremath{E_{\mathrm{T,cal}}^{\mathrm{iso}}}}
\newcommand{\ptisotrack}   {\ensuremath{p_{\mathrm{T,track}}^{\mathrm{iso}}}}

\newcommand{\Etisotruth}   {\ensuremath{E_{\mathrm{T,true}}^{\mathrm{iso}}}}
\newcommand{\ptisotruth}   {\ensuremath{p_{\mathrm{T,true}}^{\mathrm{iso}}}}

\newcommand{\meff}{\ensuremath{m_{\mathrm{eff}}}}

\newcommand{\alpgen}{{\sc alpgen}}
\newcommand{\madgraph}{{\sc MadGraph}}
\newcommand{\jimmy}{{\sc jimmy}}
\newcommand{\mcatnlo}{{\sc mc@nlo}}
\newcommand{\sherpa}{{\sc Sherpa}}
\newcommand{\hathor}{{\sc hathor}}
\newcommand{\powheg}{{\sc powheg-box}}
\newcommand{\herwig}{{\sc herwig}}
\newcommand{\pythia}{{\sc Pythia}}
\newcommand{\geant}{{\sc geant4}}
\newcommand{\fastjet}{{\sc FastJet}}
\newcommand{\antikt}{\ensuremath{\mathrm{anti}\text{-}k_{t}}}

\newcommand{\dchp}{\ensuremath{H^{\pm\pm}}}
\newcommand{\dchplh}{\ensuremath{H^{\pm\pm}_{L}}}

\newcommand{\DeltaR}{\ensuremath{\Delta R}}

\newcommand{\epsfid}{\ensuremath{\epsilon^{\mathrm{fid}}}}
\newcommand{\sigfid}{\ensuremath{\sigma^{\mathrm{fid}}}}

\newcommand{\sigmaB}{\ensuremath{\sigma\cdot\mathrm{BR}}}

\newcommand{\sigmavis}{\ensuremath{\sigma_{\mathrm{95}}^{\mathrm{vis}}}}

\newcommand{\ratiopaneldescription}{The bottom panel shows the ratio of events observed in data to those expected from background sources for each bin.}

\mathchardef\mhyphen="2D

\newcommand{\tauhad}{\ensuremath{\tau_\mathrm{had}}}

\newcommand{\tauh}    {\tauhad}

\newcommand{\threeL}  {\ensuremath{\geq3 e/\mu}}
\newcommand{\twoLtau} {\ensuremath{2 e/\mu + \geq1 \tauh}}
\newcommand{\twoLoneT}{\twoLtau}
\newcommand{\twoL}    {\ensuremath{2 e/\mu}}
\newcommand{\oneLtau} {\ensuremath{1 e/\mu + 1 \tauh}}
\newcommand{\oneLoneT}{\oneLtau}

\newcommand{\intLdt}  {\int L\mathrm{d}t}

\begin{document}

\preprint{CERN-PH-EP-2012-310}
\preprint{To be submitted to Physical Review D}

\title{Search for new phenomena in events with three charged
  leptons at $\sqrt{s}=7\TeV$ with the ATLAS detector} 

\author{The ATLAS Collaboration}

\date{\today}

\begin{abstract}
A generic search for anomalous production of events with at least three charged leptons is presented.  The search uses 
a $pp$-collision data sample at a center-of-mass energy of $\sqrt{s} = 7\TeV$ corresponding to 4.6~\ifb\ of integrated luminosity 
collected in 2011 by the ATLAS detector at the CERN Large Hadron Collider.  Events are required to contain at least two 
electrons or muons, while the third lepton may either be an additional electron or muon, or a hadronically decaying tau lepton.  
Events are categorized by the presence or absence of a reconstructed tau-lepton or $Z$-boson candidate 
decaying to leptons.  No significant excess above backgrounds 
expected from Standard Model processes is observed.  Results are presented as upper limits on event yields from non-Standard-Model 
processes producing at least three prompt, isolated leptons, given as functions of lower bounds on several kinematic variables.
Fiducial efficiencies for model testing are also provided.  The use of the results is illustrated by
setting upper limits on the production of doubly-charged Higgs bosons decaying to same-sign lepton pairs.
\end{abstract}


\maketitle

\newpage

\section{Introduction}
\label{sec:Introduction}
Events with more than two energetic, prompt, and isolated
 charged leptons are rarely produced at hadron colliders.  
Such events offer a clean probe of electroweak processes at high 
center-of-mass energies, and their production at enhanced rates above Standard Model
predictions would constitute evidence for new phenomena.  Models predicting 
events with multiple leptons in the final state include 
excited neutrino models~\cite{excitednu,excitednu2}, fourth-generation quark models~\cite{bprime}, the Zee--Babu neutrino mass model~\cite{zee1,zee2,babu},
supersymmetry~\cite{Miyazawa:1966,Ramond:1971gb,Golfand:1971iw,Neveu:1971rx,Neveu:1971iv,Gervais:1971ji,Volkov:1973ix,Wess:1973kz,Wess:1974tw}, and models with doubly-charged Higgs bosons~\cite{DCH0,DCH1}, including Higgs triplet models~\cite{DCH3,DCH2}.

The production of multilepton events in the Standard Model is dominated by 
 $WZ$ and $ZZ$ production, where both bosons decay leptonically.
Smaller contributions come from events with top-quark pairs produced in 
association with a $W$ or $Z$ boson, and from triboson production.  
Isolated but non-prompt lepton candidates misidentified as prompt arise in Drell--Yan events
produced in association with a photon 
that converts in the detector and is reconstructed as an electron.  
Prompt but non-isolated leptons misidentified as isolated can arise from Dalitz decays~\cite{dalitz,kroll}.  
Additional non-prompt, non-isolated leptons arise from heavy-flavor decays and from mesons 
that decay in flight. 
Fake leptons can arise from hadrons that satisfy the lepton identification criteria.

This paper presents a search for the anomalous production of events with 
at least three charged leptons in the final state.  The search uses a data 
set  collected in 2011 by the ATLAS detector 
at the CERN Large Hadron Collider (LHC) 
corresponding to 4.6~\ifb\ of 
$pp$ collisions at a center-of-mass energy of $\sqrt{s} = 7~\TeV$.  
Events are required to have at least two isolated electrons or muons, or one of each, while the third lepton
may be either an additional electron or muon or a hadronically decaying tau lepton (\tauh).

Searches for new phenomena at the LHC are challenged by large cross-section Standard Model processes
that overwhelm any events from rare interactions.  Such backgrounds must be reduced by triggers before storing 
event data for future study; these triggers should be highly efficient at selecting processes of interest while 
reducing the overall rate of events by orders of magnitude.  Additional requirements made on either leptons or 
event kinematics must likewise have both large background rejection factors and high efficiencies for events with 
real leptons.  The reconstruction and identification of \tauh\ candidates in a busy hadronic environment is
particularly challenging, requiring the use of sophisticated analysis techniques to reduce backgrounds from
parton-initiated jets.  The analysis presented here attempts to reduce the backgrounds from Standard Model
processes as much as possible, while retaining events that are potentially interesting for broad classes 
of new physics models.

Selected events are grouped into four categories 
by the presence or absence of
a \tauh\ candidate and by the presence or absence of a combination of leptons consistent with 
a $Z$-boson decay.
The search is carried out separately in each category by inspecting several variables of interest. 
The results of the search are presented 
as model-independent limits.  Efficiencies
for selecting leptons within the fiducial volume are also presented in order to aid the interpretation
of the results in the context of specific models of new phenomena.

Related searches for new phenomena in events with multilepton final states have not shown any
significant deviation from Standard Model expectations.
The CMS Collaboration has conducted a search similar to the one presented here using 4.98~\ifb\ 
of 7~\TeV\ data~\cite{cmsML}.  
The ATLAS Collaboration has performed a search for supersymmetry
in final states with three leptons~\cite{atlasSUSYML}, as have experiments at the Tevatron~\cite{cdfML,dzeroML}.  
The search presented here complements the previous searches
by providing limits outside of the context of a specific model of new phenomena.

This paper is organized as follows: the ATLAS detector is described in
Section~\ref{sec:Detector}, followed by a description of the samples and
event selection in Sections~\ref{sec:Samples} and~\ref{sec:Selection}, respectively.
The categorization of events and definition of signal regions is presented
in Section~\ref{sec:SignalRegions}.
The background estimation techniques and the results of the
application of those techniques in control regions are described in
Section~\ref{sec:Backgrounds}.
Systematic uncertainties are discussed in Section~\ref{sec:Systematics}.  The
results of the search are presented in Section~\ref{sec:Results}.
Fiducial efficiencies for model testing are provided in Section~\ref{sec:ModelTesting}, and are used to set upper limits
on the pair-production of doubly-charged Higgs bosons.

\section{The ATLAS detector}
\label{sec:Detector}
The ATLAS experiment~\cite{atlas} is a multipurpose particle physics
detector with a forward-backward symmetric cylindrical geometry and nearly 
4$\pi$ coverage in solid angle~\footnote{ATLAS uses a right-handed
  coordinate system with its origin at the nominal interaction point (IP)
  in the center of the detector and the $z$-axis along the beam pipe. The
  $x$-axis points from the IP to the center of the LHC ring, and the 
  $y$-axis points upward. Cylindrical coordinates $(r,\phi)$ are used in the
  transverse plane, $\phi$ being the azimuthal angle around the beam
  pipe. The pseudorapidity is defined in terms of the polar angle $\theta$
  as $\eta=-\ln\tan(\theta/2)$.  The variable $\DeltaR$ is used to evaluate
  the distance between objects, and is defined as: $\DeltaR =
  \sqrt{(\Delta\phi)^2 + (\Delta\eta)^2}$\label{geometryfootnote}}. The inner tracking detector
covers the pseudorapidity range \abseta $<$ 2.5, and consists of a silicon
pixel detector, a silicon microstrip detector (SCT), and, for \abseta $<$
2.0, a straw tube transition radiation tracker. The inner detector is surrounded by a
thin superconducting solenoid providing a 2 T magnetic field. 
The calorimeter system covers the pseudorapidity range $|\eta|<4.9$. Within the region $|\eta|<3.2$, 
electromagnetic calorimetry is provided by barrel and end-cap high-granularity lead liquid-argon (LAr) 
electromagnetic calorimeters, with an additional thin LAr presampler covering $|\eta|<1.8$, to correct 
for energy loss in material upstream of the calorimeters. Hadronic calorimetry is provided by the 
steel/scintillating-tile calorimeter, segmented into three barrel structures within $|\eta|<1.7$, and 
two copper/LAr hadronic endcap calorimeters. The solid angle coverage is completed with forward copper/LAr 
and tungsten/LAr calorimeter modules optimized for electromagnetic and hadronic measurements respectively. 
The muon
spectrometer surrounds the calorimeters. It consists of three large
air-core superconducting toroid systems with eight coils each and stations of precision tracking and trigger 
chambers providing accurate muon tracking for $\abseta<2.7$.
A three-level trigger system~\cite{trigger} is used to select events for further analysis offline.

\section{Monte Carlo simulation and data sets}
\label{sec:Samples}
Monte Carlo (MC) simulation samples are used to estimate backgrounds from events with three
prompt leptons. The ATLAS detector is simulated using \geant~\cite{geant},
and simulated events are reconstructed using the same software as that used for collision data.
Small post-reconstruction corrections are applied to account for differences
in efficiency, momentum resolution and scale, and energy resolution and scale between
data and simulation~\cite{elecperf,muonperf}.  

The largest Standard Model backgrounds with at least three prompt leptons are $WZ$ and $ZZ$
production where the bosons decay leptonically.  These processes are modeled with \sherpa\ 1.4.1~\cite{sherpa}.  
These samples include the case where 
the $Z$ boson (or $\gamma^{*}$) 
is off-shell, and the $\gamma^{*}$ has an invariant mass above twice the muon (tau) mass for 
$\gamma^{*}\to\mu\mu$ ($\gamma^{*}\to\tau\tau$), and above 100 MeV for $\gamma^{*}\to ee$.
Diagrams where a $\gamma^{*}$ is produced as radiation from 
a final-state lepton and decays to additional leptons, {\it i.e.} $W\to\ell^{*}\nu\to\ell\gamma^{*}\nu\to\ell\ell'\ell'\nu$
and $Z\to\ell\ell^{*}\to\ell\ell\gamma^{*}\to\ell\ell\ell'\ell'$, where $\ell$ and $\ell'$ need not have the same flavor, 
are also included.
The leading-order predictions from \sherpa\ are cross-checked with next-to-leading-order calculations
from \powheg\ 1.0~\cite{powheg}.  Diagrams including a Standard Model Higgs boson have negligible contributions
in all signal regions under study.

The production of $\ttbar+W/Z$ processes (also denoted $\ttbar+V$) is simulated with 
\madgraph\ 5.1.3.28~\cite{madgraph} for the matrix
element and \pythia\ 6.425~\cite{pythia} for the parton shower and fragmentation.  Corrections to the
normalization from higher-order effects for these samples are 20\% 
for $\ttbar+W$~\cite{ttW} and 30\% for $\ttbar+Z$~\cite{ttZ}.
Leptons from Drell--Yan processes produced in association with a photon that converts 
in the detector (denoted $Z+\gamma$ in the following) are modeled with \pythia.  
Additional samples are used to model dilepton backgrounds for 
control regions with fewer than three leptons.
Events from $\ttbar$ production are simulated with \mcatnlo\ 4.01~\cite{mcatnlo}, with \herwig\ 6.520~\cite{herwig} 
for the parton shower and fragmentation, and \jimmy\ 4.31~\cite{jimmy} for 
the underlying event.  Events from $W$+jets and $W+\gamma$ production are simulated 
with \alpgen\ 2.13~\cite{alpgen} for the matrix element, \herwig\ for the 
parton shower and fragmentation, and \jimmy\ for the underlying event.

Simulated samples of pair-produced doubly-charged Higgs bosons~\cite{DCH0,DCH1,DCH2}
are used to illustrate the results of this search in the context of a specific scenario.  
The doubly-charged Higgs bosons decay to pairs of same-sign leptons, producing up to
four energetic, prompt, isolated charged leptons in the final state.
The doubly-charged Higgs bosons are simulated with masses ranging from 100~\GeV\ to 500~\GeV.
A sample of pair-produced fourth-generation down-type quarks~\cite{bprime}
is also considered when estimating fiducial efficiencies and potential contributions from non-Standard-Model processes.
In this model, the heavy quarks decay to top quarks and $W$ bosons, producing four $W$ bosons and two bottom quarks.
This analysis is sensitive to the subset of such events in which at least three 
of the $W$ bosons decay leptonically.
The heavy quark is assumed to have a mass of 500~\GeV, corresponding to the approximate expected experimental limit.  
The normalization for this sample is provided at approximately 
next-to-next-to-leading-order accuracy by \hathor\ 1.2~\cite{hathor}.  
Both the doubly-charged Higgs boson and fourth-generation quark samples are generated with \pythia.

The parton distribution functions
for the \sherpa\ and \powheg\ samples are taken from CT10~\cite{ct10}, and from MRST2007 LO$^{**}$~\cite{mrst} 
for the \pythia\ and \herwig\ samples.
The \madgraph\ and \alpgen\ samples use CTEQ6L1~\cite{cteq6l1}.  The \mcatnlo\ sample uses CTEQ6.6~\cite{cteq66}.

Additional $pp$ interactions (pileup) in the same or nearby bunch crossings are modeled 
with \pythia.  Simulated events are reweighted to reproduce the distribution of $pp$ interactions per
crossing observed in data over the course of the 2011 run.  
The mean number of interactions per bunch crossing for the data was ten.
The luminosity has been measured with an uncertainty of $\pm$3.9\%~\cite{luminositypaper}.

\section{Event Selection}
\label{sec:Selection}
  Events are required to have
fired at least one single-electron or single-muon trigger.  The electron trigger requires a minimum threshold on the momentum
transverse to the beamline (\pt) of 20~\GeV\ for 
data collected in the early part of 2011, and 22~\GeV\ for data collected later in the year.  The muon 
\pt\ threshold is 18 \GeV\ for the full data set.
The efficiency of the trigger requirements for events satisfying all selection criteria ranges from 95\% to 99\% depending
on the signal region, and is evaluated with simulated $WZ$ events.  
In order to ensure that the efficiency is independent of the \pt\ of the leptons, the offline event selection requires 
that at least one lepton (electron or muon) has $\pt\geq25\GeV$.
At least one such lepton must also be consistent with having fired the relevant single-lepton trigger.
A muon associated with the trigger must lie within $|\eta|<2.4$ due to the limited acceptance of the muon trigger, while 
triggered electrons must lie within $|\eta|<2.47$, excluding the calorimeter barrel/end-cap transition region ($1.37 \leq |\eta| < 1.52$).
Additional muons in the event must lie within $|\eta|<2.5$ and have $\pt \geq 10$~\GeV.  
Additional electrons must satisfy the same $\eta$ requirements as triggered electrons, and must have 
$\pt\geq 10$~\GeV.
The third lepton in the event may be an additional electron or muon satisfying the same requirements
as the second lepton, or a \tauh\ with 
$\ptvis \geq 15$~\GeV\ and $|\etavis| < 2.5$, where $\ptvis$ and $\etavis$ denote the 
\pt\ and $\eta$ of the visible products of the tau decay, with no corrections for
the momentum carried by neutrinos.  Throughout this paper the four-momenta of tau candidates are defined
only by the visible decay products.

All parts of the detector are required to have been operating properly
for the events under study.  
Events must have a reconstructed primary vertex candidate with at
least three associated tracks, where each track must have $\pt>0.4\GeV$.  
In events with multiple primary vertex candidates, the primary vertex is 
chosen to be the one  with the largest value of $\Sigma{p_{\mathrm{T}}^{2}}$, where the 
sum is taken over all reconstructed tracks associated with the vertex.  Events
with pairs of leptons that are of the same flavor but opposite sign and have an invariant
mass below 20~\GeV\ are excluded to avoid contributions from low-mass hadronic resonances.

The lepton selection includes requirements to reduce the contributions
from non-prompt or fake lepton candidates.  These requirements exploit 
the transverse and longitudinal impact parameters of their tracks with respect to the primary
vertex, the isolation of the lepton candidates from nearby hadronic activity, and
in the case of electron and \tauh\ candidates, the lateral and longitudinal profiles
of the shower in the electromagnetic calorimeter.  There are also requirements for electrons on the quality
of the reconstructed track and its match to the cluster in the calorimeter.  These requirements 
are described in more detail below.

Electron candidates are required to satisfy the ``tight'' identification criteria described in Ref.~\cite{elecperf}, updated
for the increased pileup in the 2011 data set.  
Muons must have tracks with hits in both the inner tracking detector and muon spectrometer, and must
satisfy criteria on track quality described in Ref.~\cite{muonperf}.

The transverse impact parameter significance is defined as $|d_{0}/\sigma(d_{0})|$, where 
$d_0$ is the transverse impact parameter of the reconstructed track with respect to the primary vertex and
$\sigma(d_0)$ is the estimated uncertainty on $d_0$.  This quantity must be less than 3.0 
for muon candidates.  Electrons must satisfy a looser cut of $|d_{0}/\sigma(d_{0})|<10$, since 
interactions with material in the inner tracking detector often reduce the quality of the reconstructed track.
The longitudinal impact parameter $z_{0}$ must satisfy $|z_{0} \sin(\theta)|<1$~mm for both electrons
and muons.

Electrons and muons are required to be isolated through the use of two variables sensitive to
the amount of hadronic activity near the candidate.  The first,
\ptisotrack, is the 
scalar sum of the transverse momenta of all tracks with $\pt \geq 1$~\GeV\ in a cone
of $\Delta R<0.3$ around the lepton axis.
The sum excludes the 
track associated with the lepton candidate, and also excludes tracks inconsistent with originating from  
the primary vertex.  
The second, \Etisocal, is the sum of the transverse energies of 
cells in the electromagnetic
and hadronic calorimeters in a cone of the same size.  For electron candidates 
this sum excludes a rectangular region around the candidate axis 
of $0.125\times 0.172$ in $\eta\times\phi$ (corresponding to $5\times7$ cells in the main sampling
layer of the electromagnetic calorimeter) and is 
corrected for the imperfect containment of the electron transverse energy 
within the excluded region.  
For muons, the sum
only includes cells above a certain threshold in order to suppress noise, and does not include
cells with energy deposits from the muon candidate.  For both electrons and muons,
the value of \Etisocal\ is corrected for the expected effects of pileup
interactions.  Muon candidates are required to have $\ptisotrack/\pt < 0.13$ and
$\Etisocal/\pt < 0.14$, while electron candidates are required to have $\ptisotrack/\pt < 0.15$
and $\Etisocal/\pt < 0.14$; see Ref.~\cite{atlasHWW} for the optimization of these requirements.

Jets in the event are reconstructed using the \fastjet~\cite{fastjet} implementation of
 the \antikt\ algorithm~\cite{antikt}, with distance parameter $R=0.4$.  The jet four-momenta are corrected
for the non-compensating nature of the calorimeter, for inactive material in front of the
calorimeters, and for pileup~\cite{jetcorr,JER}.
Jets used in this analysis are required to have $\pt\geq25$~\GeV\ and lie within $|\eta|<4.9$.  
Jets within the acceptance of the inner tracking detector must fulfill a requirement, based on tracking information, that they
originate from the primary vertex.
The missing transverse momentum, \ptmiss, is defined as the negative vector sum of the transverse momenta 
of reconstructed
jets, leptons, and any remaining calorimeter clusters unassociated with reconstructed objects.  The magnitude
of \ptmiss\ is denoted \met.

Tau leptons decaying to an electron (muon) and neutrinos
are selected with the nominal identification criteria described above, and
are classified as electrons (muons).
Hadronically decaying tau candidates are constructed from jet candidates and are then selected using a boosted decision tree (BDT), which is
trained to distinguish hadronically decaying tau leptons from quark-  and gluon-initiated jets~\cite{tauid}.
The BDT is trained separately for tau candidates with one and three charged decay products, referred to as ``one-prong'' and ``three-prong'' taus, respectively.
In this analysis, only one-prong $\tauh$ candidates satisfying the ``tight'' working point criteria are considered.
This working point is roughly 35\% efficient for one-prong $\tauh$ candidates originating 
from $W$-boson or $Z$-boson decays, and has a jet rejection factor of roughly 300.
Additional requirements to remove \tauh\ candidates initiated by prompt 
electrons or muons are also imposed.  
A BDT trained to discriminate between electron-initiated \tauh\ candidates and true \tauh\ candidates provides
a factor of roughly 400 in rejection at 90\% efficiency.  Muon-initiated \tauh\ candidates are identified with a cut-based method,
which achieves a factor of two in rejection at 96\% efficiency.  The identification of both electron- and muon-initiated \tauh\ candidates
is discussed further in Ref.~\cite{tauid}.

Since lepton and jet candidates can be reconstructed as multiple objects, the following logic is applied to remove overlaps.
If two electrons are separated by $\Delta R<0.1$, the
candidate with lower \pt\ is neglected.  If a jet lies within $\Delta R=0.2$ of an electron or $\tauh$ 
candidate, the jet is neglected, while if the separation of the jet from an electron candidate satisfies $0.2\leq\Delta R<0.4$,
the electron is neglected.  In addition, electrons within $\Delta R=0.1$ of a muon are also neglected, as
are $\tauh$ candidates within $\Delta R=0.2$ of electron or muon candidates.  Finally, muon
candidates with a jet within $\Delta R=0.4$ are neglected.

\section{Signal Regions}
\label{sec:SignalRegions}
Events satisfying all selection criteria are classified into four categories.  Events in which at least three of the lepton candidates 
are electrons or muons are selected first, followed by events with two electrons or muons, or one of each, 
and at least one \tauh\ candidate.
These two categories are referred to 
as \threeL\ and \twoLtau\ respectively.
Next, events in each of those two categories
are sub-divided by the presence or absence of a reconstructed $Z$-boson candidate,
which is defined as an opposite-sign
same-flavor pair of lepton candidates with a total invariant mass within $\pm$20~\GeV\ of 
the $Z$-boson mass~\cite{pdg}.  An additional electron may also be included in the combination with the same-flavor opposite-sign pair to satisfy
the invariant mass requirement, to handle cases where an energetic photon from final-state radiation converts in the detector and is reconstructed
as a prompt electron.
Events with a reconstructed $Z$-boson candidate are referred to as on-$Z$, and those without such a candidate
are referred to as off-$Z$.  
The resulting four categories are mutually exclusive, and
are chosen to isolate the contributions from backgrounds such as
jets faking $\tauh$ candidates, and events with $Z$ bosons produced in 
association with a jet that fakes a prompt lepton.  In order to remain independent of
the $Z$+jets control region described in Section~\ref{sec:Backgrounds}, the on-$Z$ regions have
a minimum \met\ requirement of 20~\GeV.

Several kinematic variables are used to characterize the events that satisfy all selection criteria.
The variable \Htlep\ is defined as the scalar sum of transverse momenta, or \ptvis\ for $\tauh$ candidates,
of the three leading leptons.
The variable \Htjets\ is defined as the sum of transverse momenta of all selected jets in the event.  
The ``effective mass'', \St, is the scalar sum of \met, \Htjets, and the
transverse momenta of all identified leptons in the event.  

Subsets of selected events are defined based on  kinematic properties.  The \Htlep\ distribution
is considered for all events in each category.  The \met\ distribution is considered 
separately for events with \Htjets\ below and above 100~\GeV, which serves to separate
events produced through weak and strong interactions.  The \St\ distribution is
considered  for events with  and without a requirement of $\met \geq 75$~\GeV.  Increasing lower bounds
on the value of each kinematic variable define signal regions; the lower bounds are shown in 
Table~\ref{tab:signal_regions}.

\begin{table}[tbp]
  \begin{center}
    \caption{Kinematic signal regions defined in the analysis.  The on-$Z$ regions have an additional requirement of 
    $\met~>~20~\GeV$.}
    \label{tab:signal_regions}
    \begin{tabular}{l l l}
      \hline
      \hline
      Variable     &Lower Bounds [GeV]               &Additional Requirement\\
      \hline       
      \htlep       &0, 100, 150, 200, 300            &\\
      \met         &0, 50, 75                        &$\Ht < 100$ \GeV\\
      \met         &0, 50, 75                        &$\Ht \geq 100$ \GeV\\
      \st          &0, 150, 300, 500                 &\\
      \st          &0, 150, 300, 500                 &$\met{} \geq 75$ \GeV\\
      \hline
      \hline
    \end{tabular}
  \end{center}
\end{table}

\section{Background estimation}
\label{sec:Backgrounds}
Standard Model processes that produce events with three lepton candidates fall into three classes.  The first 
consists of events in which prompt leptons are produced in the hard
interaction, including the $WZ$, $ZZ$, and $\ttbar+W/Z$ processes.
A second class of events includes Drell--Yan production in association
with an energetic $\gamma$, which then converts in the detector to produce
a single reconstructed electron.  A third class of events
arises from non-prompt, non-isolated, or fake lepton candidates satisfying the identification
criteria described in Section~\ref{sec:Selection}.

The first class of backgrounds is dominated by 
$WZ\to\ell\nu\ell'\ell'$ and 
$ZZ\to\ell\ell\ell'\ell'$ events.  Smaller contributions
come from $\ttbar+W \to b\bar{b}\ell\nu\ell'\nu\ell''\nu$ and 
$\ttbar+Z \to b\bar{b} \ell\nu\ell'\nu\ell''\ell''$ events.  Contributions
from triboson events, such as $WWW\to\ell\nu\ell'\nu\ell''\nu$ production, are negligible.  All such
processes are modeled with the dedicated MC samples described in Section~\ref{sec:Samples}.  
Reconstructed leptons in the simulated samples are required
to be  consistent with the decay of a vector boson or tau lepton from 
the hard interaction.
The second class of backgrounds, from Drell--Yan production in association
with a hard photon, is also modeled with MC simulation.

The class of events that includes non-prompt or fake leptons, referred to here
as the reducible background, is estimated
using {\it in-situ} techniques which rely minimally on simulation.  
Such backgrounds for muons arise from semi-leptonic $b$- or $c$-hadron decays, from in-flight decays 
of pions or kaons, and from energetic jets that reach the muon spectrometer.
Electron candidates can also arise from misidentified hadrons or jets.
Hadronically decaying taus have large backgrounds from narrow, low-track-multiplicity jets that mimic $\tauh$ signatures.

Relaxed criteria are defined for each lepton flavor.  These criteria, in
combination with a requirement that candidates fail the nominal identification
criteria, produce samples of lepton candidates that are rich in background
with minimal contributions from misidentified prompt leptons.  For electrons
and muons, the isolation criteria are relaxed to accept non-isolated leptons.
Electrons are also allowed to fail the ``tight'' electron identification criteria,
provided they satisfy the ``medium'' criteria~\cite{elecperf}. The relaxed $\tauh$ identification loosens the
 requirement on the BDT score.

These samples of events are used to measure the ratio of the number of leptons
satisfying the nominal identification criteria to the number that fail the nominal criteria but satisfy the relaxed criteria.
This ratio can then be applied as a scale factor -- referred to here as a ``fake factor'' --
to multilepton events satisfying the relaxed criteria
to estimate the background in signal regions.
For electron and muon candidates, the sample used to measure the fake factor
consists of events that pass the high-\pt\ single-lepton triggers described in Section~\ref{sec:Selection}.  
Events with more than one
selected lepton are removed from the sample to avoid overlap with the signal region, and to
reduce the contamination from Drell--Yan processes.
Muons must also fail the
nominal requirement on $|d_{0}/\sigma(d_{0})|$ to further remove prompt contributions.
Finally, events where the transverse mass (\mt) of the electron combined with the
\met\ is larger than 25 \GeV\ are also rejected to avoid contamination from $W$+jets, where \mt\ is
defined as:
\begin{equation}
  \mt\ \equiv \sqrt{(E_{\mathrm{T}}^{\ell}+\met)^{2} - \vert\mathbf{p}_{\mathrm{T}}^{\ell}+\ptmiss\vert^{2}}.
\end{equation}
For events with muons\ the transverse mass requirement is relaxed to 40~\GeV\, since the inversion of the
$|d_{0}/\sigma(d_{0})|$ requirement is sufficient to remove most of the contributions from $W$-boson decays.

For $\tauh$ candidates, a sample of $\gamma$+jet events is used to measure the fake factors.
The production of prompt photons in $pp$ collisions is dominated by the Compton process
$qg\to q\gamma$, yielding a sample of $\gamma$+jet events that is rich in quark-initiated jets.
In the events considered here, an energetic photon is used to tag the event, and the $\tauh$ candidate is 
the away-side jet.  The photon is required to have $\pt>40$~\GeV, and satisfy 
the ``tight'' identification criteria~\cite{atlasphoton}.  The photon
candidate is also required to have $\Etisocal<5$~\GeV.  These criteria have been shown
to yield a mostly pure sample of photon candidates, with the remainder largely consisting
of events in which a jet fragments into a leading $\pi^{0}$ that then decays to two photons.
The resulting sample suffers from minimal contamination from true $\tauh$ candidates, with
the largest contribution from $W(\to\tauh\nu)$+jet events, where the jet is identified as a photon,
contributing less than 1\% to the total sample.

The fake factors for all flavors are parameterized as functions of the \pt\ and $|\eta|$
of the candidates, to account for changes in the composition of the nominal and relaxed
samples in different kinematic ranges.  For 
electrons and muons with $\pt>100$~\GeV, the fake factor is computed from a
linear extrapolation of the fake factors between 35~\GeV\ and 100~\GeV.
An additional parameterization is added to
account for the heavy-flavor content of the event based on the output of the MV1 $b$-tagging algorithm.
The MV1 algorithm uses a neural network to identify $b$-jets based on the outputs of several secondary-vertex and
three-dimensional-impact-parameter taggers, which are described in detail in Ref~\cite{mv1}.
The largest MV1 score associated with any jet in the event is used to parameterize the fake factors.
The correlation
of this variable with the use of the inverted $|d_{0}/\sigma(d_{0})|$ requirement when estimating the muon fake factors 
leads to a bias in events with large MV1 scores, causing the muon fake factors to be underestimated
by a factor of two.  This bias is corrected using MC simulated samples. 

Contributions from prompt leptons can bias the reducible background estimates in two ways.  The first arises 
when prompt leptons populate either the tight or relaxed regions when deriving the fake factors.
The second arises when prompt leptons populate the relaxed region when applying the fake factors.  In 
all cases, the effects of prompt leptons on the reducible background estimates are evaluated and corrected using
MC simulation.  

The background estimates are tested in several control regions.
A control region rich in events with a $Z$ boson produced in association with a jet
is defined to test the reducible background
estimates.  Events in this region have three identified lepton candidates,
with the requirement that a pair of opposite-sign, same-flavor leptons has
an invariant mass within $\pm$20~\GeV\ of the $Z$-boson mass.  The additional requirement
that the \met\ does not exceed 20~\GeV\ avoids overlap with the signal regions.  This is referred
to as the $Z$+jets region.

A second control region, also consisting of events with three lepton candidates, 
is defined using the low-mass Drell--Yan events rejected by
the requirement that no opposite-sign, same-flavor lepton pair have $m(\ell^+\ell^-)<20\GeV$.
This region is referred to as the low-mass Drell--Yan region.

A third region is defined in order to probe the estimates of backgrounds from non-prompt and non-isolated sources 
in events rich in heavy-flavor decays.  Events are required to have exactly two same-sign leptons and 
$\met > 40$~\GeV.   Events are further required to have a $b$-jet candidate selected by the MV1 tagging algorithm,
using a working point that is 60\% efficient and that has a light-jet mis-tag rate of less than 1\% for jets with
$\pt<100\GeV$.
This sample is estimated to be primarily composed of lepton+jets
$\ttbar$ events.   The same-sign requirement suppresses events where both $W$ bosons decay leptonically,
and enhances the contributions from events where one lepton candidate originates from semileptonic $b$ decay.
This region is referred to as the $\ttbar$ region.  
An upper limit on \Htjets\ of 300~\GeV\ reduces potential contamination from new phenomena.

Good agreement between the expected and observed
event yields is seen in all control regions as  shown in Table~\ref{tab:controlregions}.  
Figure~\ref{fig:Z_emu_OffZMT} shows the \mt\ distribution of the \met\ and the lepton not associated with the $Z$ boson candidate
in the \threeL\ channel of the $Z$+jets region.  
Figure~\ref{fig:Z_SubsublPt} shows the \pt\ distribution for the 
third lepton candidate in the $Z$+jets region.  
The \pt\ distribution for the subleading lepton ($\tauh$ candidate) in the \ttbar\ region is shown in 
Fig.~\ref{fig:CR_Top_subleadpt}.  The \pt\ distribution for the third lepton 
in the low-mass Drell--Yan region is shown in 
Fig.~\ref{fig:CR_upsilon_subsublpt}.  The \Htlep, \met, and \meff\ distributions are not shown here, but also are in good agreement in the control regions.
The contributions from new phenomena in the control regions are estimated with doubly-charged Higgs and fourth-generation quark events.  
An example of
such contamination is shown with fourth-generation quark events in 
Fig.~\ref{fig:CR_Top_emu_subleadpt}, where the contamination is small.  The contributions in all other control regions 
from events with pair-produced 
doubly-charged Higgs bosons or fourth-generation quarks are negligible.

\begin{table}[tbp]
\begin{center}
  \caption{The predicted and observed number of events in the $Z$+jets, low-mass Drell--Yan, and $\ttbar$
  control regions.  The $Z$+jets and low-mass Drell--Yan regions are populated by trilepton events,
  while the $\ttbar$ region is composed of same-sign dilepton events.  Statistical and systematic uncertainties on the expected
  event yields are combined as described in Section~\ref{sec:Systematics}.}
 \label{tab:controlregions} 
 \begin{tabular}{l c c c c}
   \hline\hline
   Channel     &Irreducible                            &Reducible                              &Total                                  &Observed\\
   \hline
   \multicolumn{5}{c}{$Z$+jets}\\
   \hline
   \threeL     &165~$\pm$~26                         &160~$\pm$~50                  &320~$\pm$~60                 &359\\
   \twoLtau    &3.0~$\pm$~0.6                        &1480~$\pm$~360                &1480~$\pm$~360               &1696\\
   \hline
   \multicolumn{5}{c}{Low-mass Drell--Yan}\\
   \hline
   \threeL     &55~$\pm$~9                           &34~$\pm$~12                   &89~$\pm$~15                   &101\\
   \twoLoneT   &0.5~$\pm$~0.1                       &91~$\pm$~23                   &92~$\pm$~23                   &96\\
   \hline
   \multicolumn{5}{c}{$\ttbar$}\\
   \hline
   \twoL       & 25 $\pm$ 4                            &58 $\pm$ 23                            & 83 $\pm$ 23                           & 87\\
   \oneLoneT   & 1.9 $\pm$ 0.4                         &107 $\pm$ 27                           & 109 $\pm$ 27                          & 103\\
   \hline 
   \hline
 \end{tabular} 
 \end{center} 
\end{table} 

\begin{figure}[tbp]
  \begin{center}
    \includegraphics[width=\columnwidth]{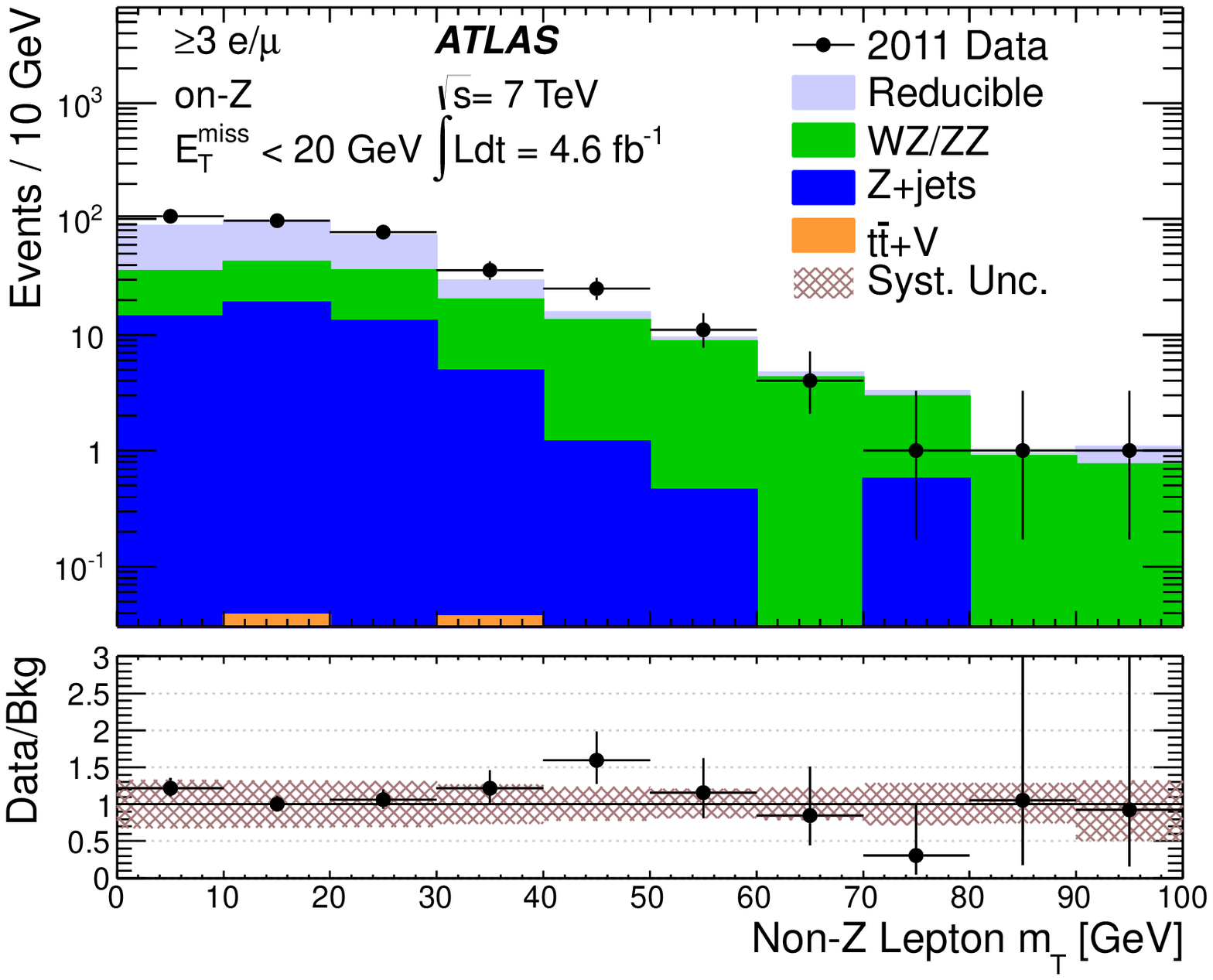}
    \caption{The $\mt$ distribution of the \met\ and the lepton not associated with the $Z$-boson candidate decay in 
      $\geq 3 e/\mu$ events in the $Z$+jets control region.  The last bin shows the integral of events
    above 90~\GeV.  \ratiopaneldescription}
    \label{fig:Z_emu_OffZMT}
  \end{center}
\end{figure}

\begin{figure*}[tbp]
  \begin{center}
    \subfigure[]{\includegraphics[width=.48\textwidth]{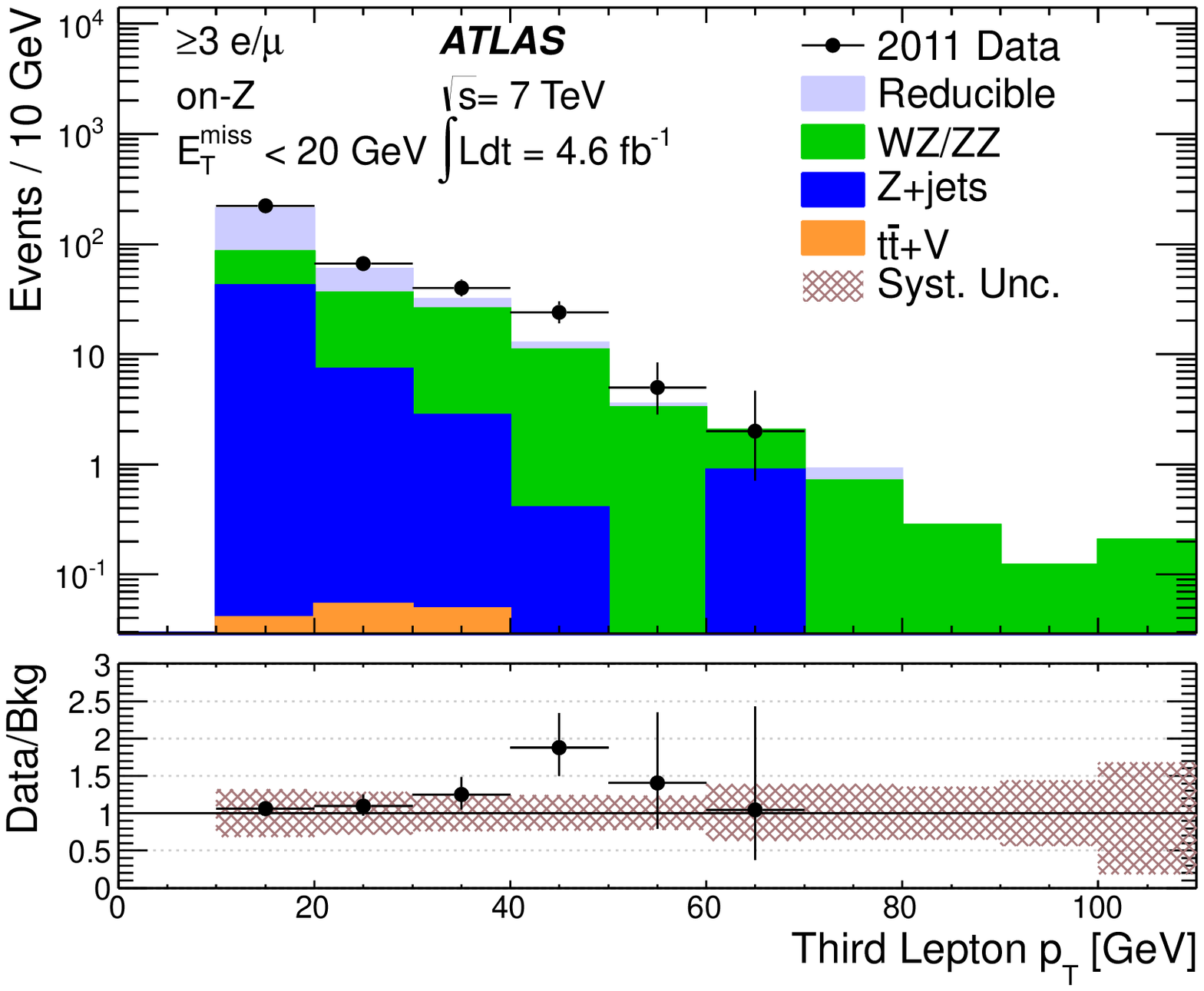}\label{fig:Z_emu_SubsublPt}}
    \subfigure[]{\includegraphics[width=.48\textwidth]{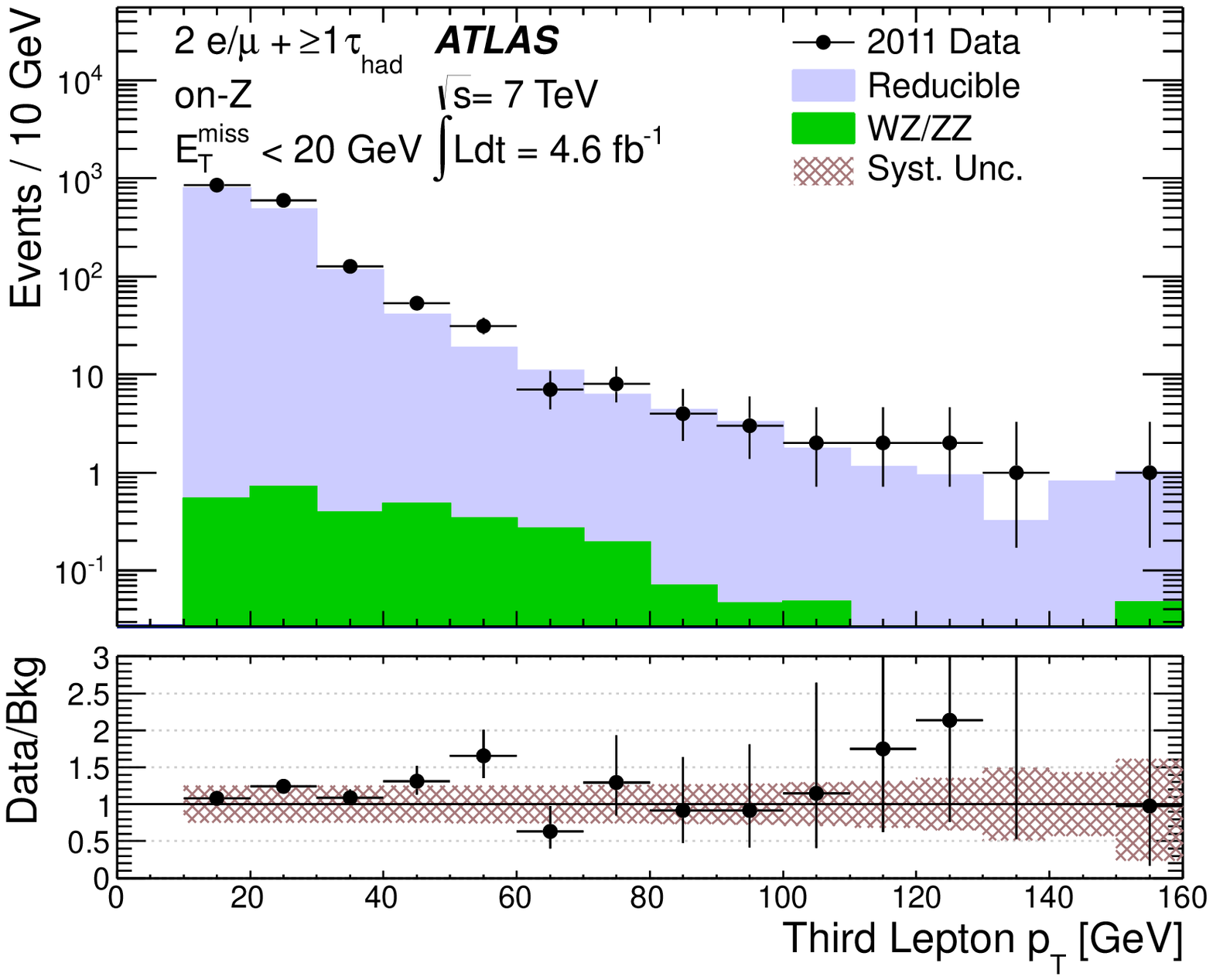}\label{fig:Z_emu_tau_SubsublPt}}
    \caption{The $\pt$ distribution of the third lepton candidate in (a) \threeL\ events and (b) \twoLoneT\ events in the $Z$+jets control region.  
      The last bin in the left (right) plot shows the integral of events
    above 100~\GeV\ (150~\GeV).  \ratiopaneldescription}
    \label{fig:Z_SubsublPt}
  \end{center}
\end{figure*}

\begin{figure*}[tbp]
  \begin{center}
    \subfigure[]{\includegraphics[width=\columnwidth]{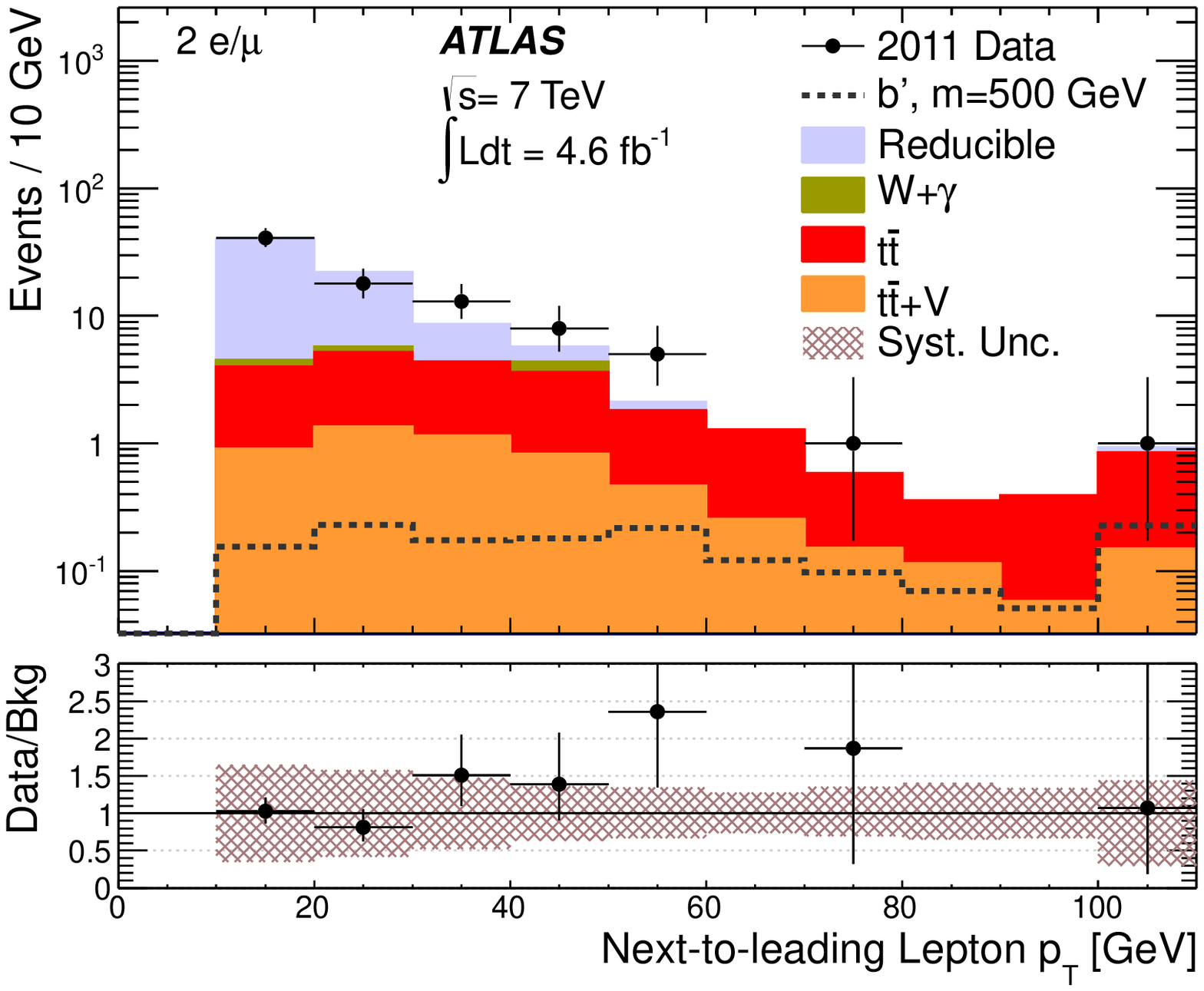}\label{fig:CR_Top_emu_subleadpt}}
    \subfigure[]{\includegraphics[width=\columnwidth]{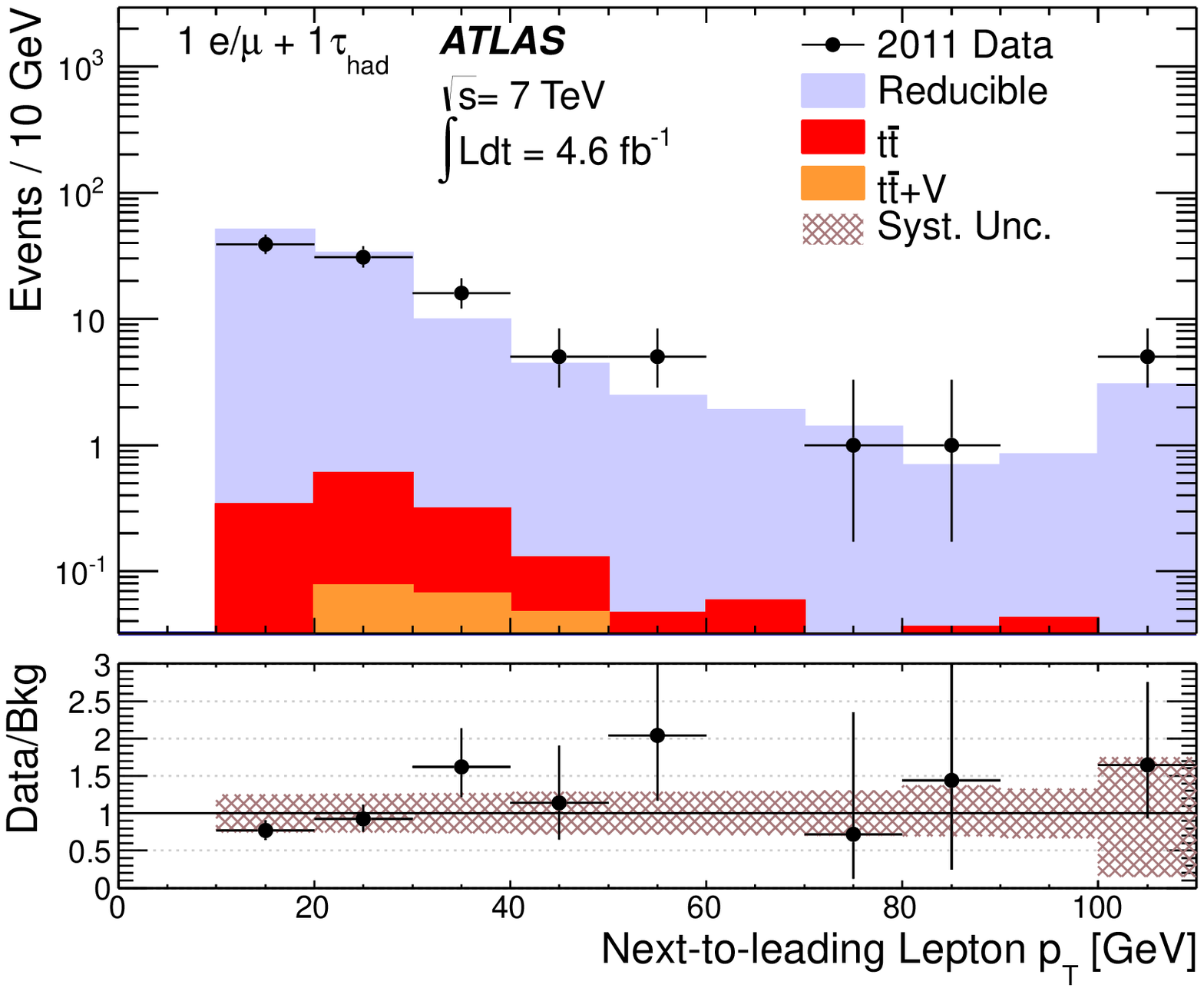}\label{fig:CR_Top_tau_subleadpt}}
    \caption{The $\pt$ distribution of the (a) subleading lepton in $2 e/\mu$ events and (b) \tauh\ in $1 e/\mu + 1 \tauh$ events in the $\ttbar$ control region.  The expected 
    contribution from non-Standard-Model processes is illustrated in the left figure by events with fourth-generation
    down-type quarks ($b'$).  The contribution from $b'$ events in the right figure is negligible.  
    The last bin in each plot shows the integral of events above 100~\GeV.  \ratiopaneldescription}
    \label{fig:CR_Top_subleadpt}
  \end{center}
\end{figure*}

\begin{figure*}[tbp]
  \begin{center}
    \subfigure[]{\includegraphics[width=\columnwidth]{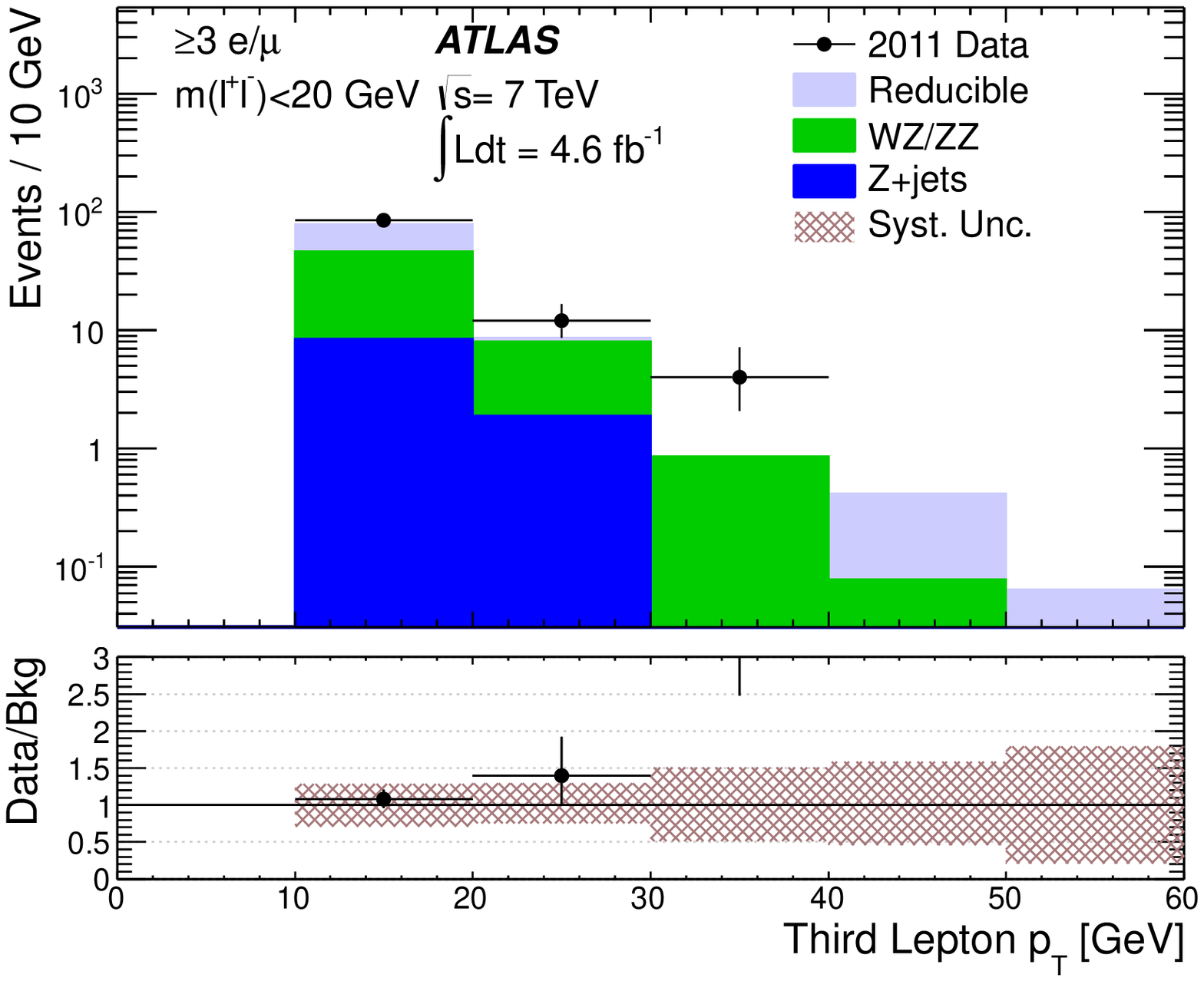}\label{fig:CR_upsilon_emu_subsublpt}}
    \subfigure[]{\includegraphics[width=\columnwidth]{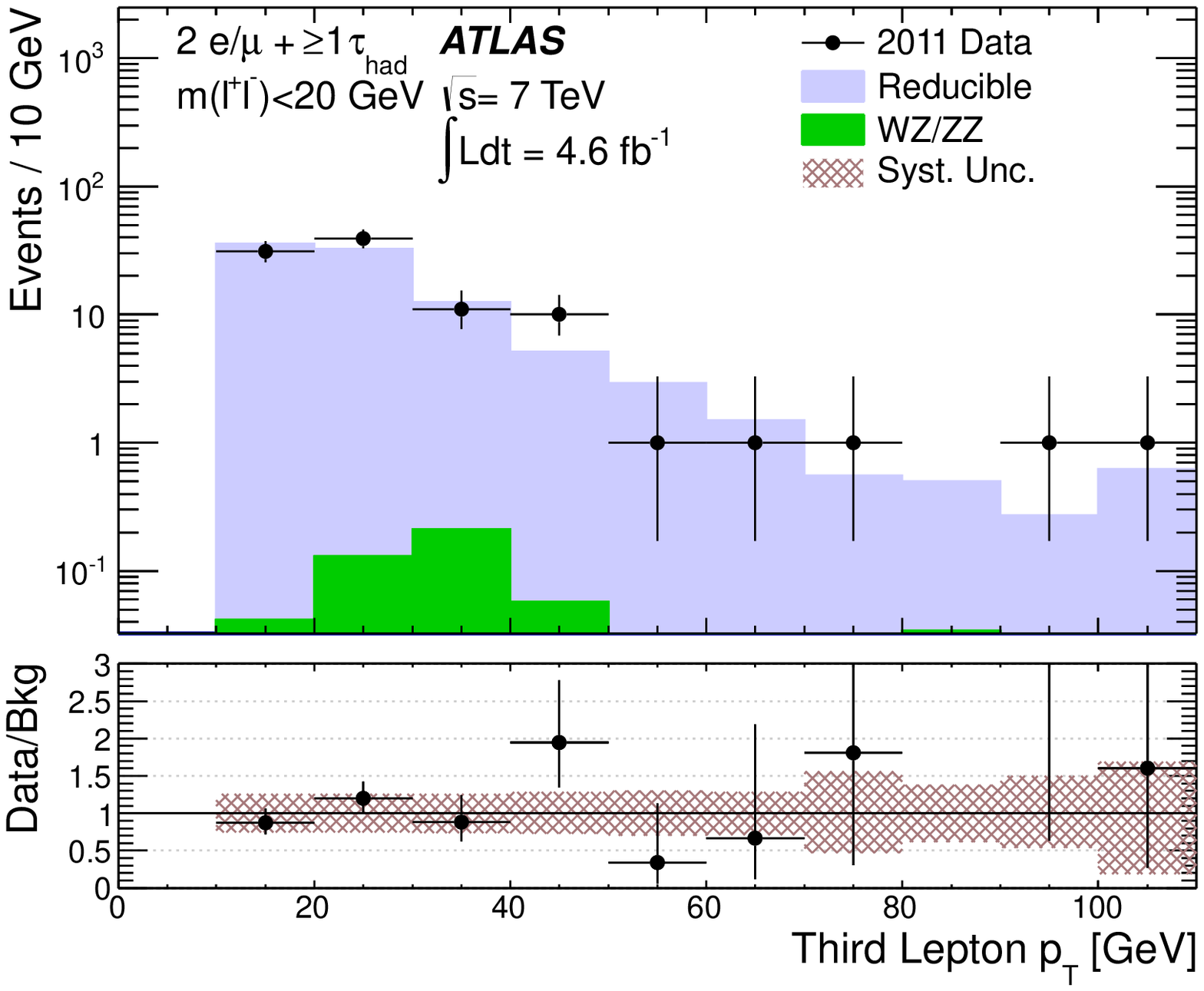}\label{fig:CR_upsilon_tau_subsublpt}}
    \caption{The $\pt$ distribution of the (a) third lepton candidate in \threeL\ events and (b) \tauh\ in \twoLoneT\ events in the low-mass Drell--Yan control region.  
      The last bin shows the integral of events
    above 50~\GeV\ (100~\GeV) in the left (right) figure.  \ratiopaneldescription}
    \label{fig:CR_upsilon_subsublpt}
  \end{center}
\end{figure*}

\section{Systematic uncertainties}
\label{sec:Systematics}
Systematic uncertainties on the predicted backgrounds come from several sources.  These uncertainties are summarized in 
Table~\ref{tab:MCsysts},
presented as ranges of relative uncertainties on the total expected background yields across all signal 
regions and channels.

\begin{table}[htbp]
  \begin{center}
  \caption{The range of systematic uncertainties originating from different sources,
    presented as the relative uncertainty on the total expected background yield in all 
    signal regions under study.  In cases where a source of uncertainty contributes 
    less than 1\% of total uncertainty in any of the signal regions, the minimum is presented 
    as ``$\leq 1$\%''.}
  \label{tab:MCsysts}
  \begin{tabular}{l r r}
    \hline
    \hline
    Source of uncertainty                 &\multicolumn{2}{c}{Uncertainty}\\
    \hline
    Trigger efficiency                    &$(\leq 1)$ -- &1\%\\
    \hline
    Electron energy scale                 &$(\leq 1)$ -- &13\%\\
    Electron energy resolution            &$(\leq 1)$ -- &1\%\\
    Electron identification               &$(\leq 1)$ -- &3\%\\
    Electron non-prompt/fake backgrounds  &$(\leq 1)$ -- &13\%\\
    \hline
    Muon momentum scale                   &$(\leq 1)$ -- &1\%\\
    Muon momentum resolution              &$(\leq 1)$ -- &7\%\\
    Muon identification                   &$(\leq 1)$ -- &1\%\\
    Muon non-prompt/fake backgrounds      &$(\leq 1)$ -- &51\%\\
    \hline
    Tau energy scale                      &$(\leq 1)$ -- &4\%\\
    Tau identification                    &$(\leq 1)$ -- &4\%\\
    Tau non-prompt/fake backgrounds       &$(\leq 1)$ -- &24\%\\
    \hline
    Jet energy scale                      &$(\leq 1)$ -- &6\%\\
    Jet energy resolution                 &$(\leq 1)$ -- &3\%\\
    \hline
    Soft \met\ terms                      &$(\leq 1)$ -- &14\%\\
    \hline
    Luminosity                            &\multicolumn{2}{r}{3.9\%}\\
    \hline
    Cross-section uncertainties           &$(\leq 1)$ -- &14\%\\
    \hline
    Statistical uncertainties             &1 -- &25\%\\
    \hline
    \hline
    Total uncertainty                     &11 -- &56\%\\
    \hline
    \hline
  \end{tabular}
  \end{center}
\end{table}

The backgrounds modeled
with simulated samples have uncertainties associated with trigger efficiencies, lepton efficiencies, lepton momentum scales and
resolution, and jet energy scales and resolution.  The uncertainty on the \met\ in simulation is computed from varying the
inputs to the \met\ calculation within their uncertainties on the energy/momentum scale and resolution,
and is thus strongly correlated with the other uncertainties and not presented separately.  Contributions to the 
\met\ from soft activity not associated with high-\pt\ objects are presented separately.  Uncertainties on the jet energy scale and resolution
are significant in regions requiring large values of \Htjets\ or \St, and are small otherwise.

Uncertainties on the cross sections of the different Standard Model processes modeled by simulation
are also considered.  
The \sherpa\ predictions of the $WZ$ and $ZZ$ processes are cross-checked with the next-to-leading-order 
predictions from \powheg\ in a kinematic region similar to the signal regions considered in this search,
resulting in 10\% and 25\% uncertainties in the normalization, respectively.  
Uncertainties from renormalization and factorization scale variations, as well as the 
variation of the parton distribution functions, contribute an additional 10\% and 7\% respectively,
taken from Ref.~\cite{samesign}.
The $\ttbar+W$ and $\ttbar+Z$ backgrounds carry a total uncertainty of 50\%\ based on parton distribution function and scale variations, and
on large higher-order corrections~\cite{ttW,ttZ}.  The Drell--Yan samples have a total 
uncertainty of 7\%~\cite{Zprime}.

The reducible background estimates carry large uncertainties from several sources.  A 40\%
uncertainty is assigned to the fake factors used to estimate the reducible electron and muon backgrounds,
based on closure studies in MC samples and cross-checks in control regions.
For electrons and muons 
with $\pt>100\GeV$, where the fake factors are extrapolated from the values at lower
\pt, a 100\% uncertainty is assigned.  A 100\% uncertainty is also assigned to the fake factors for muons with 
high $b$-tagging scores, due to the large correction taken from MC simulation to remove the bias between
the $b$-tagging algorithm and the inverted $d_{0}$ requirement.  
For the $\tauh$ fake estimates, a 25\% uncertainty on the fake factors is determined by altering the 
composition of the relaxed sample.  In signal regions where the relaxed
samples are poorly populated, statistical uncertainties on the reducible background estimates
become significant, especially in regions with high \met\ or \Htjets\ requirements.

In all of the signal regions under study, the dominant systematic uncertainties on the total background estimate 
arise from the uncertainty associated
with the reducible background estimates or from the uncertainty on the cross sections used for backgrounds taken
from MC simulation.

Uncertainties on the efficiency for potential sources of new phenomena include contributions from lepton trigger and identification efficiencies, 
and lepton momentum scale and resolution.  Larger uncertainties on the signal efficiency are assigned based on variations observed 
between several simulated samples, including pair-production of doubly-charged Higgs bosons and of fourth-generation down-type quarks, and are 10\% for the \threeL\ channels, and 20\% for the \twoLoneT\ channels.

\section{Results}
\label{sec:Results}
Event yields for the most inclusive signal regions in each search channel are presented
in Table~\ref{tab:event_yield_summary}.
No significant deviation from the expected background is observed.  The yields for all signal regions
are presented in Tables~\ref{t:Result_All_HTLep}--\ref{t:Result_HighMET_ST} of Appendix~\ref{app:yields}.

\begin{table}[tbp]
  \begin{center}
    \caption{The expected and observed event yields for all inclusive signal channels.  The expected yields are presented with two uncertainties, 
      the first is the statistical uncertainty, and the second is the systematic uncertainty.}
    \label{tab:event_yield_summary}
    \begin{tabular}{c c r r r@{\hskip 0.3in}  c}
      \hline
      \hline
      Flavor Chan.  &\multicolumn{1}{c}{$Z$ Chan.}          &\multicolumn{3}{c}{Expected}	         &\multicolumn{1}{c}{Observed}\\
      \hline
      \threeL         &off-$Z$	            & 107 $\pm$ & 7 $\pm$ & 24	 & 99\\
      \threeL         &on-$Z$	            & 510 $\pm$ &10 $\pm$ & 70	 &588\\
      \twoLoneT       &off-$Z$              & 220 $\pm$ & 5 $\pm$ & 50	 &226\\
      \twoLoneT       &on-$Z$               &1060 $\pm$ &10 $\pm$ &260   &914\\
      \hline
    \end{tabular}
  \end{center}
\end{table}

The \Htlep\ distributions for the two off-$Z$ signal channels are shown in 
Fig.~\ref{fig:htlep}, and the \met\ distributions for the same channels are shown in
Fig.~\ref{fig:MET}.  The \St\ distributions for the two on-$Z$ channels are shown in 
Fig.~\ref{fig:ST}.  The \St\ distribution for the on-$Z$, \threeL\ channel in Fig.~\ref{fig:ST3L}
has 4 events with $\St>1\TeV$ where a total of 2.2 events are expected.

\begin{figure*}[tbp]
  \begin{center}
    \subfigure[]  {\includegraphics[width=.48\textwidth]{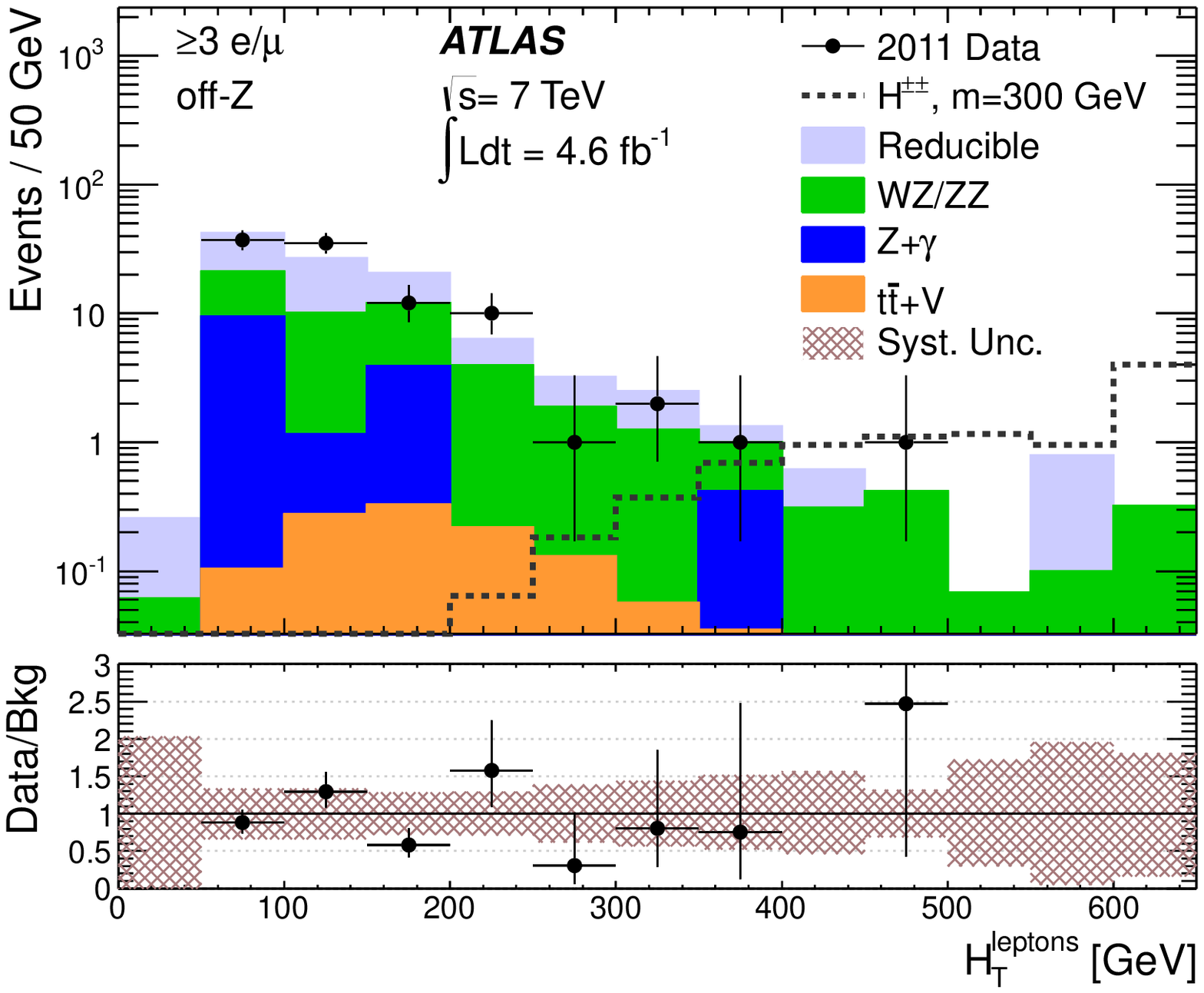}}
    \subfigure[]{\includegraphics[width=.48\textwidth]{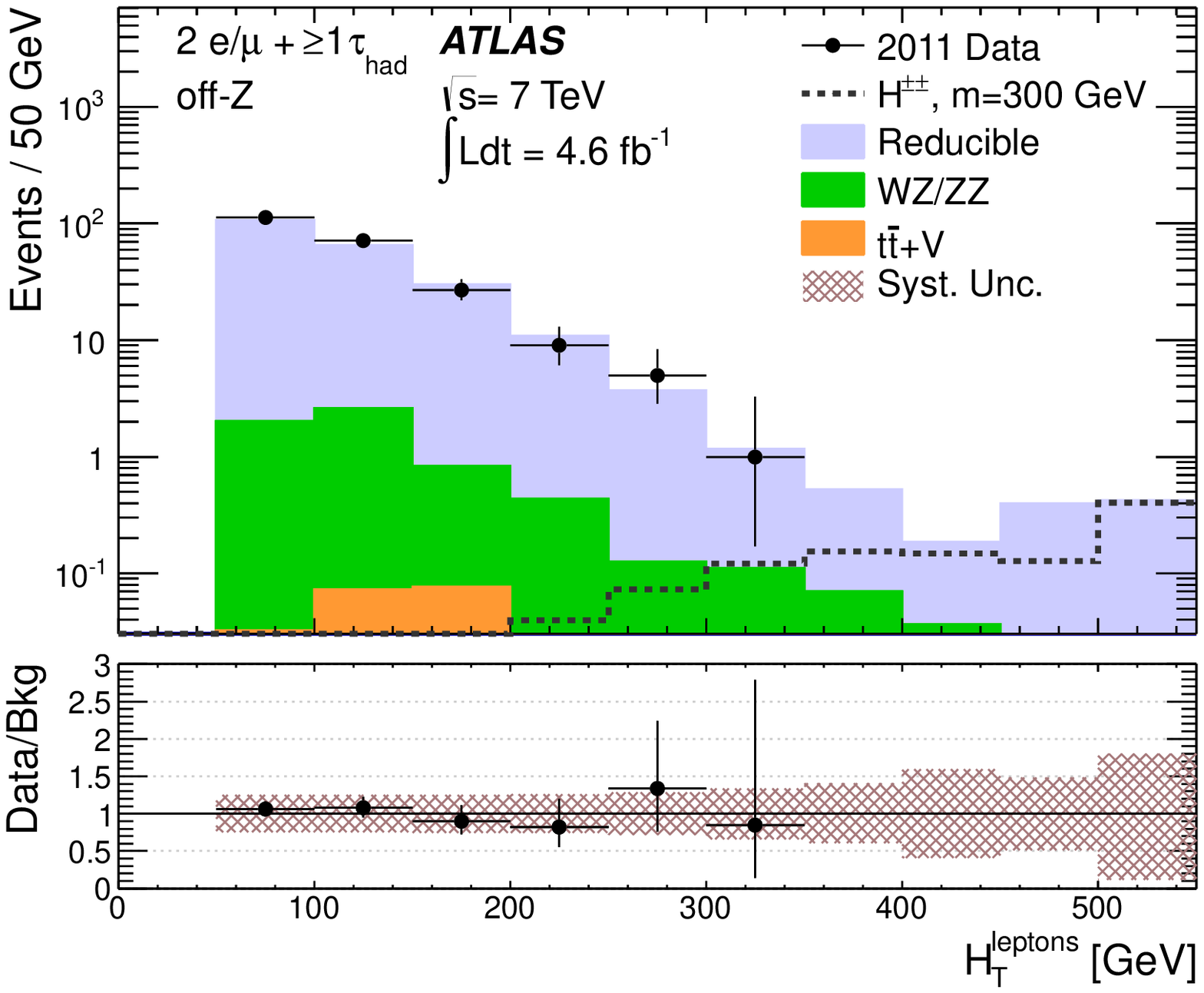}}
    \caption{The \Htlep\ distribution for the off-$Z$ (a) \threeL\ and (b) \twoLtau\ signal channels.   The dashed lines represent the expected
      contributions from events with pair-produced doubly-charged Higgs bosons with masses of 300~\GeV.  The last bin in the left (right) figure shows 
      the integral of events above 600~\GeV\ (500~\GeV).  \ratiopaneldescription}
    \label{fig:htlep}
  \end{center}
\end{figure*}

\begin{figure*}[tbp]
  \begin{center}
    \subfigure[]  {\includegraphics[width=\columnwidth]{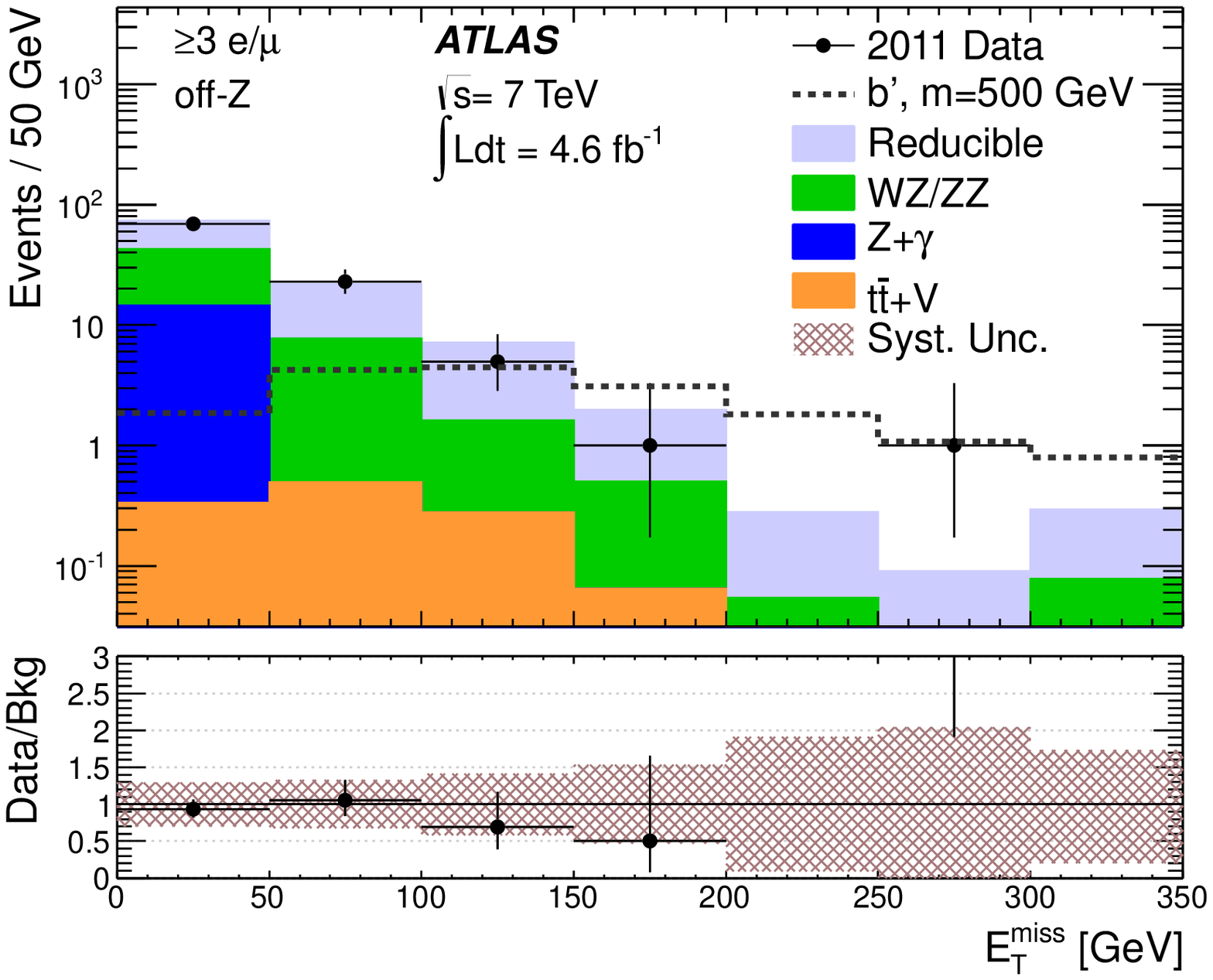}}
    \subfigure[]{\includegraphics[width=\columnwidth]{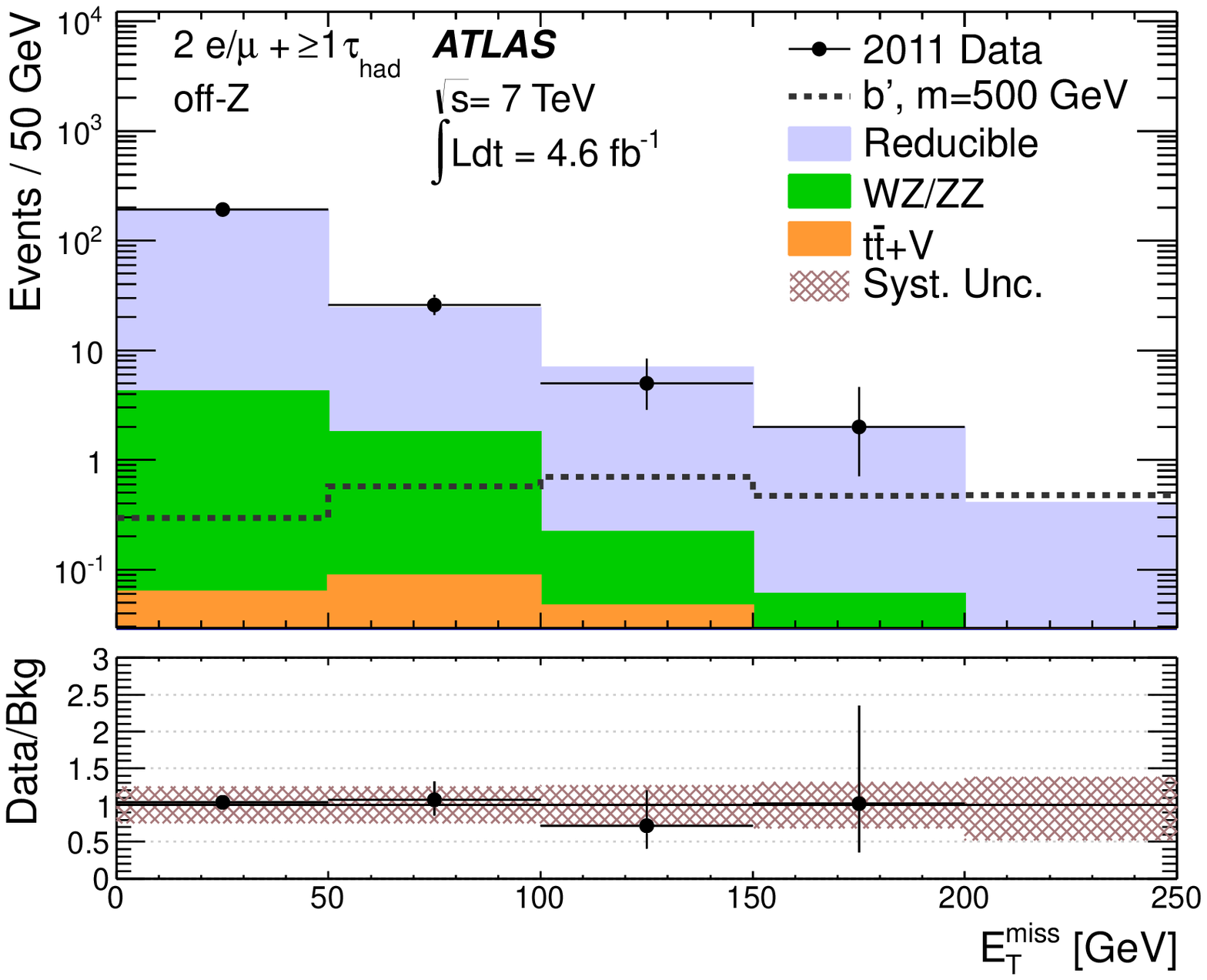}}
    \caption{The \met\ distribution for the off-$Z$ (a) \threeL\ and (b) \twoLoneT\ signal channels.  The dashed lines represent the expected
      contributions from events with fourth-generation down-type quarks with masses of 500~\GeV.  The last bin in the left (right) figure shows 
    the integral of events above 300~\GeV\ (200~\GeV).  \ratiopaneldescription}
    \label{fig:MET}
  \end{center}
\end{figure*}

\begin{figure*}[tbp]
  \begin{center}
    \subfigure[]  {\includegraphics[width=\columnwidth]{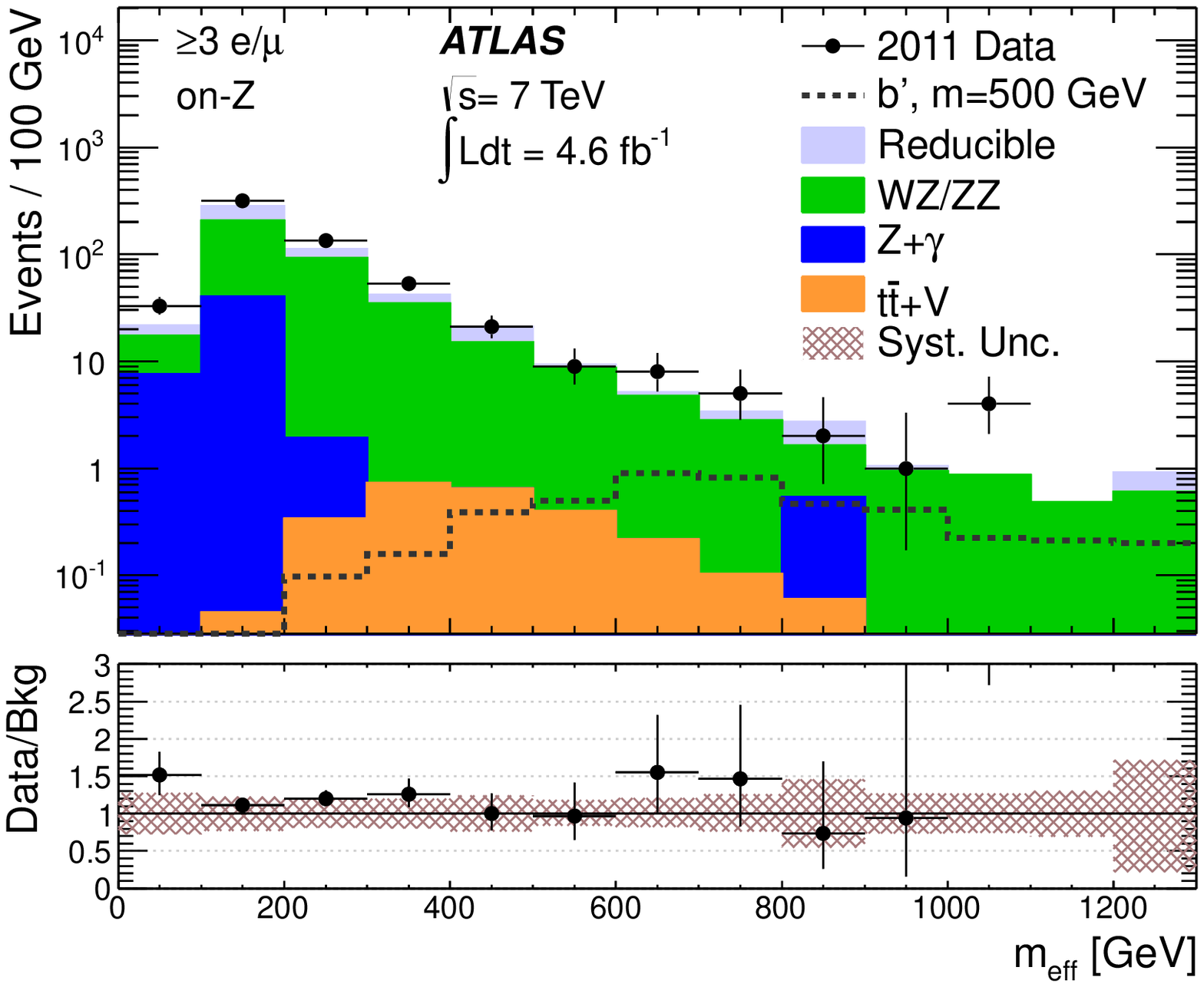}\label{fig:ST3L}}
    \subfigure[]{\includegraphics[width=\columnwidth]{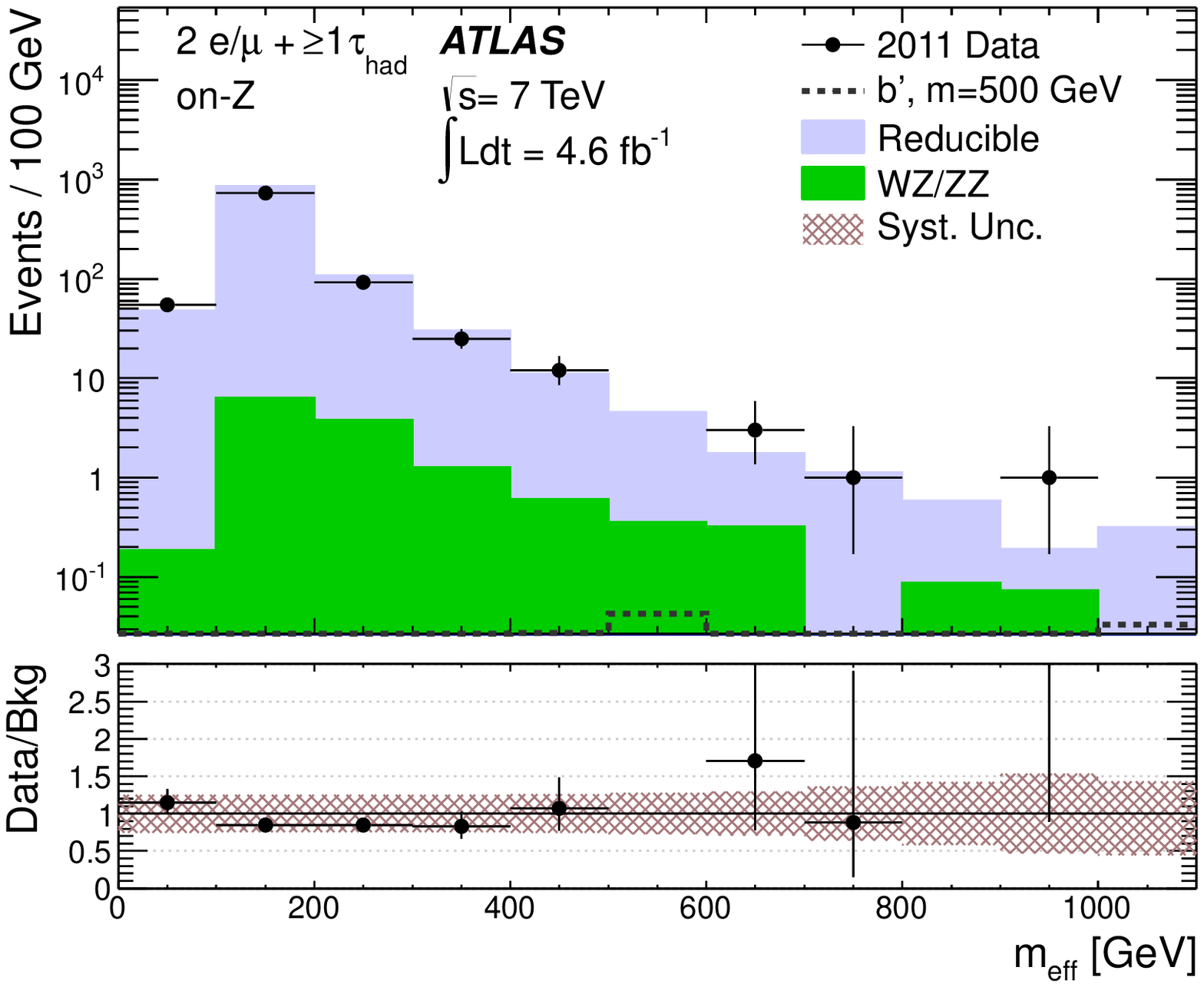}}
    \caption{The \St\ distribution for the on-$Z$ (a) \threeL\ and (b) \twoLoneT\ signal channels.  The dashed lines represent the expected
      contributions from events with fourth-generation down-type quarks with masses of 500~\GeV.  The last bin in the left (right) figure shows 
    the integral of events above 1.2~\TeV\ (1~\TeV).  In the \threeL\ channel, a total of 2.13 events are expected for $\St > 1\TeV$, and 4 events
    are observed.  \ratiopaneldescription}
    \label{fig:ST}
  \end{center}
\end{figure*}

The observed event yields in different signal regions are used to constrain contributions from new phenomena.
The 95\% confidence level (CL) upper limits on the number of events from non-Standard-Model sources ($N_{95}$) are calculated
using the CL$_{s}$ method~\cite{cls}.  All statistical and systematic uncertainties on estimated backgrounds are
incorporated into the limit-setting procedure, with correlations taken into account where appropriate.  Systematic
uncertainties on the signal efficiency are also included as described in Section~\ref{sec:Systematics}. 
The $N_{95}$ limits are then converted into limits on the ``visible cross section'' (\sigmavis) 
using the relationship $\sigmavis=N_{95}/\intLdt$.

Figures~\ref{fig:Limit_All_HTLep}--\ref{fig:Limit_HighMET_ST} show the 
resulting observed limits, along with the median expected limits with $\pm1\sigma$ and $\pm2\sigma$ uncertainties.
Observed and expected limits are also presented in Tables~\ref{t:lim_htlep}--\ref{t:lim_st_highmet} of Appendix~\ref{app:limits}.
The most inclusive signal regions for the \Htlep\ and \St\ variables are composed of the same events within each channel,
leading to identical limits.

\begin{figure}[tbp]
  \begin{center}
    \includegraphics[width=\columnwidth]{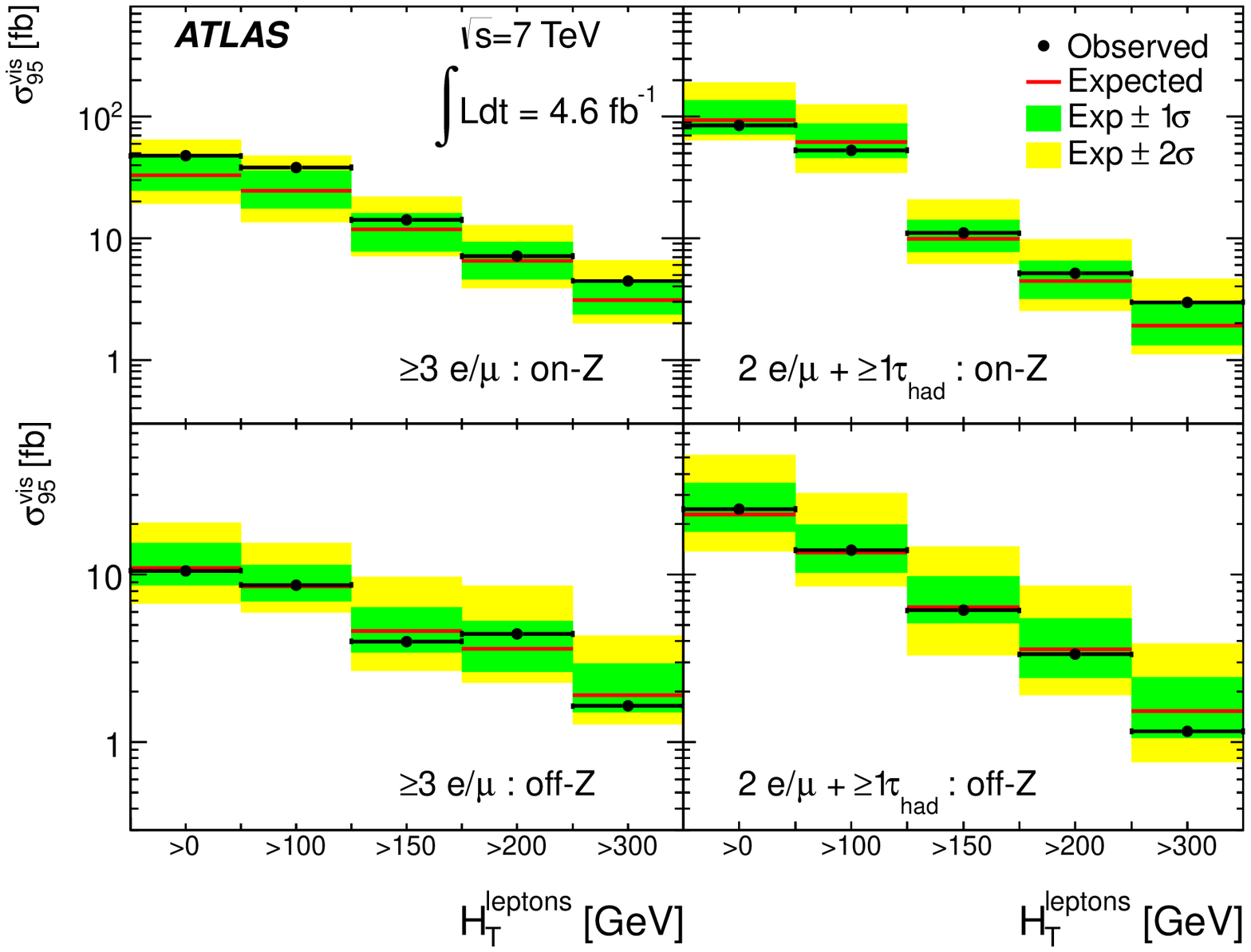}
    \caption{The observed- and median-expected 95\% CL limit on the visible cross section (\sigmavis)
      in the different signal channels, as functions of increasing lower bounds on \Htlep.  The
      $\pm1\sigma$ and $\pm2\sigma$ uncertainties on the median expected limit are indicated by green and 
      yellow bands, respectively.}
    \label{fig:Limit_All_HTLep}
  \end{center}
\end{figure}

\begin{figure}[tbp]
  \begin{center}
    \includegraphics[width=\columnwidth]{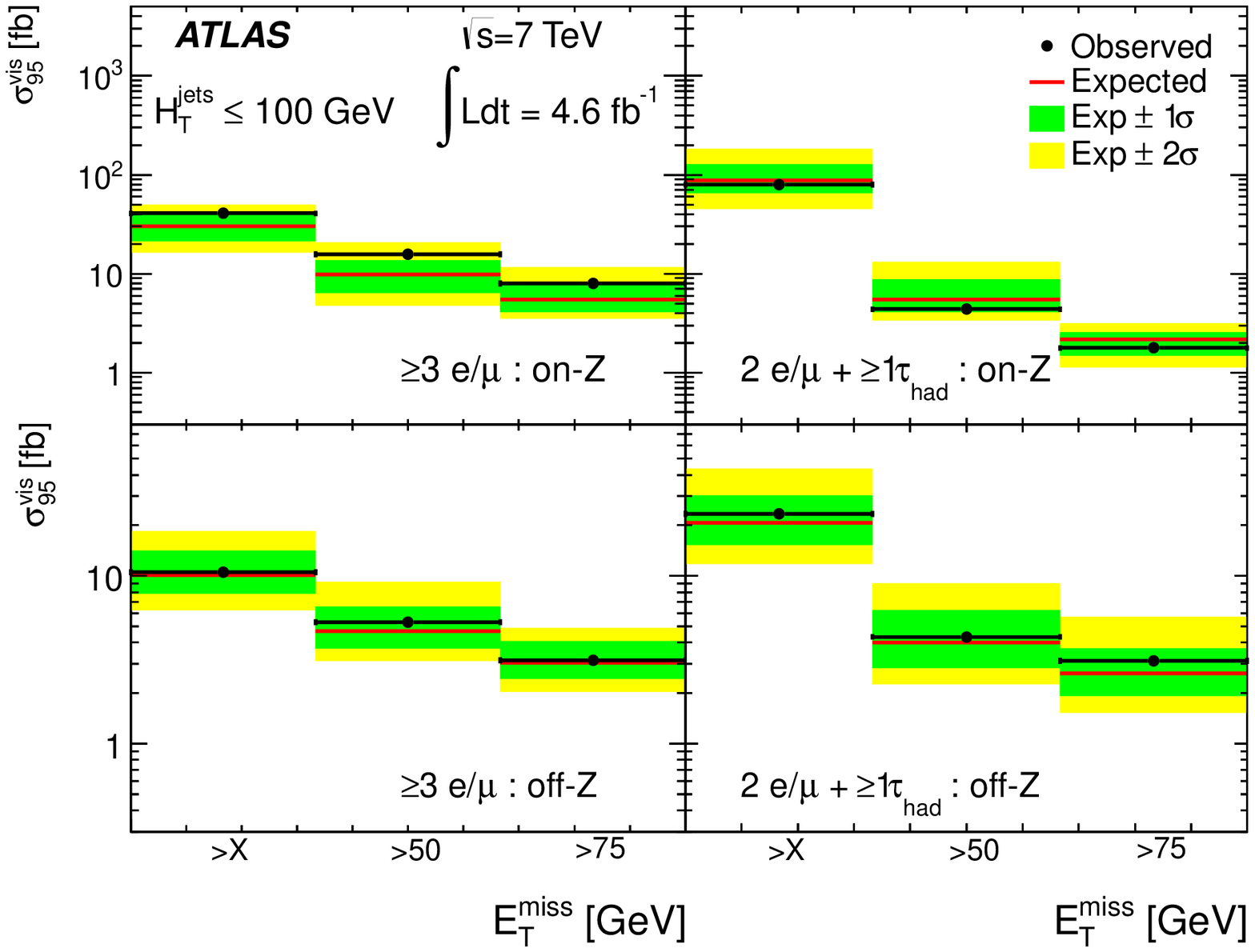}
    \caption{The observed- and median-expected 95\% CL limit on the visible cross section (\sigmavis)
      in the different signal channels, as functions of increasing lower bounds on \met, for
      events with $\Htjets < 100$~\GeV.  The lowest bin boundary $X$ is 0~\GeV{} for the off-$Z$ channels, and 20~\GeV{} for
      the on-$Z$ channels.  The
      $\pm1\sigma$ and $\pm2\sigma$ uncertainties on the median expected limit are indicated by green and 
      yellow bands, respectively.}
    \label{fig:Limit_MET_Weak}
  \end{center}
\end{figure}

\begin{figure}[tbp]
  \begin{center}
    \includegraphics[width=\columnwidth]{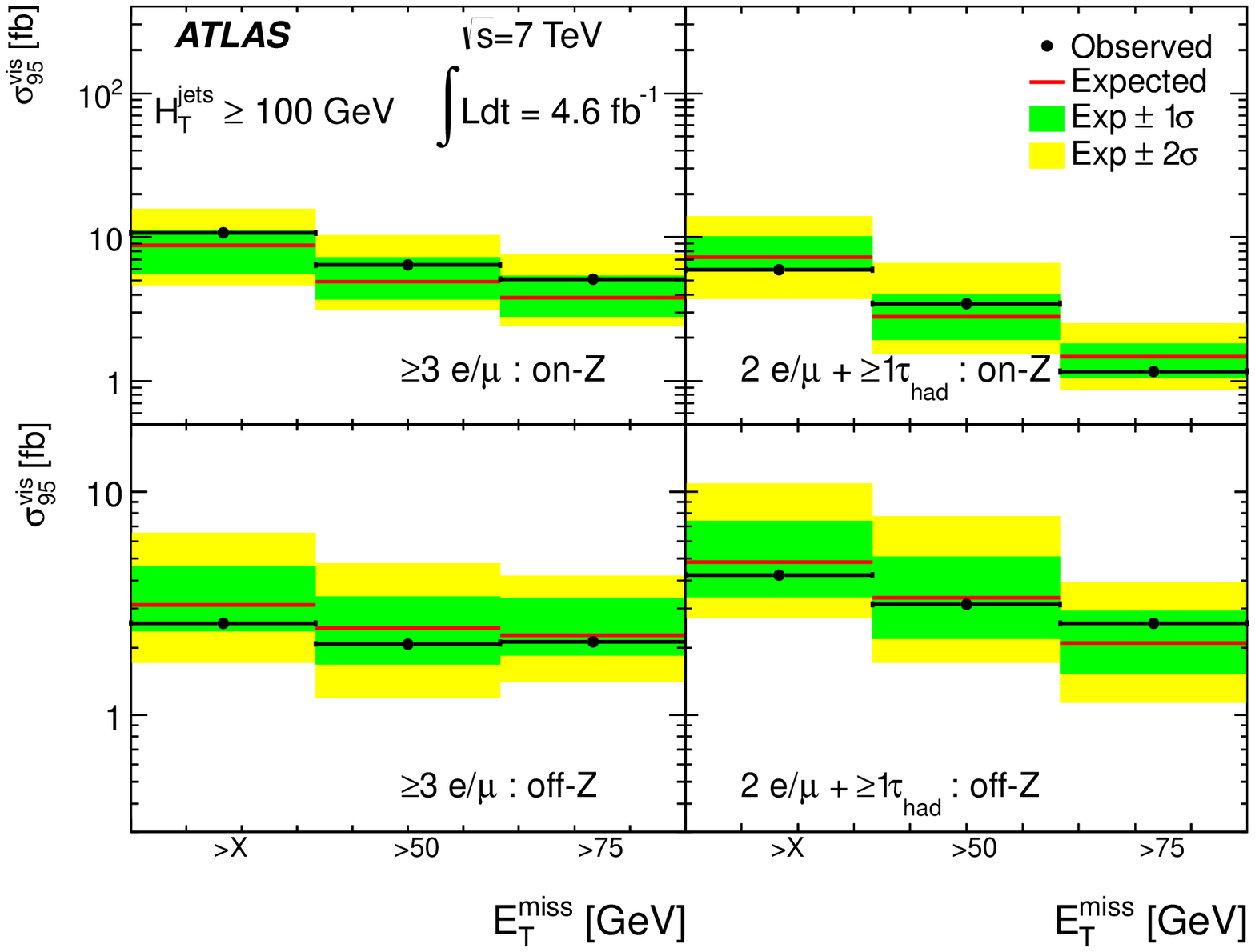}
    \caption{The observed- and median-expected 95\% CL limit on the visible cross section (\sigmavis)
      in the different signal channels, as functions of increasing lower bounds on \met, for
      events with $\Htjets \geq\ 100$~\GeV.  The lowest bin boundary $X$ is 0~\GeV{} for the off-$Z$ channels, and 20~\GeV{} for
      the on-$Z$ channels.  The
      $\pm1\sigma$ and $\pm2\sigma$ uncertainties on the median expected limit are indicated by green and 
      yellow bands, respectively.}
    \label{fig:Limit_MET_Strong}
  \end{center}
\end{figure}

\begin{figure}[tbp]
  \begin{center}
    \includegraphics[width=\columnwidth]{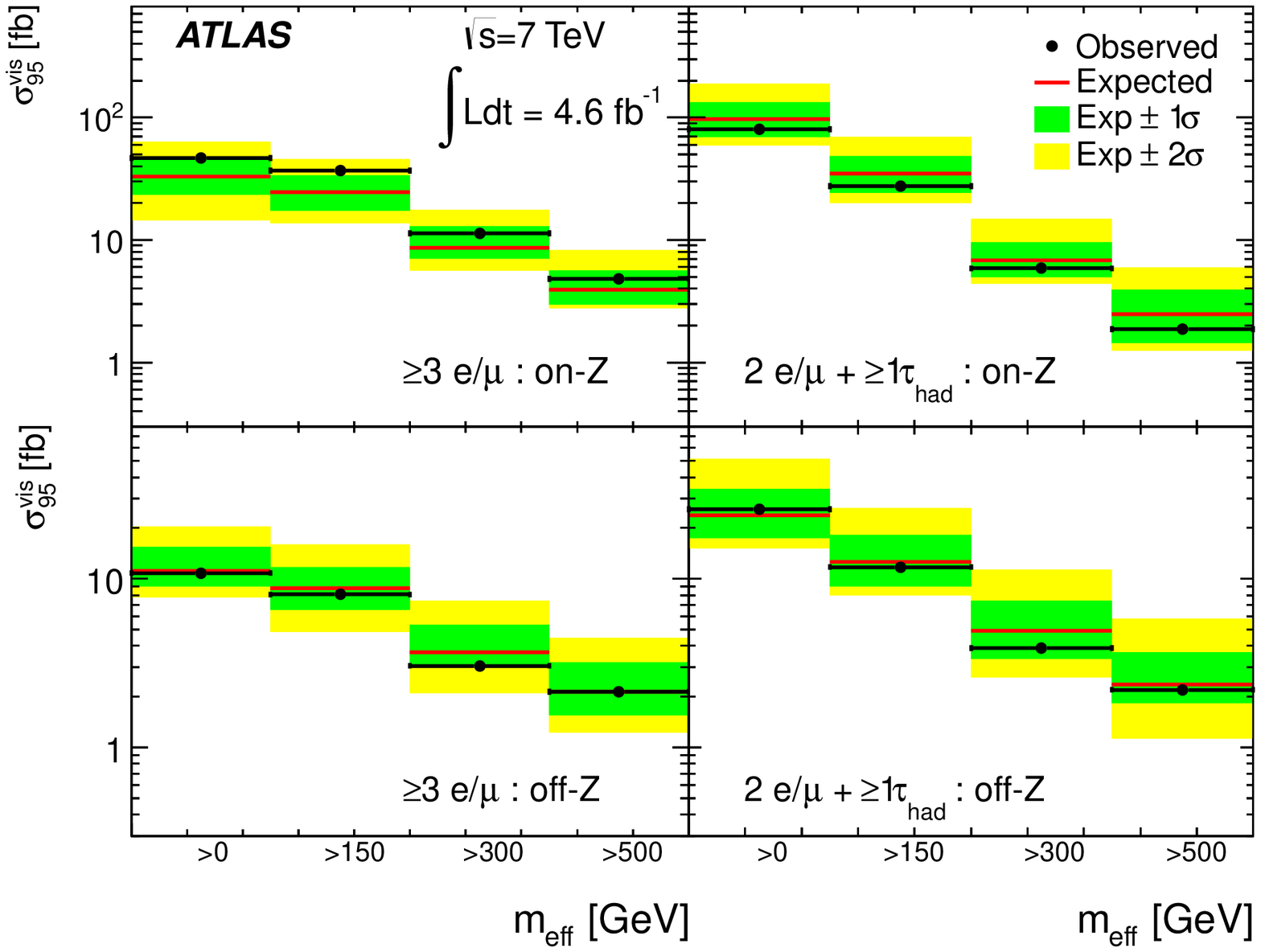}
    \caption{The observed- and median-expected 95\% CL limit on the visible cross section (\sigmavis)
       in the different signal channels, as functions of increasing lower bounds on \St.  The
      $\pm1\sigma$ and $\pm2\sigma$ uncertainties on the median expected limit are indicated by green and 
      yellow bands, respectively.}
    \label{fig:Limit_ST}
  \end{center}
\end{figure}

\begin{figure}[tbp]
  \begin{center}
    \includegraphics[width=\columnwidth]{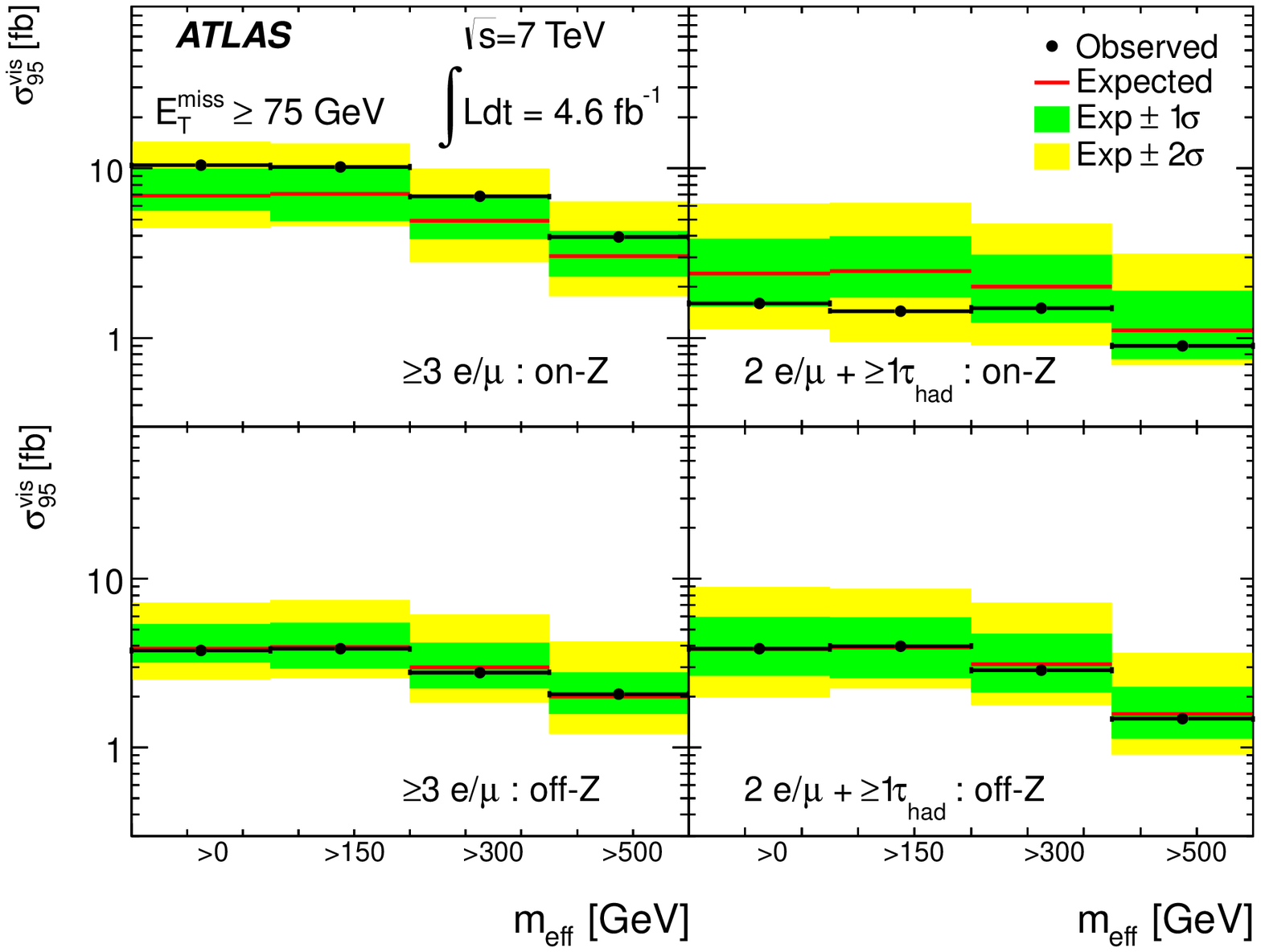}
    \caption{The observed- and median-expected 95\% CL limit on the visible cross section (\sigmavis)
      in the different signal channels, as functions of increasing lower bounds on \St, for events
      with $\met > 75$~\GeV.  The
      $\pm1\sigma$ and $\pm2\sigma$ uncertainties on the median expected limit are indicated by green and 
      yellow bands, respectively.}
    \label{fig:Limit_HighMET_ST}
  \end{center}
\end{figure}

\section{Model testing}
\label{sec:ModelTesting}
The $\sigmavis$ limits 
can be converted into 
upper limits on the cross section of a specific model as follows:
\begin{itemize}
\item Events from the new model are examined at the particle (MC-generator) level and 
kinematic requirements on the particles are applied.
These include the $\pt$ and $\eta$ requirements for leptons and jets, 
and isolation requirements for the leptons. No special treatment for pileup is necessary.
\item The number of events passing this selection determines the cross section for the model
given the fiducial constraints, $\sigfid$.
\item A correction factor must be applied to take into account detector effects. This correction factor,
called $\epsfid$, is model-dependent, and is subject to 
uncertainties from detector resolution, reconstruction efficiency, pileup, and vertex selection.
This correction factor represents the ratio of the number of events satisfying the selection
criteria after reconstruction to all those satisfying the fiducial acceptance criteria at the particle level.
As this correction factor accounts for detector effects, no unfolding of the reconstructed distributions is
necessary.
\item A 95\% CL upper-limit on the cross section in the new model is then given by:
\end{itemize}

\begin{equation}
  \sigma_{95}^{\mathrm{fid}} = \frac{N_{95}}{\epsfid \intLdt} = \frac{\sigmavis}{\epsfid}.
\end{equation}

The value of $\epsfid$ in the \threeL\  channels
ranges from roughly 0.50 for fourth-generation quark models to over 0.70 for 
doubly-charged Higgs models producing 
up to four high-\pt\ leptons.  In the \twoLtau\  channels, $\epsfid$
is roughly 0.10 for a variety
of models.  Finite momentum resolution in the detector 
can cause particles with
true momenta outside the kinematic acceptance (\eg\ muons with $\pt< 10$~\GeV) to be accepted
after reconstruction.  The fraction of such events after selection is at most 3\% for the \threeL\  channels 
and 4\% for the \twoLtau\  channels.

In order to determine $\epsfid$ for unexplored models of new phenomena producing at 
least
three prompt, isolated, and charged leptons in the final state, per-lepton efficiencies 
parameterized by the lepton
kinematics are provided here.  While the experimental results are based on reconstructed quantities,
all requirements in the following are defined at the particle level.
The per-lepton efficiencies attempt to emulate the ATLAS detector
response, thereby allowing a comparison of the yields from
particle-level event samples with the cross-section limits provided
above without the need for a detector simulation.

Electrons at the particle level are required
to have $\pt \geq 10$~\GeV, and to satisfy  $|\eta|<2.47$ and $|\eta|\notin(1.37,1.52)$.  Particle-level muons are required
to have $\pt \geq 10$~\GeV, and to have $|\eta|<2.5$.  Electrons and muons are both required to be prompt,
and not associated with a secondary vertex, unless they are the product of tau-lepton decays.
Leptonically decaying tau candidates are required to
produce electrons
or muons that satisfy the criteria above.
Hadronically decaying tau candidates are required to have $\ptvis \geq 10 \GeV$ and
 $|\etavis|<2.5$, where the visible products of the tau decay include 
all particles except neutrinos.  As with reconstructed tau candidates, the tau 
four-momentum at the particle level is defined only by the visible decay products.

Generated electrons and muons are further required to be isolated.  A track isolation energy
at the particle
level corresponding to \ptisotrack, denoted \ptisotruth, is defined as the scalar sum of transverse momenta of 
charged particles within a cone of
$\Delta R < 0.3$ around the lepton axis.  
Particles used in the sum are included after hadronization and must have $\pt>1\GeV$.
A fiducial isolation energy corresponding to \Etisocal, denoted
\Etisotruth, is
defined as the sum of all particles inside the annulus $0.1 < \Delta R < 0.3$ around the lepton axis.  
Neutrinos and other stable, weakly-interacting particles are excluded from both \ptisotruth\ and \Etisotruth; muons are excluded
from \Etisotruth.  Electrons must
satisfy $\ptisotruth/\pt < 0.13$ and $\Etisotruth/\pt < 0.2$, while muons must satisfy
 $\ptisotruth/\pt < 0.15$ and $\Etisotruth/\pt < 0.2$.

Events with at least three leptons as defined above must have at least two electrons and/or muons,
at least one of which has $\pt\geq25$~\GeV.  The third lepton is allowed to be an electron or muon, in which
case the event is classified as a \threeL\ event, or a hadronically decaying tau lepton, in which case it is a \twoLtau\ event.

A simulated sample of $WZ$ events is used to determine
the per-lepton efficiencies $\epsilon_{\ell}$.  The leptons above are matched to reconstructed lepton candidates that satisfy
the selection criteria defined in Section~\ref{sec:Selection}, with $\epsilon_{\ell}$ defined as the ratio of 
the number of reconstructed leptons satisfying all selection criteria to the number of generated leptons satisfying the fiducial criteria.  Separate
values of $\epsilon_{\ell}$ are measured for each lepton flavor.  In the case of electrons and muons, $\epsilon_{\ell}$
is determined separately for leptons from tau decays.  

All efficiencies are measured as functions of the lepton \pt\ and $\eta$. The efficiencies for
electrons and taus are shown in Tables~\ref{t:fideff_num1} and \ref{t:fideff_num2}.
The $\eta$
dependence of the muon efficiencies is treated by separate \pt\ efficiency measurements for muons with
$|\eta|<0.1$ and those with $|\eta|\geq0.1$, and is shown in Table~\ref{t:fideff_num3}. 
For taus, the efficiency tables include the efficiency for
taus generated with $\ptvis<15\GeV$ but reconstructed with $\ptvis\geq15\GeV$, due to resolution effects.
The corresponding efficiencies for electrons and muons generated below 10~\GeV\ are much smaller, and are not included here.
The final per-lepton efficiency for electrons and taus is obtained as $\epsilon_{\ell} = \epsilon(\pt)\cdot\epsilon(\eta) / \langle \epsilon \rangle$, where
$\langle \epsilon \rangle$ is $0.69$ for prompt electrons, $0.53$ for electrons from tau decays, and
$0.17$ for hadronically decaying taus.

\begin{table}[tbp]
\begin{center}
\caption{The fiducial efficiency for electrons and taus in different \pt\ ranges.  For tau candidates, $\pt\equiv\ptvis$.}
 \label{t:fideff_num1} 
 \begin{tabular}{lccc} 
 \hline \hline 
\pt{} [\GeV]& Prompt $e$& $\tau\rightarrow e$& $\tauh$ \\ \hline 
$10\text{--}15$& 0.394$\pm$0.003& 0.381$\pm$0.004& 0.025$\pm$0.002\\ 
$15\text{--}20$& 0.510$\pm$0.003& 0.515$\pm$0.005& 0.147$\pm$0.004\\ 
$20\text{--}25$& 0.555$\pm$0.003& 0.542$\pm$0.006& 0.225$\pm$0.005\\ 
$25\text{--}30$& 0.626$\pm$0.002& 0.601$\pm$0.007& 0.229$\pm$0.006\\ 
$30\text{--}40$& 0.691$\pm$0.002& 0.673$\pm$0.006& 0.215$\pm$0.005\\ 
$40\text{--}50$& 0.738$\pm$0.002& 0.729$\pm$0.008& 0.206$\pm$0.006\\ 
$50\text{--}60$& 0.774$\pm$0.002& 0.76$\pm$0.01& 0.202$\pm$0.008\\ 
$60\text{--}80$& 0.796$\pm$0.002& 0.77$\pm$0.01& 0.198$\pm$0.008\\ 
$ 80\text{--}100$& 0.830$\pm$0.002& 0.82$\pm$0.02& 0.21$\pm$0.01\\ 
$100\text{--}200$& 0.850$\pm$0.003& 0.81$\pm$0.02& 0.23$\pm$0.02\\ 
$200\text{--}400$& 0.878$\pm$0.009& 0.85$\pm$0.07& 0.19$\pm$0.05\\ 
 \hline \hline 
 \end{tabular} 
 \end{center} 
 \end{table} 

\begin{table}[tbp]
\begin{center}
\caption{The fiducial efficiency for electrons and taus in different $\eta$ ranges.  For tau candidates, $\eta\equiv\etavis$.}
 \label{t:fideff_num2} 
 \begin{tabular}{lccc} 
 \hline \hline 
$|\eta|$ & Prompt $e$& $\tau\rightarrow e$& $\tauh$ \\ \hline 
$0.0\text{--}0.1$& 0.675$\pm$0.003& 0.52$\pm$0.01& 0.210$\pm$0.009\\ 
$0.1\text{--}0.5$& 0.757$\pm$0.001& 0.595$\pm$0.005& 0.195$\pm$0.004\\ 
$0.5\text{--}1.0$& 0.747$\pm$0.001& 0.581$\pm$0.005& 0.179$\pm$0.004\\ 
$1.0\text{--}1.5$& 0.666$\pm$0.002& 0.494$\pm$0.006& 0.138$\pm$0.004\\ 
$1.5\text{--}2.0$& 0.607$\pm$0.002& 0.465$\pm$0.006& 0.170$\pm$0.004\\ 
$2.0\text{--}2.5$& 0.591$\pm$0.002& 0.475$\pm$0.007& 0.163$\pm$0.005\\ 
 \hline \hline 
 \end{tabular} 
 \end{center} 
 \end{table} 

\begin{table}[tbp]
\begin{center}
\caption{The fiducial efficiency for muons in different $\pt$ ranges.}
 \label{t:fideff_num3} 
 \begin{tabular}{lcccc} 
 \hline \hline 
\multicolumn{1}{c}{\pt} &\multicolumn{2}{c}{Prompt $\mu$}    & \multicolumn{2}{c}{$\tau\rightarrow\mu$} \\ 
$[$\GeV$]$                 & $|\eta| > 0.1$ &$|\eta| < 0.1$     &$|\eta| > 0.1$ &$|\eta| < 0.1$            \\ 
 \hline 
 $ 10\text{--}15$& 0.852$\pm$0.002& 0.47$\pm$0.02& 0.66$\pm$0.004& 0.36$\pm$0.02\\ 
 $ 15\text{--}20$& 0.896$\pm$0.002& 0.51$\pm$0.01& 0.71$\pm$0.005& 0.38$\pm$0.02\\ 
 $ 20\text{--}25$& 0.912$\pm$0.001& 0.52$\pm$0.01& 0.734$\pm$0.005& 0.43$\pm$0.03\\ 
 $ 25\text{--}30$& 0.921$\pm$0.001& 0.50$\pm$0.01& 0.750$\pm$0.006& 0.39$\pm$0.03\\ 
 $ 30\text{--}40$& 0.927$\pm$0.001& 0.507$\pm$0.007& 0.779$\pm$0.005& 0.46$\pm$0.03\\ 
 $ 40\text{--}50$& 0.928$\pm$0.001& 0.513$\pm$0.008& 0.784$\pm$0.007& 0.45$\pm$0.04\\ 
 $ 50\text{--}60$& 0.932$\pm$0.001& 0.532$\pm$0.009& 0.79$\pm$0.01& 0.37$\pm$0.05\\ 
 $ 60\text{--}80$& 0.932$\pm$0.001& 0.524$\pm$0.009& 0.81$\pm$0.01& 0.43$\pm$0.06\\ 
 $ 80\text{--}100$& 0.932$\pm$0.002& 0.51$\pm$0.01& 0.77$\pm$0.02& 0.53$\pm$0.09\\ 
 $100\text{--}200$& 0.930$\pm$0.002& 0.50$\pm$0.01& 0.83$\pm$0.02& 0.47$\pm$0.12\\ 
 $200\text{--}400$& 0.919$\pm$0.007& 0.45$\pm$0.05& 0.59$\pm$0.11& -\\ 
  \hline \hline 
 \end{tabular} 
 \end{center} 
 \end{table}

The resulting per-lepton efficiencies are then combined to yield a selection efficiency for a given event satisfying
the fiducial acceptance criteria.  For events with exactly three leptons,
the total efficiency for the event is the product of the individual lepton efficiencies.  For events
with more than three leptons, the additional leptons in order of descending \pt\ only contribute to 
the total efficiency when a lepton with higher \pt\ is not selected, leading to terms like 
$\epsilon_1\epsilon_2\epsilon_4(1-\epsilon_3)$, where $\epsilon_i$ denotes the fiducial efficiency for
the $i^{th}$ \pt-ordered lepton.  The method can be extended to cover the number of leptons
expected by the model under consideration.

Jets at the particle level are reconstructed from all stable particles, excluding muons and neutrinos, 
with the \antikt\ algorithm using a distance parameter $R=0.4$.  
Overlaps between jets and leptons are removed as described in Section~\ref{sec:Selection}.
\met\ is defined as the magnitude of the vector sum of the transverse momenta of all stable, weakly-interacting particles, including
those produced in models of new phenomena.  The kinematic variables used for limit setting are defined as before:
\Htlep\ is the scalar sum of the transverse momenta, or \ptvis\ for $\tauh$ candidates, of the three leptons that define the event;
\Htjets\ is the scalar sum of all jets surviving overlap removal; \met\ is as defined above, and \St\ is
the sum of \met, \Htjets, and all transverse momenta of selected leptons in the event.

Predictions of the rate and kinematic properties of events with multiple leptons made with the method described above 
agree well with the same quantities after detector simulation for a variety of models of new phenomena.  
Uncertainties, 
based on the level of agreement seen across a variety of models, are estimated 
at 10\% for the \threeL\  channels, and 20\% for the \twoLtau\  channels.  These uncertainties are included
in the limits presented in Section~\ref{sec:Results}.

As an example of the application of the method described in this section,
the $\sigmavis$ limits can be used to constrain models predicting the pair-production of 
doubly-charged Higgs bosons. The constraints from dedicated and optimized
analyses by ATLAS~\cite{atlasDCH2012} and CMS~\cite{cmsDCH2012ya} are expected to be stronger than the constraints obtained
here, but these numbers serve to benchmark the results presented in this
paper. 

Assuming a branching ratio of 100\% for
the decay $\dchp\rightarrow\mu^{\pm}\mu^{\pm}$, the acceptance of
the fiducial selection is 91\% and \epsfid{} is 71\%
for $m(\dchp)=300\GeV$. The resulting 95\% CL upper limit on the
cross section times branching ratio ($\sigmaB$) is 2.5 fb.  The observed and median expected 
upper limits are shown in Fig.~\ref{f:lim_dch_mumu}, along with the observed upper limit from the dedicated search by ATLAS~\cite{atlasDCH2012}. 
These results are obtained using 
 the $\Htlep\ge300\GeV$ signal region in the \threeL, off-$Z$ channel. 
The theoretical cross section for $\dchp$ coupling to left-handed fermions ($\dchplh$) implies that
$\dchplh$ masses below 330$\GeV$ are excluded at 95\% CL for BR($\dchp\rightarrow\mu^{\pm}\mu^{\pm}$)=100\%.

\begin{figure*}[tbp]
  \centering
   \subfigure[$\dchp\to\mu^{\pm}\mu^{\pm}$]{\includegraphics[width=\columnwidth]{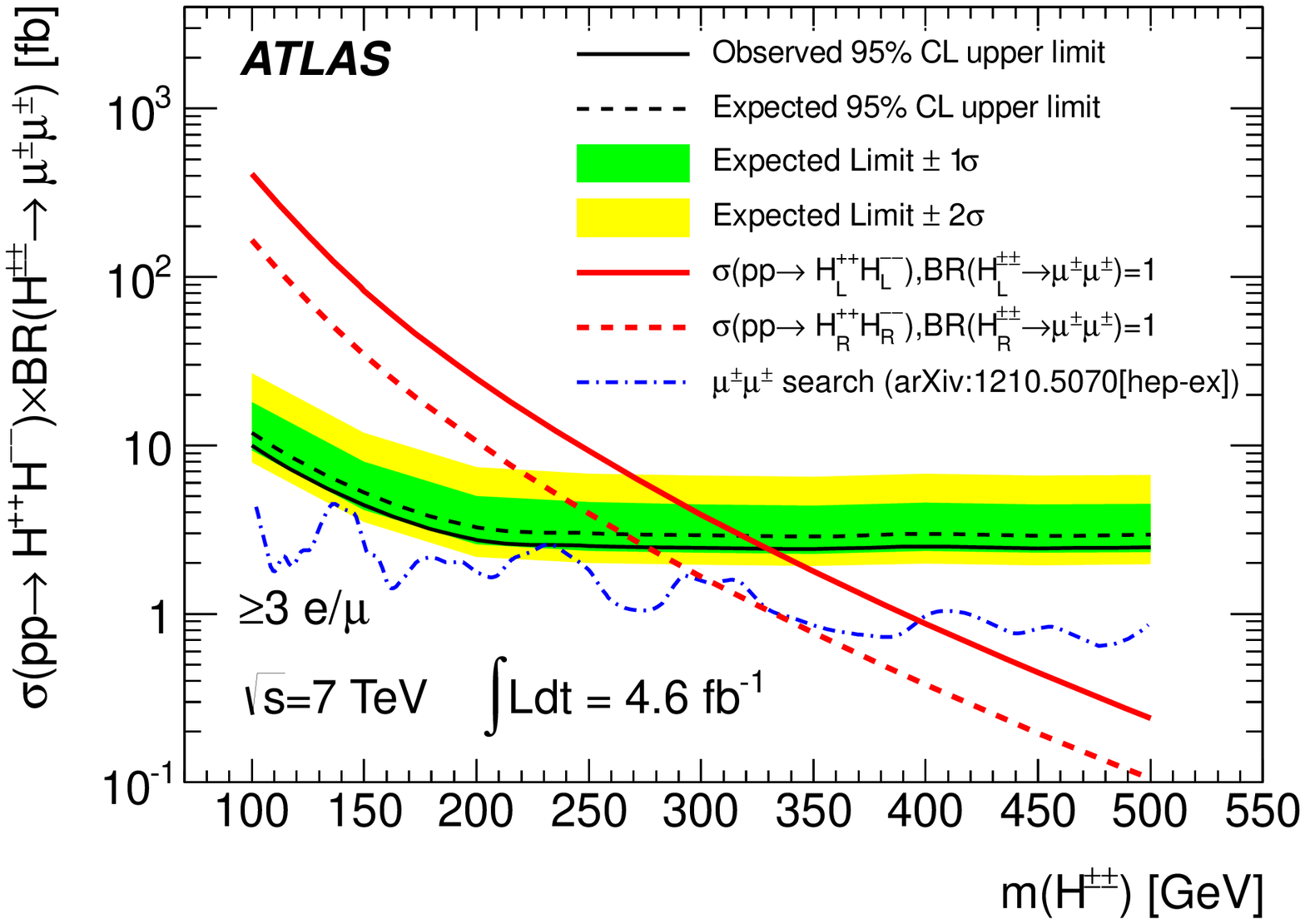}\label{f:lim_dch_mumu}}
   \subfigure[$\dchp\to\mu^{\pm}\tau^{\pm}$]{\includegraphics[width=\columnwidth]{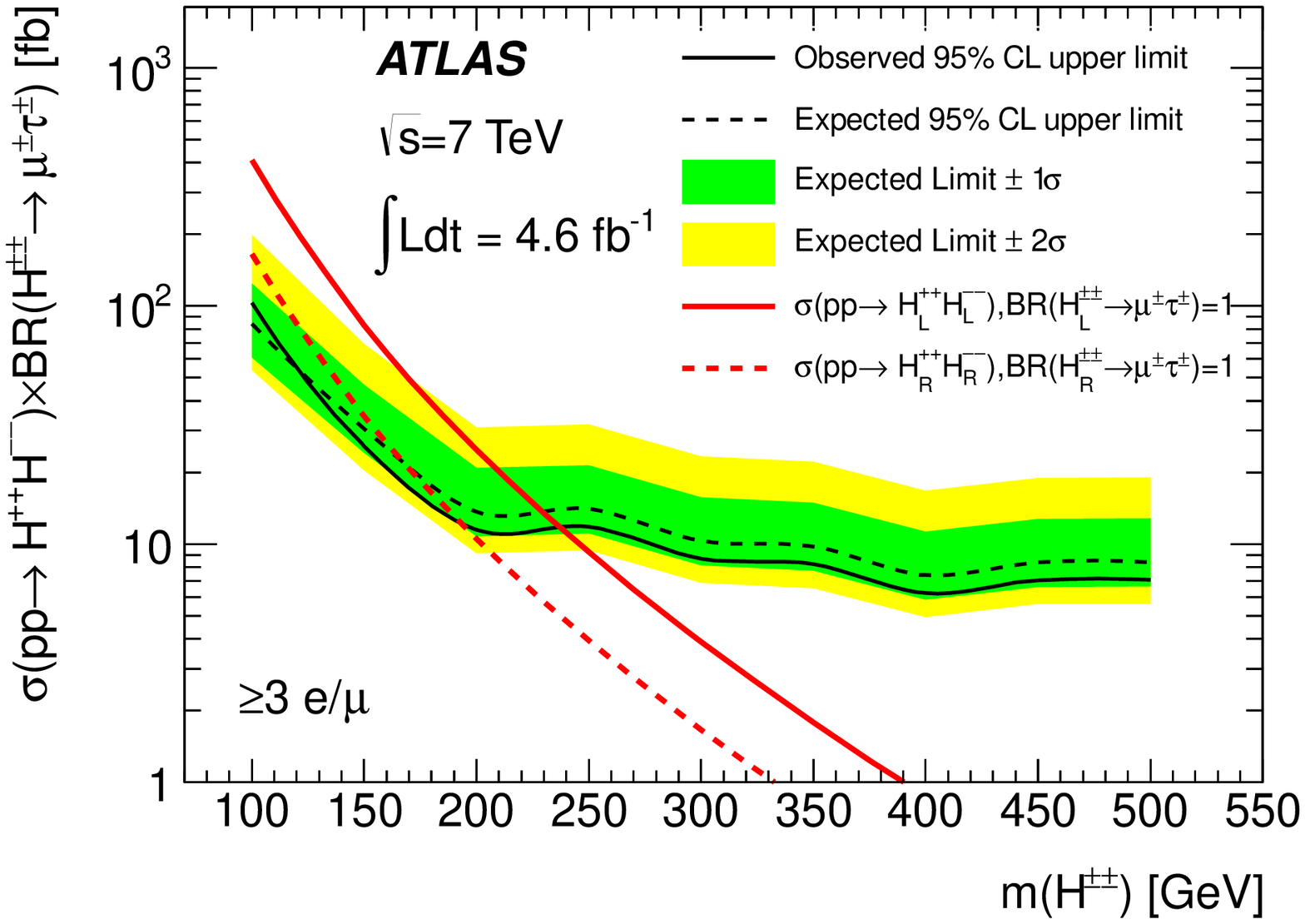}\label{f:lim_dch_mutau}}
   \caption{The expected and observed 95\% confidence level upper limits
     on the cross section times branching ratio of the (a) $\dchp\to\mu^{\pm}\mu^{\pm}$ and (b) $\dchp\to\mu^{\pm}\tau^{\pm}$ final states
     as a function of the \dchp{} mass for the \threeL\ channel. For $\dchp\to\mu^{\pm}\mu^{\pm}$, the median expected limit on the \dchplh{}
     mass is 319 \GeV\ and the corresponding observed limit is 330 \GeV; for $\dchp\to\mu^{\pm}\tau^{\pm}$, the median expected limit is 229 \GeV\ and 
     the corresponding observed limit is 237 \GeV. Results from the dedicated ATLAS search for 
     $\dchp{}\to\mu^{\pm}\mu^{\pm}$~\cite{atlasDCH2012} are also shown.}
  \label{f:lim_dch}
\end{figure*}

For the case with BR($\dchp\rightarrow\mu^{\pm}\tau^{\pm}$)=100\%, the
acceptance for the \threeL\ (\twoLtau) channel is 24\% (49\%), and \epsfid{} is 59\% (13\%) for $m(\dchp)=200\GeV$. The
corresponding upper limit on the cross section is 12 (19) fb, with $m(\dchplh)<237\GeV$ $(220\GeV)$ excluded at 95\% CL.
In this case, the off-$Z$ $\Htlep\ge300\GeV$ signal region is used to calculate the expected limits for all \dchp{} masses
except for m(\dchp)=$100\GeV$ where the off-$Z$, $\Htlep\ge200\GeV$ signal region is used.
The observed and median expected limits from the \threeL\ channel are shown in Fig.~\ref{f:lim_dch_mutau}.

\section{Conclusion}
\label{sec:Conclusion}
A generic search for new phenomena in events with at least three energetic, charged, prompt, and isolated leptons has been presented,
using a data sample corresponding to an integrated luminosity of 4.6~\ifb\ of $pp$ collision data collected by the ATLAS experiment.  
The search was conducted in separate channels based on the presence or absence 
of a hadronically decaying tau lepton or reconstructed $Z$ boson, and yielded no
significant deviation from background yields expected from the Standard Model.
Upper limits at 95\% confidence level on event yields due to non-Standard-Model processes were
placed as a function of lower bounds on several kinematic variables.  Additional information
on the fiducial selection of events populating the signal regions under study has been provided.
The use of this information in the interpretation of the results in the context of models of new phenomena 
has been illustrated by setting upper limits on the production of doubly-charged Higgs bosons decaying
to same-sign lepton pairs.

\acknowledgments

We thank CERN for the very successful operation of the LHC, as well as the
support staff from our institutions without whom ATLAS could not be
operated efficiently.

We acknowledge the support of ANPCyT, Argentina; YerPhI, Armenia; ARC,
Australia; BMWF and FWF, Austria; ANAS, Azerbaijan; SSTC, Belarus; CNPq and FAPESP,
Brazil; NSERC, NRC and CFI, Canada; CERN; CONICYT, Chile; CAS, MOST and NSFC,
China; COLCIENCIAS, Colombia; MSMT CR, MPO CR and VSC CR, Czech Republic;
DNRF, DNSRC and Lundbeck Foundation, Denmark; EPLANET, ERC and NSRF, European Union;
IN2P3-CNRS, CEA-DSM/IRFU, France; GNSF, Georgia; BMBF, DFG, HGF, MPG and AvH
Foundation, Germany; GSRT and NSRF, Greece; ISF, MINERVA, GIF, DIP and Benoziyo Center,
Israel; INFN, Italy; MEXT and JSPS, Japan; CNRST, Morocco; FOM and NWO,
Netherlands; BRF and RCN, Norway; MNiSW, Poland; GRICES and FCT, Portugal; MERYS
(MECTS), Romania; MES of Russia and ROSATOM, Russian Federation; JINR; MSTD,
Serbia; MSSR, Slovakia; ARRS and MVZT, Slovenia; DST/NRF, South Africa;
MICINN, Spain; SRC and Wallenberg Foundation, Sweden; SER, SNSF and Cantons of
Bern and Geneva, Switzerland; NSC, Taiwan; TAEK, Turkey; STFC, the Royal
Society and Leverhulme Trust, United Kingdom; DOE and NSF, United States of
America.

The crucial computing support from all WLCG partners is acknowledged
gratefully, in particular from CERN and the ATLAS Tier-1 facilities at
TRIUMF (Canada), NDGF (Denmark, Norway, Sweden), CC-IN2P3 (France),
KIT/GridKA (Germany), INFN-CNAF (Italy), NL-T1 (Netherlands), PIC (Spain),
ASGC (Taiwan), RAL (UK) and BNL (USA) and in the Tier-2 facilities
worldwide.

\bibliography{multilepton_paper}

\begin{thebibliography}{61}%
\makeatletter
\providecommand \@ifxundefined [1]{%
 \@ifx{#1\undefined}
}%
\providecommand \@ifnum [1]{%
 \ifnum #1\expandafter \@firstoftwo
 \else \expandafter \@secondoftwo
 \fi
}%
\providecommand \@ifx [1]{%
 \ifx #1\expandafter \@firstoftwo
 \else \expandafter \@secondoftwo
 \fi
}%
\providecommand \natexlab [1]{#1}%
\providecommand \enquote  [1]{``#1''}%
\providecommand \bibnamefont  [1]{#1}%
\providecommand \bibfnamefont [1]{#1}%
\providecommand \citenamefont [1]{#1}%
\providecommand \href@noop [0]{\@secondoftwo}%
\providecommand \href [0]{\begingroup \@sanitize@url \@href}%
\providecommand \@href[1]{\@@startlink{#1}\@@href}%
\providecommand \@@href[1]{\endgroup#1\@@endlink}%
\providecommand \@sanitize@url [0]{\catcode `\\12\catcode `\$12\catcode
  `\&12\catcode `\#12\catcode `\^12\catcode `\_12\catcode `\%12\relax}%
\providecommand \@@startlink[1]{}%
\providecommand \@@endlink[0]{}%
\providecommand \url  [0]{\begingroup\@sanitize@url \@url }%
\providecommand \@url [1]{\endgroup\@href {#1}{\urlprefix }}%
\providecommand \urlprefix  [0]{URL }%
\providecommand \Eprint [0]{\href }%
\providecommand \doibase [0]{http://dx.doi.org/}%
\providecommand \selectlanguage [0]{\@gobble}%
\providecommand \bibinfo  [0]{\@secondoftwo}%
\providecommand \bibfield  [0]{\@secondoftwo}%
\providecommand \translation [1]{[#1]}%
\providecommand \BibitemOpen [0]{}%
\providecommand \bibitemStop [0]{}%
\providecommand \bibitemNoStop [0]{.\EOS\space}%
\providecommand \EOS [0]{\spacefactor3000\relax}%
\providecommand \BibitemShut  [1]{\csname bibitem#1\endcsname}%
\let\auto@bib@innerbib\@empty
\bibitem [{\citenamefont {Matsumoto}\ \emph {et~al.}(2010)\citenamefont
  {Matsumoto}, \citenamefont {Nabeshima},\ and\ \citenamefont
  {Yoshioka}}]{excitednu}%
  \BibitemOpen
  \bibfield  {author} {\bibinfo {author} {\bibfnamefont {S.}~\bibnamefont
  {Matsumoto}}, \bibinfo {author} {\bibfnamefont {T.}~\bibnamefont
  {Nabeshima}}, \ and\ \bibinfo {author} {\bibfnamefont {K.}~\bibnamefont
  {Yoshioka}},\ }\href {\doibase 10.1007/JHEP06(2010)058} {\bibfield  {journal}
  {\bibinfo  {journal} {J. High Energy Phys.}\ }\textbf {\bibinfo {volume}
  {1006}},\ \bibinfo {pages} {058} (\bibinfo {year} {2010})},\ \Eprint
  {http://arxiv.org/abs/1004.3852} {arXiv:1004.3852 [hep-ph]} \BibitemShut
  {NoStop}%
\bibitem [{\citenamefont {Belyaev}\ \emph {et~al.}(2005)\citenamefont
  {Belyaev}, \citenamefont {Leroy},\ and\ \citenamefont
  {Mehdiyev}}]{excitednu2}%
  \BibitemOpen
  \bibfield  {author} {\bibinfo {author} {\bibfnamefont {A.}~\bibnamefont
  {Belyaev}}, \bibinfo {author} {\bibfnamefont {C.}~\bibnamefont {Leroy}}, \
  and\ \bibinfo {author} {\bibfnamefont {R.~R.}\ \bibnamefont {Mehdiyev}},\
  }\href {\doibase 10.1140/epjcd/s2005-02-006-0} {\bibfield  {journal}
  {\bibinfo  {journal} {Eur. Phys. J. C}\ }\textbf {\bibinfo {volume} {41S2}},\
  \bibinfo {pages} {1} (\bibinfo {year} {2005})},\ \Eprint
  {http://arxiv.org/abs/hep-ph/0401066} {arXiv:hep-ph/0401066} \BibitemShut
  {NoStop}%
\bibitem [{\citenamefont {Frampton}\ \emph {et~al.}(2000)\citenamefont
  {Frampton}, \citenamefont {Hung},\ and\ \citenamefont {Sher}}]{bprime}%
  \BibitemOpen
  \bibfield  {author} {\bibinfo {author} {\bibfnamefont {P.~H.}\ \bibnamefont
  {Frampton}}, \bibinfo {author} {\bibfnamefont {P.}~\bibnamefont {Hung}}, \
  and\ \bibinfo {author} {\bibfnamefont {M.}~\bibnamefont {Sher}},\ }\href
  {\doibase 10.1016/S0370-1573(99)00095-2} {\bibfield  {journal} {\bibinfo
  {journal} {Phys. Rept.}\ }\textbf {\bibinfo {volume} {330}},\ \bibinfo
  {pages} {263} (\bibinfo {year} {2000})},\ \Eprint
  {http://arxiv.org/abs/hep-ph/9903387} {arXiv:hep-ph/9903387} \BibitemShut
  {NoStop}%
\bibitem [{\citenamefont {Zee}(1985)}]{zee1}%
  \BibitemOpen
  \bibfield  {author} {\bibinfo {author} {\bibfnamefont {A.}~\bibnamefont
  {Zee}},\ }\href
  {http://www.sciencedirect.com/science/article/pii/0370269385906252}
  {\bibfield  {journal} {\bibinfo  {journal} {Phys. Lett. B}\ }\textbf
  {\bibinfo {volume} {161}},\ \bibinfo {pages} {141} (\bibinfo {year}
  {1985})}\BibitemShut {NoStop}%
\bibitem [{\citenamefont {Zee}(1986)}]{zee2}%
  \BibitemOpen
  \bibfield  {author} {\bibinfo {author} {\bibfnamefont {A.}~\bibnamefont
  {Zee}},\ }\href {http://dx.doi.org/10.1016/0370-2693(88)91584-5} {\bibfield
  {journal} {\bibinfo  {journal} {Nucl. Phys. B}\ }\textbf {\bibinfo {volume}
  {264}},\ \bibinfo {pages} {99} (\bibinfo {year} {1986})}\BibitemShut
  {NoStop}%
\bibitem [{\citenamefont {Babu}(1988)}]{babu}%
  \BibitemOpen
  \bibfield  {author} {\bibinfo {author} {\bibfnamefont {K.~S.}\ \bibnamefont
  {Babu}},\ }\href
  {http://www.sciencedirect.com/science/article/pii/0370269388915845}
  {\bibfield  {journal} {\bibinfo  {journal} {Phys. Lett. B}\ }\textbf
  {\bibinfo {volume} {203}},\ \bibinfo {pages} {132} (\bibinfo {year}
  {1988})}\BibitemShut {NoStop}%
\bibitem [{\citenamefont {Miyazawa}(1966)}]{Miyazawa:1966}%
  \BibitemOpen
  \bibfield  {author} {\bibinfo {author} {\bibfnamefont {H.}~\bibnamefont
  {Miyazawa}},\ }\href {\doibase 10.1143/PTP.36.1266} {\bibfield  {journal}
  {\bibinfo  {journal} {Prog. Theor. Phys.}\ }\textbf {\bibinfo {volume} {36
  (6)}},\ \bibinfo {pages} {1266} (\bibinfo {year} {1966})}\BibitemShut
  {NoStop}%
\bibitem [{\citenamefont {Ramond}(1971)}]{Ramond:1971gb}%
  \BibitemOpen
  \bibfield  {author} {\bibinfo {author} {\bibfnamefont {P.}~\bibnamefont
  {Ramond}},\ }\href {\doibase 10.1103/PhysRevD.3.2415} {\bibfield  {journal}
  {\bibinfo  {journal} {Phys. Rev. D}\ }\textbf {\bibinfo {volume} {3}},\
  \bibinfo {pages} {2415} (\bibinfo {year} {1971})}\BibitemShut {NoStop}%
\bibitem [{\citenamefont {Gol'fand}\ and\ \citenamefont
  {Likhtman}(1971)}]{Golfand:1971iw}%
  \BibitemOpen
  \bibfield  {author} {\bibinfo {author} {\bibfnamefont {Y.~A.}\ \bibnamefont
  {Gol'fand}}\ and\ \bibinfo {author} {\bibfnamefont {E.~P.}\ \bibnamefont
  {Likhtman}},\ }\href@noop {} {\bibfield  {journal} {\bibinfo  {journal} {JETP
  Lett.}\ }\textbf {\bibinfo {volume} {13}},\ \bibinfo {pages} {323} (\bibinfo
  {year} {1971})},\ \bibinfo {note} {[Pisma
  Zh.Eksp.Teor.Fiz.13:452-455,1971]}\BibitemShut {NoStop}%
\bibitem [{\citenamefont {Neveu}\ and\ \citenamefont
  {Schwarz}(1971{\natexlab{a}})}]{Neveu:1971rx}%
  \BibitemOpen
  \bibfield  {author} {\bibinfo {author} {\bibfnamefont {A.}~\bibnamefont
  {Neveu}}\ and\ \bibinfo {author} {\bibfnamefont {J.~H.}\ \bibnamefont
  {Schwarz}},\ }\href {\doibase 10.1016/0550-3213(71)90448-2} {\bibfield
  {journal} {\bibinfo  {journal} {Nucl. Phys. B}\ }\textbf {\bibinfo {volume}
  {31}},\ \bibinfo {pages} {86} (\bibinfo {year}
  {1971}{\natexlab{a}})}\BibitemShut {NoStop}%
\bibitem [{\citenamefont {Neveu}\ and\ \citenamefont
  {Schwarz}(1971{\natexlab{b}})}]{Neveu:1971iv}%
  \BibitemOpen
  \bibfield  {author} {\bibinfo {author} {\bibfnamefont {A.}~\bibnamefont
  {Neveu}}\ and\ \bibinfo {author} {\bibfnamefont {J.~H.}\ \bibnamefont
  {Schwarz}},\ }\href {\doibase 10.1103/PhysRevD.4.1109} {\bibfield  {journal}
  {\bibinfo  {journal} {Phys. Rev. D}\ }\textbf {\bibinfo {volume} {4}},\
  \bibinfo {pages} {1109} (\bibinfo {year} {1971}{\natexlab{b}})}\BibitemShut
  {NoStop}%
\bibitem [{\citenamefont {Gervais}\ and\ \citenamefont
  {Sakita}(1971)}]{Gervais:1971ji}%
  \BibitemOpen
  \bibfield  {author} {\bibinfo {author} {\bibfnamefont {J.}~\bibnamefont
  {Gervais}}\ and\ \bibinfo {author} {\bibfnamefont {B.}~\bibnamefont
  {Sakita}},\ }\href {\doibase 10.1016/0550-3213(71)90351-8} {\bibfield
  {journal} {\bibinfo  {journal} {Nucl. Phys. B}\ }\textbf {\bibinfo {volume}
  {34}},\ \bibinfo {pages} {632} (\bibinfo {year} {1971})}\BibitemShut
  {NoStop}%
\bibitem [{\citenamefont {Volkov}\ and\ \citenamefont
  {Akulov}(1973)}]{Volkov:1973ix}%
  \BibitemOpen
  \bibfield  {author} {\bibinfo {author} {\bibfnamefont {D.~V.}\ \bibnamefont
  {Volkov}}\ and\ \bibinfo {author} {\bibfnamefont {V.~P.}\ \bibnamefont
  {Akulov}},\ }\href {\doibase 10.1016/0370-2693(73)90490-5} {\bibfield
  {journal} {\bibinfo  {journal} {Phys. Lett. B}\ }\textbf {\bibinfo {volume}
  {46}},\ \bibinfo {pages} {109} (\bibinfo {year} {1973})}\BibitemShut
  {NoStop}%
\bibitem [{\citenamefont {Wess}\ and\ \citenamefont
  {Zumino}(1974{\natexlab{a}})}]{Wess:1973kz}%
  \BibitemOpen
  \bibfield  {author} {\bibinfo {author} {\bibfnamefont {J.}~\bibnamefont
  {Wess}}\ and\ \bibinfo {author} {\bibfnamefont {B.}~\bibnamefont {Zumino}},\
  }\href {\doibase 10.1016/0370-2693(74)90578-4} {\bibfield  {journal}
  {\bibinfo  {journal} {Phys. Lett. B}\ }\textbf {\bibinfo {volume} {49}},\
  \bibinfo {pages} {52} (\bibinfo {year} {1974}{\natexlab{a}})}\BibitemShut
  {NoStop}%
\bibitem [{\citenamefont {Wess}\ and\ \citenamefont
  {Zumino}(1974{\natexlab{b}})}]{Wess:1974tw}%
  \BibitemOpen
  \bibfield  {author} {\bibinfo {author} {\bibfnamefont {J.}~\bibnamefont
  {Wess}}\ and\ \bibinfo {author} {\bibfnamefont {B.}~\bibnamefont {Zumino}},\
  }\href {\doibase 10.1016/0550-3213(74)90355-1} {\bibfield  {journal}
  {\bibinfo  {journal} {Nucl. Phys. B}\ }\textbf {\bibinfo {volume} {70}},\
  \bibinfo {pages} {39} (\bibinfo {year} {1974}{\natexlab{b}})}\BibitemShut
  {NoStop}%
\bibitem [{\citenamefont {Rizzo}(1982)}]{DCH0}%
  \BibitemOpen
  \bibfield  {author} {\bibinfo {author} {\bibfnamefont {T.~G.}\ \bibnamefont
  {Rizzo}},\ }\href {\doibase 10.1103/PhysRevD.25.1355} {\bibfield  {journal}
  {\bibinfo  {journal} {Phys. Rev. D}\ }\textbf {\bibinfo {volume} {25}},\
  \bibinfo {pages} {1355} (\bibinfo {year} {1982})}\BibitemShut {NoStop}%
\bibitem [{\citenamefont {Georgi}\ and\ \citenamefont {Machacek}(1985)}]{DCH1}%
  \BibitemOpen
  \bibfield  {author} {\bibinfo {author} {\bibfnamefont {H.}~\bibnamefont
  {Georgi}}\ and\ \bibinfo {author} {\bibfnamefont {M.}~\bibnamefont
  {Machacek}},\ }\href {\doibase 10.1016/0550-3213(85)90325-6} {\bibfield
  {journal} {\bibinfo  {journal} {Nucl. Phys. B}\ }\textbf {\bibinfo {volume}
  {262}},\ \bibinfo {pages} {463 } (\bibinfo {year} {1985})}\BibitemShut
  {NoStop}%
\bibitem [{\citenamefont {Montalvo}\ \emph {et~al.}(2006)\citenamefont
  {Montalvo}, \citenamefont {Jr.}, \citenamefont {Borges},\ and\ \citenamefont
  {Tonasse}}]{DCH3}%
  \BibitemOpen
  \bibfield  {author} {\bibinfo {author} {\bibfnamefont {J.~C.}\ \bibnamefont
  {Montalvo}}, \bibinfo {author} {\bibfnamefont {N.~V.~C.}\ \bibnamefont
  {Jr.}}, \bibinfo {author} {\bibfnamefont {J.~S.}\ \bibnamefont {Borges}}, \
  and\ \bibinfo {author} {\bibfnamefont {M.~D.}\ \bibnamefont {Tonasse}},\
  }\href {\doibase 10.1016/j.nuclphysb.2006.08.013} {\bibfield  {journal}
  {\bibinfo  {journal} {Nucl. Phys. B}\ }\textbf {\bibinfo {volume} {756}},\
  \bibinfo {pages} {1 } (\bibinfo {year} {2006})},\ \bibinfo {note} {erratum:
  ibid. {\bf 796} 422 (2008)}\BibitemShut {NoStop}%
\bibitem [{\citenamefont {Gunion}\ \emph {et~al.}(1990)\citenamefont {Gunion},
  \citenamefont {Vega},\ and\ \citenamefont {Wudka}}]{DCH2}%
  \BibitemOpen
  \bibfield  {author} {\bibinfo {author} {\bibfnamefont {J.~F.}\ \bibnamefont
  {Gunion}}, \bibinfo {author} {\bibfnamefont {R.}~\bibnamefont {Vega}}, \ and\
  \bibinfo {author} {\bibfnamefont {J.}~\bibnamefont {Wudka}},\ }\href
  {\doibase 10.1103/PhysRevD.42.1673} {\bibfield  {journal} {\bibinfo
  {journal} {Phys. Rev. D}\ }\textbf {\bibinfo {volume} {42}},\ \bibinfo
  {pages} {1673} (\bibinfo {year} {1990})}\BibitemShut {NoStop}%
\bibitem [{\citenamefont {Dalitz}(1951)}]{dalitz}%
  \BibitemOpen
  \bibfield  {author} {\bibinfo {author} {\bibfnamefont {R.}~\bibnamefont
  {Dalitz}},\ }\href@noop {} {\bibfield  {journal} {\bibinfo  {journal} {Proc.
  Phys. Soc. A}\ }\textbf {\bibinfo {volume} {64}},\ \bibinfo {pages} {667}
  (\bibinfo {year} {1951})}\BibitemShut {NoStop}%
\bibitem [{\citenamefont {Kroll}\ and\ \citenamefont {Wada}(1955)}]{kroll}%
  \BibitemOpen
  \bibfield  {author} {\bibinfo {author} {\bibfnamefont {N.~M.}\ \bibnamefont
  {Kroll}}\ and\ \bibinfo {author} {\bibfnamefont {W.}~\bibnamefont {Wada}},\
  }\href {\doibase 10.1103/PhysRev.98.1355} {\bibfield  {journal} {\bibinfo
  {journal} {Phys. Rev.}\ }\textbf {\bibinfo {volume} {98}},\ \bibinfo {pages}
  {1355} (\bibinfo {year} {1955})}\BibitemShut {NoStop}%
\bibitem [{\citenamefont {{CMS Collaboration}}(2012{\natexlab{a}})}]{cmsML}%
  \BibitemOpen
  \bibfield  {author} {\bibinfo {author} {\bibnamefont {{CMS Collaboration}}},\
  }\href {\doibase 10.1007/JHEP06(2012)169} {\bibfield  {journal} {\bibinfo
  {journal} {J. High Energy Phys.}\ }\textbf {\bibinfo {volume} {1206}},\
  \bibinfo {pages} {169} (\bibinfo {year} {2012}{\natexlab{a}})},\ \Eprint
  {http://arxiv.org/abs/1204.5341} {arXiv:1204.5341 [hep-ex]} \BibitemShut
  {NoStop}%
\bibitem [{\citenamefont {{ATLAS
  Collaboration}}(2012{\natexlab{a}})}]{atlasSUSYML}%
  \BibitemOpen
  \bibfield  {author} {\bibinfo {author} {\bibnamefont {{ATLAS
  Collaboration}}},\ }\href@noop {} {\bibfield  {journal} {\bibinfo  {journal}
  {Submitted to Phys. Lett. B}\ } (\bibinfo {year} {2012}{\natexlab{a}})},\
  \Eprint {http://arxiv.org/abs/1208.3144} {arXiv:1208.3144 [hep-ex]}
  \BibitemShut {NoStop}%
\bibitem [{\citenamefont {{CDF Collaboration}}(2008)}]{cdfML}%
  \BibitemOpen
  \bibfield  {author} {\bibinfo {author} {\bibnamefont {{CDF Collaboration}}},\
  }\href {\doibase 10.1103/PhysRevLett.101.251801} {\bibfield  {journal}
  {\bibinfo  {journal} {Phys. Rev. Lett.}\ }\textbf {\bibinfo {volume} {101}},\
  \bibinfo {pages} {251801} (\bibinfo {year} {2008})},\ \Eprint
  {http://arxiv.org/abs/0808.2446} {arXiv:0808.2446 [hep-ex]} \BibitemShut
  {NoStop}%
\bibitem [{\citenamefont {{D0 Collaboration}}(2009)}]{dzeroML}%
  \BibitemOpen
  \bibfield  {author} {\bibinfo {author} {\bibnamefont {{D0 Collaboration}}},\
  }\href {\doibase 10.1016/j.physletb.2009.08.011} {\bibfield  {journal}
  {\bibinfo  {journal} {Phys. Lett. B}\ }\textbf {\bibinfo {volume} {680}},\
  \bibinfo {pages} {34} (\bibinfo {year} {2009})},\ \Eprint
  {http://arxiv.org/abs/0901.0646} {arXiv:0901.0646 [hep-ex]} \BibitemShut
  {NoStop}%
\bibitem [{\citenamefont {{ATLAS Collaboration}}(2008)}]{atlas}%
  \BibitemOpen
  \bibfield  {author} {\bibinfo {author} {\bibnamefont {{ATLAS
  Collaboration}}},\ }\href {\doibase 10.1088/1748-0221/3/08/S08003} {\bibfield
   {journal} {\bibinfo  {journal} {JINST}\ }\textbf {\bibinfo {volume} {3}},\
  \bibinfo {pages} {S08003} (\bibinfo {year} {2008})}\BibitemShut {NoStop}%
\bibitem [{Note1()}]{Note1}%
  \BibitemOpen
  \bibinfo {note} {ATLAS uses a right-handed coordinate system with its origin
  at the nominal interaction point (IP) in the center of the detector and the
  $z$-axis along the beam pipe. The $x$-axis points from the IP to the center
  of the LHC ring, and the $y$-axis points upward. Cylindrical coordinates
  $(r,\phi )$ are used in the transverse plane, $\phi $ being the azimuthal
  angle around the beam pipe. The pseudorapidity is defined in terms of the
  polar angle $\theta $ as $\protect \ensuremath {\etaa }=-\mathop {\mathgroup
  \symoperators ln}\nolimits \mathop {\mathgroup \symoperators tan}\nolimits
  (\theta /2)$. The variable $\protect \ensuremath {\Delta R}$ is used to
  evaluate the distance between objects, and is defined as: $\protect
  \ensuremath {\Delta R}= \protect \sqrt {(\Delta \phi )^2 + (\Delta \protect
  \ensuremath {\etaa })^2}$\label {geometryfootnote}}\BibitemShut {NoStop}%
\bibitem [{\citenamefont {{ATLAS
  Collaboration}}(2012{\natexlab{b}})}]{trigger}%
  \BibitemOpen
  \bibfield  {author} {\bibinfo {author} {\bibnamefont {{ATLAS
  Collaboration}}},\ }\href {\doibase 10.1140/epjc/s10052-011-1849-1}
  {\bibfield  {journal} {\bibinfo  {journal} {Eur. Phys. J. C}\ }\textbf
  {\bibinfo {volume} {72}},\ \bibinfo {pages} {1849} (\bibinfo {year}
  {2012}{\natexlab{b}})},\ \Eprint {http://arxiv.org/abs/1110.1530}
  {arXiv:1110.1530 [hep-ex]} \BibitemShut {NoStop}%
\bibitem [{\citenamefont {Agostinelli}\ \emph {et~al.}(2003)\citenamefont
  {Agostinelli} \emph {et~al.}}]{geant}%
  \BibitemOpen
  \bibfield  {author} {\bibinfo {author} {\bibfnamefont {S.}~\bibnamefont
  {Agostinelli}} \emph {et~al.},\ }\href {\doibase
  10.1016/S0168-9002(03)01368-8} {\bibfield  {journal} {\bibinfo  {journal}
  {Nucl. Instrum. Methods}\ }\textbf {\bibinfo {volume} {506}},\ \bibinfo
  {pages} {250 } (\bibinfo {year} {2003})}\BibitemShut {NoStop}%
\bibitem [{\citenamefont {{ATLAS
  Collaboration}}(2012{\natexlab{c}})}]{elecperf}%
  \BibitemOpen
  \bibfield  {author} {\bibinfo {author} {\bibnamefont {{ATLAS
  Collaboration}}},\ }\href {\doibase 10.1140/epjc/s10052-012-1909-1}
  {\bibfield  {journal} {\bibinfo  {journal} {Eur. Phys. J. C}\ }\textbf
  {\bibinfo {volume} {72}},\ \bibinfo {pages} {1909} (\bibinfo {year}
  {2012}{\natexlab{c}})},\ \Eprint {http://arxiv.org/abs/1110.3174}
  {arXiv:1110.3174 [hep-ex]} \BibitemShut {NoStop}%
\bibitem [{\citenamefont {{ATLAS
  Collaboration}}(2011{\natexlab{a}})}]{muonperf}%
  \BibitemOpen
  \bibfield  {author} {\bibinfo {author} {\bibnamefont {{ATLAS
  Collaboration}}},\ }\href {https://cdsweb.cern.ch/record/1345743} {\bibfield
  {journal} {\bibinfo  {journal} {ATLAS-CONF-2011-063}\ } (\bibinfo {year}
  {2011}{\natexlab{a}})},\ \bibinfo {note}
  {\url{https://cdsweb.cern.ch/record/1345743}}\BibitemShut {NoStop}%
\bibitem [{\citenamefont {Gleisberg}\ \emph {et~al.}(2009)\citenamefont
  {Gleisberg}, \citenamefont {H{\"o}che}, \citenamefont {Krauss}, \citenamefont
  {Sch{\"o}nherr}, \citenamefont {Schumann} \emph {et~al.}}]{sherpa}%
  \BibitemOpen
  \bibfield  {author} {\bibinfo {author} {\bibfnamefont {T.}~\bibnamefont
  {Gleisberg}}, \bibinfo {author} {\bibfnamefont {S.}~\bibnamefont
  {H{\"o}che}}, \bibinfo {author} {\bibfnamefont {F.}~\bibnamefont {Krauss}},
  \bibinfo {author} {\bibfnamefont {M.}~\bibnamefont {Sch{\"o}nherr}}, \bibinfo
  {author} {\bibfnamefont {S.}~\bibnamefont {Schumann}},  \emph {et~al.},\
  }\href {\doibase 10.1088/1126-6708/2009/02/007} {\bibfield  {journal}
  {\bibinfo  {journal} {J. High Energy Phys.}\ }\textbf {\bibinfo {volume}
  {0902}},\ \bibinfo {pages} {007} (\bibinfo {year} {2009})},\ \Eprint
  {http://arxiv.org/abs/0811.4622} {arXiv:0811.4622 [hep-ph]} \BibitemShut
  {NoStop}%
\bibitem [{\citenamefont {Melia}\ \emph {et~al.}(2011)\citenamefont {Melia},
  \citenamefont {Nason}, \citenamefont {Rontsch},\ and\ \citenamefont
  {Zanderighi}}]{powheg}%
  \BibitemOpen
  \bibfield  {author} {\bibinfo {author} {\bibfnamefont {T.}~\bibnamefont
  {Melia}}, \bibinfo {author} {\bibfnamefont {P.}~\bibnamefont {Nason}},
  \bibinfo {author} {\bibfnamefont {R.}~\bibnamefont {Rontsch}}, \ and\
  \bibinfo {author} {\bibfnamefont {G.}~\bibnamefont {Zanderighi}},\ }\href
  {\doibase 10.1007/JHEP11(2011)078} {\bibfield  {journal} {\bibinfo  {journal}
  {J. High Energy Phys.}\ }\textbf {\bibinfo {volume} {1111}},\ \bibinfo
  {pages} {078} (\bibinfo {year} {2011})},\ \Eprint
  {http://arxiv.org/abs/1107.5051} {arXiv:1107.5051 [hep-ph]} \BibitemShut
  {NoStop}%
\bibitem [{\citenamefont {Alwall}\ \emph {et~al.}(2011)\citenamefont {Alwall},
  \citenamefont {Herquet}, \citenamefont {Maltoni}, \citenamefont {Mattelaer},\
  and\ \citenamefont {Stelzer}}]{madgraph}%
  \BibitemOpen
  \bibfield  {author} {\bibinfo {author} {\bibfnamefont {J.}~\bibnamefont
  {Alwall}}, \bibinfo {author} {\bibfnamefont {M.}~\bibnamefont {Herquet}},
  \bibinfo {author} {\bibfnamefont {F.}~\bibnamefont {Maltoni}}, \bibinfo
  {author} {\bibfnamefont {O.}~\bibnamefont {Mattelaer}}, \ and\ \bibinfo
  {author} {\bibfnamefont {T.}~\bibnamefont {Stelzer}},\ }\href {\doibase
  10.1007/JHEP06(2011)128} {\bibfield  {journal} {\bibinfo  {journal} {J. High
  Energy Phys.}\ }\textbf {\bibinfo {volume} {1106}},\ \bibinfo {pages} {128}
  (\bibinfo {year} {2011})},\ \Eprint {http://arxiv.org/abs/1106.0522}
  {arXiv:1106.0522 [hep-ph]} \BibitemShut {NoStop}%
\bibitem [{\citenamefont {Sj{\"o}strand}\ \emph {et~al.}(2001)\citenamefont
  {Sj{\"o}strand}, \citenamefont {Ed{\'e}n}, \citenamefont {Friberg},
  \citenamefont {L{\"o}nnblad}, \citenamefont {Miu}, \citenamefont {Mrenna},\
  and\ \citenamefont {Emanuel}}]{pythia}%
  \BibitemOpen
  \bibfield  {author} {\bibinfo {author} {\bibfnamefont {T.}~\bibnamefont
  {Sj{\"o}strand}}, \bibinfo {author} {\bibfnamefont {P.}~\bibnamefont
  {Ed{\'e}n}}, \bibinfo {author} {\bibfnamefont {C.}~\bibnamefont {Friberg}},
  \bibinfo {author} {\bibfnamefont {L.}~\bibnamefont {L{\"o}nnblad}}, \bibinfo
  {author} {\bibfnamefont {G.}~\bibnamefont {Miu}}, \bibinfo {author}
  {\bibfnamefont {S.}~\bibnamefont {Mrenna}}, \ and\ \bibinfo {author}
  {\bibfnamefont {N.}~\bibnamefont {Emanuel}},\ }\href {\doibase
  10.1016/S0010-4655(00)00236-8} {\bibfield  {journal} {\bibinfo  {journal}
  {Comput. Phys. Comm.}\ }\textbf {\bibinfo {volume} {135}},\ \bibinfo {pages}
  {238} (\bibinfo {year} {2001})},\ \Eprint
  {http://arxiv.org/abs/hep-ph/0010017} {arXiv:hep-ph/0010017} \BibitemShut
  {NoStop}%
\bibitem [{\citenamefont {Campbell}\ and\ \citenamefont {Ellis}(2012)}]{ttW}%
  \BibitemOpen
  \bibfield  {author} {\bibinfo {author} {\bibfnamefont {J.~M.}\ \bibnamefont
  {Campbell}}\ and\ \bibinfo {author} {\bibfnamefont {R.~K.}\ \bibnamefont
  {Ellis}},\ }\href {\doibase 10.1007/JHEP07(2012)052} {\bibfield  {journal}
  {\bibinfo  {journal} {J. High Energy Phys.}\ }\textbf {\bibinfo {volume}
  {1207}},\ \bibinfo {pages} {052} (\bibinfo {year} {2012})},\ \Eprint
  {http://arxiv.org/abs/1204.5678} {arXiv:1204.5678 [hep-ph]} \BibitemShut
  {NoStop}%
\bibitem [{\citenamefont {Garzelli}\ \emph {et~al.}(2012)\citenamefont
  {Garzelli}, \citenamefont {Kardos}, \citenamefont {Papadopoulos},\ and\
  \citenamefont {Trocsanyi}}]{ttZ}%
  \BibitemOpen
  \bibfield  {author} {\bibinfo {author} {\bibfnamefont {M.}~\bibnamefont
  {Garzelli}}, \bibinfo {author} {\bibfnamefont {A.}~\bibnamefont {Kardos}},
  \bibinfo {author} {\bibfnamefont {C.}~\bibnamefont {Papadopoulos}}, \ and\
  \bibinfo {author} {\bibfnamefont {Z.}~\bibnamefont {Trocsanyi}},\ }\href
  {\doibase 10.1103/PhysRevD.85.074022} {\bibfield  {journal} {\bibinfo
  {journal} {Phys. Rev. D}\ }\textbf {\bibinfo {volume} {85}},\ \bibinfo
  {pages} {074022} (\bibinfo {year} {2012})},\ \Eprint
  {http://arxiv.org/abs/1111.1444} {arXiv:1111.1444 [hep-ph]} \BibitemShut
  {NoStop}%
\bibitem [{\citenamefont {Frixione}\ and\ \citenamefont
  {Webber}(2002)}]{mcatnlo}%
  \BibitemOpen
  \bibfield  {author} {\bibinfo {author} {\bibfnamefont {S.}~\bibnamefont
  {Frixione}}\ and\ \bibinfo {author} {\bibfnamefont {B.~R.}\ \bibnamefont
  {Webber}},\ }\href@noop {} {\bibfield  {journal} {\bibinfo  {journal} {J.
  High Energy Phys.}\ }\textbf {\bibinfo {volume} {06}},\ \bibinfo {pages}
  {029} (\bibinfo {year} {2002})},\ \Eprint
  {http://arxiv.org/abs/hep-ph/0204244} {arXiv:hep-ph/0204244} \BibitemShut
  {NoStop}%
\bibitem [{\citenamefont {Corcella}\ \emph {et~al.}(2001)\citenamefont
  {Corcella}, \citenamefont {Knowles}, \citenamefont {Marchesini},
  \citenamefont {Moretti}, \citenamefont {Odagiri} \emph {et~al.}}]{herwig}%
  \BibitemOpen
  \bibfield  {author} {\bibinfo {author} {\bibfnamefont {G.}~\bibnamefont
  {Corcella}}, \bibinfo {author} {\bibfnamefont {I.}~\bibnamefont {Knowles}},
  \bibinfo {author} {\bibfnamefont {G.}~\bibnamefont {Marchesini}}, \bibinfo
  {author} {\bibfnamefont {S.}~\bibnamefont {Moretti}}, \bibinfo {author}
  {\bibfnamefont {K.}~\bibnamefont {Odagiri}},  \emph {et~al.},\ }\href@noop {}
  {\bibfield  {journal} {\bibinfo  {journal} {J. High Energy Phys.}\ }\textbf
  {\bibinfo {volume} {0101}},\ \bibinfo {pages} {010} (\bibinfo {year}
  {2001})},\ \Eprint {http://arxiv.org/abs/hep-ph/0011363}
  {arXiv:hep-ph/0011363} \BibitemShut {NoStop}%
\bibitem [{\citenamefont {Butterworth}\ \emph {et~al.}(1996)\citenamefont
  {Butterworth}, \citenamefont {Forshaw},\ and\ \citenamefont
  {Seymour}}]{jimmy}%
  \BibitemOpen
  \bibfield  {author} {\bibinfo {author} {\bibfnamefont {J.}~\bibnamefont
  {Butterworth}}, \bibinfo {author} {\bibfnamefont {J.~R.}\ \bibnamefont
  {Forshaw}}, \ and\ \bibinfo {author} {\bibfnamefont {M.}~\bibnamefont
  {Seymour}},\ }\href {\doibase 10.1007/s002880050286} {\bibfield  {journal}
  {\bibinfo  {journal} {Z. Phys. C}\ }\textbf {\bibinfo {volume} {72}},\
  \bibinfo {pages} {637} (\bibinfo {year} {1996})},\ \Eprint
  {http://arxiv.org/abs/hep-ph/9601371} {arXiv:hep-ph/9601371} \BibitemShut
  {NoStop}%
\bibitem [{\citenamefont {Mangano}\ \emph {et~al.}(2003)\citenamefont
  {Mangano}, \citenamefont {Moretti}, \citenamefont {Piccinini}, \citenamefont
  {Pittau},\ and\ \citenamefont {Polosa}}]{alpgen}%
  \BibitemOpen
  \bibfield  {author} {\bibinfo {author} {\bibfnamefont {M.~L.}\ \bibnamefont
  {Mangano}}, \bibinfo {author} {\bibfnamefont {M.}~\bibnamefont {Moretti}},
  \bibinfo {author} {\bibfnamefont {F.}~\bibnamefont {Piccinini}}, \bibinfo
  {author} {\bibfnamefont {R.}~\bibnamefont {Pittau}}, \ and\ \bibinfo {author}
  {\bibfnamefont {A.~D.}\ \bibnamefont {Polosa}},\ }\href@noop {} {\bibfield
  {journal} {\bibinfo  {journal} {J. High Energy Phys.}\ }\textbf {\bibinfo
  {volume} {0307}},\ \bibinfo {pages} {001} (\bibinfo {year} {2003})},\ \Eprint
  {http://arxiv.org/abs/hep-ph/0206293} {arXiv:hep-ph/0206293} \BibitemShut
  {NoStop}%
\bibitem [{\citenamefont {Aliev}\ \emph {et~al.}(2011)\citenamefont {Aliev},
  \citenamefont {Lacker}, \citenamefont {Langenfeld}, \citenamefont {Moch},
  \citenamefont {Uwer} \emph {et~al.}}]{hathor}%
  \BibitemOpen
  \bibfield  {author} {\bibinfo {author} {\bibfnamefont {M.}~\bibnamefont
  {Aliev}}, \bibinfo {author} {\bibfnamefont {H.}~\bibnamefont {Lacker}},
  \bibinfo {author} {\bibfnamefont {U.}~\bibnamefont {Langenfeld}}, \bibinfo
  {author} {\bibfnamefont {S.}~\bibnamefont {Moch}}, \bibinfo {author}
  {\bibfnamefont {P.}~\bibnamefont {Uwer}},  \emph {et~al.},\ }\href {\doibase
  10.1016/j.cpc.2010.12.040} {\bibfield  {journal} {\bibinfo  {journal}
  {Comput. Phys. Commun.}\ }\textbf {\bibinfo {volume} {182}},\ \bibinfo
  {pages} {1034} (\bibinfo {year} {2011})},\ \Eprint
  {http://arxiv.org/abs/1007.1327} {arXiv:1007.1327 [hep-ph]} \BibitemShut
  {NoStop}%
\bibitem [{\citenamefont {Lai}\ \emph {et~al.}(2010)\citenamefont {Lai},
  \citenamefont {Guzzi}, \citenamefont {Huston}, \citenamefont {Li},
  \citenamefont {Nadolsky} \emph {et~al.}}]{ct10}%
  \BibitemOpen
  \bibfield  {author} {\bibinfo {author} {\bibfnamefont {H.-L.}\ \bibnamefont
  {Lai}}, \bibinfo {author} {\bibfnamefont {M.}~\bibnamefont {Guzzi}}, \bibinfo
  {author} {\bibfnamefont {J.}~\bibnamefont {Huston}}, \bibinfo {author}
  {\bibfnamefont {Z.}~\bibnamefont {Li}}, \bibinfo {author} {\bibfnamefont
  {P.~M.}\ \bibnamefont {Nadolsky}},  \emph {et~al.},\ }\href {\doibase
  10.1103/PhysRevD.82.074024} {\bibfield  {journal} {\bibinfo  {journal} {Phys.
  Rev. D}\ }\textbf {\bibinfo {volume} {82}},\ \bibinfo {pages} {074024}
  (\bibinfo {year} {2010})},\ \Eprint {http://arxiv.org/abs/1007.2241}
  {arXiv:1007.2241 [hep-ph]} \BibitemShut {NoStop}%
\bibitem [{\citenamefont {Sherstnev}\ and\ \citenamefont
  {Thorne}(2008)}]{mrst}%
  \BibitemOpen
  \bibfield  {author} {\bibinfo {author} {\bibfnamefont {A.}~\bibnamefont
  {Sherstnev}}\ and\ \bibinfo {author} {\bibfnamefont {R.~S.}\ \bibnamefont
  {Thorne}},\ }\href {\doibase 10.1140/epjc/s10052-008-0610-x} {\bibfield
  {journal} {\bibinfo  {journal} {Eur. Phys. J. C}\ }\textbf {\bibinfo {volume}
  {55}},\ \bibinfo {pages} {553} (\bibinfo {year} {2008})},\ \Eprint
  {http://arxiv.org/abs/0711.2473} {arXiv:0711.2473 [hep-ph]} \BibitemShut
  {NoStop}%
\bibitem [{\citenamefont {Pumplin}\ \emph {et~al.}(2002)\citenamefont
  {Pumplin}, \citenamefont {Stump}, \citenamefont {Huston}, \citenamefont
  {Lai}, \citenamefont {Nadolsky} \emph {et~al.}}]{cteq6l1}%
  \BibitemOpen
  \bibfield  {author} {\bibinfo {author} {\bibfnamefont {J.}~\bibnamefont
  {Pumplin}}, \bibinfo {author} {\bibfnamefont {D.}~\bibnamefont {Stump}},
  \bibinfo {author} {\bibfnamefont {J.}~\bibnamefont {Huston}}, \bibinfo
  {author} {\bibfnamefont {H.}~\bibnamefont {Lai}}, \bibinfo {author}
  {\bibfnamefont {P.~M.}\ \bibnamefont {Nadolsky}},  \emph {et~al.},\
  }\href@noop {} {\bibfield  {journal} {\bibinfo  {journal} {J. High Energy
  Phys.}\ }\textbf {\bibinfo {volume} {0207}},\ \bibinfo {pages} {012}
  (\bibinfo {year} {2002})},\ \Eprint {http://arxiv.org/abs/hep-ph/0201195}
  {arXiv:hep-ph/0201195} \BibitemShut {NoStop}%
\bibitem [{\citenamefont {Nadolsky}\ \emph {et~al.}(2008)\citenamefont
  {Nadolsky}, \citenamefont {Lai}, \citenamefont {Cao}, \citenamefont {Huston},
  \citenamefont {Pumplin} \emph {et~al.}}]{cteq66}%
  \BibitemOpen
  \bibfield  {author} {\bibinfo {author} {\bibfnamefont {P.~M.}\ \bibnamefont
  {Nadolsky}}, \bibinfo {author} {\bibfnamefont {H.-L.}\ \bibnamefont {Lai}},
  \bibinfo {author} {\bibfnamefont {Q.-H.}\ \bibnamefont {Cao}}, \bibinfo
  {author} {\bibfnamefont {J.}~\bibnamefont {Huston}}, \bibinfo {author}
  {\bibfnamefont {J.}~\bibnamefont {Pumplin}},  \emph {et~al.},\ }\href
  {\doibase 10.1103/PhysRevD.78.013004} {\bibfield  {journal} {\bibinfo
  {journal} {Phys. Rev. D}\ }\textbf {\bibinfo {volume} {78}},\ \bibinfo
  {pages} {013004} (\bibinfo {year} {2008})},\ \Eprint
  {http://arxiv.org/abs/0802.0007} {arXiv:0802.0007 [hep-ph]} \BibitemShut
  {NoStop}%
\bibitem [{\citenamefont {{ATLAS
  Collaboration}}(2011{\natexlab{b}})}]{luminositypaper}%
  \BibitemOpen
  \bibfield  {author} {\bibinfo {author} {\bibnamefont {{ATLAS
  Collaboration}}},\ }\href {\doibase 10.1140/epjc/s10052-011-1630-5}
  {\bibfield  {journal} {\bibinfo  {journal} {Eur. Phys. J. C}\ }\textbf
  {\bibinfo {volume} {71}},\ \bibinfo {pages} {1630} (\bibinfo {year}
  {2011}{\natexlab{b}})},\ \Eprint {http://arxiv.org/abs/1101.2185}
  {arXiv:1101.2185 [hep-ex]} \BibitemShut {NoStop}%
\bibitem [{\citenamefont {{ATLAS
  Collaboration}}(2012{\natexlab{d}})}]{atlasHWW}%
  \BibitemOpen
  \bibfield  {author} {\bibinfo {author} {\bibnamefont {{ATLAS
  Collaboration}}},\ }\href@noop {} {\bibfield  {journal} {\bibinfo  {journal}
  {Submitted to Phys. Lett. B}\ } (\bibinfo {year} {2012}{\natexlab{d}})},\
  \Eprint {http://arxiv.org/abs/1206.6074} {arXiv:1206.6074 [hep-ex]}
  \BibitemShut {NoStop}%
\bibitem [{\citenamefont {Cacciari}\ \emph {et~al.}(2012)\citenamefont
  {Cacciari}, \citenamefont {Salam},\ and\ \citenamefont {Soyez}}]{fastjet}%
  \BibitemOpen
  \bibfield  {author} {\bibinfo {author} {\bibfnamefont {M.}~\bibnamefont
  {Cacciari}}, \bibinfo {author} {\bibfnamefont {G.~P.}\ \bibnamefont {Salam}},
  \ and\ \bibinfo {author} {\bibfnamefont {G.}~\bibnamefont {Soyez}},\ }\href
  {\doibase 10.1140/epjc/s10052-012-1896-2} {\bibfield  {journal} {\bibinfo
  {journal} {Eur. Phys. J. C}\ }\textbf {\bibinfo {volume} {72}},\ \bibinfo
  {pages} {1896} (\bibinfo {year} {2012})},\ \Eprint
  {http://arxiv.org/abs/1111.6097} {arXiv:1111.6097 [hep-ph]} \BibitemShut
  {NoStop}%
\bibitem [{\citenamefont {Cacciari}\ \emph {et~al.}(2008)\citenamefont
  {Cacciari}, \citenamefont {Salam},\ and\ \citenamefont {Soyez}}]{antikt}%
  \BibitemOpen
  \bibfield  {author} {\bibinfo {author} {\bibfnamefont {M.}~\bibnamefont
  {Cacciari}}, \bibinfo {author} {\bibfnamefont {G.~P.}\ \bibnamefont {Salam}},
  \ and\ \bibinfo {author} {\bibfnamefont {G.}~\bibnamefont {Soyez}},\ }\href
  {\doibase 10.1088/1126-6708/2008/04/063} {\bibfield  {journal} {\bibinfo
  {journal} {J. High Energy Phys.}\ }\textbf {\bibinfo {volume} {04}},\
  \bibinfo {pages} {063} (\bibinfo {year} {2008})},\ \Eprint
  {http://arxiv.org/abs/0802.1189} {arXiv:0802.1189 [hep-ph]} \BibitemShut
  {NoStop}%
\bibitem [{\citenamefont {{ATLAS
  Collaboration}}(2011{\natexlab{c}})}]{jetcorr}%
  \BibitemOpen
  \bibfield  {author} {\bibinfo {author} {\bibnamefont {{ATLAS
  Collaboration}}},\ }\href@noop {} {\bibfield  {journal} {\bibinfo  {journal}
  {Submitted to Eur. Phys. J. C}\ } (\bibinfo {year} {2011}{\natexlab{c}})},\
  \Eprint {http://arxiv.org/abs/1112.6426} {arXiv:1112.6426 [hep-ex]}
  \BibitemShut {NoStop}%
\bibitem [{\citenamefont {{ATLAS Collaboration}}(2010)}]{JER}%
  \BibitemOpen
  \bibfield  {author} {\bibinfo {author} {\bibnamefont {{ATLAS
  Collaboration}}},\ }\href {https://cdsweb.cern.ch/record/1281311} {\bibfield
  {journal} {\bibinfo  {journal} {ATLAS-CONF-2010-054}\ } (\bibinfo {year}
  {2010})},\ \bibinfo {note}
  {\url{https://cdsweb.cern.ch/record/1281311}}\BibitemShut {NoStop}%
\bibitem [{\citenamefont {{ATLAS Collaboration}}(2011{\natexlab{d}})}]{tauid}%
  \BibitemOpen
  \bibfield  {author} {\bibinfo {author} {\bibnamefont {{ATLAS
  Collaboration}}},\ }\href {https://cdsweb.cern.ch/record/1398195} {\bibfield
  {journal} {\bibinfo  {journal} {ATLAS-CONF-2011-152}\ } (\bibinfo {year}
  {2011}{\natexlab{d}})},\ \bibinfo {note}
  {\url{https://cdsweb.cern.ch/record/1398195}}\BibitemShut {NoStop}%
\bibitem [{\citenamefont {Beringer}\ \emph {et~al.}(2012)\citenamefont
  {Beringer} \emph {et~al.}}]{pdg}%
  \BibitemOpen
  \bibfield  {author} {\bibinfo {author} {\bibfnamefont {J.}~\bibnamefont
  {Beringer}} \emph {et~al.} (\bibinfo {collaboration} {Particle Data Group}),\
  }\href {\doibase 10.1103/PhysRevD.86.010001} {\bibfield  {journal} {\bibinfo
  {journal} {Phys. Rev. D}\ }\textbf {\bibinfo {volume} {86}},\ \bibinfo
  {pages} {010001} (\bibinfo {year} {2012})}\BibitemShut {NoStop}%
\bibitem [{\citenamefont {{ATLAS
  Collaboration}}(2012{\natexlab{e}})}]{atlasphoton}%
  \BibitemOpen
  \bibfield  {author} {\bibinfo {author} {\bibnamefont {{ATLAS
  Collaboration}}},\ }\href {\doibase 10.1103/PhysRevD.85.092014} {\bibfield
  {journal} {\bibinfo  {journal} {Phys. Rev. D}\ }\textbf {\bibinfo {volume}
  {85}},\ \bibinfo {pages} {092014} (\bibinfo {year}
  {2012}{\natexlab{e}})}\BibitemShut {NoStop}%
\bibitem [{\citenamefont {{ATLAS Collaboration}}(2011{\natexlab{e}})}]{mv1}%
  \BibitemOpen
  \bibfield  {author} {\bibinfo {author} {\bibnamefont {{ATLAS
  Collaboration}}},\ }\href {https://cdsweb.cern.ch/record/1369219} {\bibfield
  {journal} {\bibinfo  {journal} {ATLAS-CONF-2011-102}\ } (\bibinfo {year}
  {2011}{\natexlab{e}})},\ \bibinfo {note}
  {\url{https://cdsweb.cern.ch/record/1369219}}\BibitemShut {NoStop}%
\bibitem [{\citenamefont {{ATLAS
  Collaboration}}(2012{\natexlab{f}})}]{samesign}%
  \BibitemOpen
  \bibfield  {author} {\bibinfo {author} {\bibnamefont {{ATLAS
  Collaboration}}},\ }\href {\doibase 10.1103/PhysRevD.85.032004} {\bibfield
  {journal} {\bibinfo  {journal} {Phys. Rev. D}\ }\textbf {\bibinfo {volume}
  {85}},\ \bibinfo {pages} {032004} (\bibinfo {year}
  {2012}{\natexlab{f}})}\BibitemShut {NoStop}%
\bibitem [{\citenamefont {{ATLAS Collaboration}}(2011{\natexlab{f}})}]{Zprime}%
  \BibitemOpen
  \bibfield  {author} {\bibinfo {author} {\bibnamefont {{ATLAS
  Collaboration}}},\ }\href {\doibase 10.1103/PhysRevLett.107.272002}
  {\bibfield  {journal} {\bibinfo  {journal} {Phys. Rev. Lett.}\ }\textbf
  {\bibinfo {volume} {107}},\ \bibinfo {pages} {272002} (\bibinfo {year}
  {2011}{\natexlab{f}})}\BibitemShut {NoStop}%
\bibitem [{\citenamefont {Read}(2002)}]{cls}%
  \BibitemOpen
  \bibfield  {author} {\bibinfo {author} {\bibfnamefont {A.~L.}\ \bibnamefont
  {Read}},\ }\href {http://stacks.iop.org/0954-3899/28/i=10/a=313} {\bibfield
  {journal} {\bibinfo  {journal} {J. of Phys. G}\ }\textbf {\bibinfo {volume}
  {28}},\ \bibinfo {pages} {2693} (\bibinfo {year} {2002})}\BibitemShut
  {NoStop}%
\bibitem [{\citenamefont {{ATLAS
  Collaboration}}(2012{\natexlab{g}})}]{atlasDCH2012}%
  \BibitemOpen
  \bibfield  {author} {\bibinfo {author} {\bibnamefont {{ATLAS
  Collaboration}}},\ }\href@noop {} {\bibfield  {journal} {\bibinfo  {journal}
  {Submitted to the Eur. Phys. J. C}\ } (\bibinfo {year}
  {2012}{\natexlab{g}})},\ \Eprint {http://arxiv.org/abs/1210.5070}
  {arXiv:1210.5070 [hep-ex]} \BibitemShut {NoStop}%
\bibitem [{\citenamefont {{CMS
  Collaboration}}(2012{\natexlab{b}})}]{cmsDCH2012ya}%
  \BibitemOpen
  \bibfield  {author} {\bibinfo {author} {\bibnamefont {{CMS Collaboration}}},\
  }\href@noop {} {\bibfield  {journal} {\bibinfo  {journal} {Submitted to the
  Eur. Phys. J. C}\ } (\bibinfo {year} {2012}{\natexlab{b}})},\ \Eprint
  {http://arxiv.org/abs/1207.2666} {arXiv:1207.2666 [hep-ex]} \BibitemShut
  {NoStop}%
\end{thebibliography}%

\clearpage

\appendix 

\section{Tables of expected and observed event yields}
\label{app:yields}

The expected and observed event yields for all signal regions under study are shown in Tables~\ref{t:Result_All_HTLep}-\ref{t:Result_HighMET_ST}.

\begin{table*}[tbp]\footnotesize
 \begin{center}
   \caption{Results for the $\Htlep$~signal regions.  Irreducible sources include all backgrounds estimated with MC simulation.  Results are presented in number of expected events as $N~\pm$ (statistical uncertainty) $\pm$ (systematic uncertainty).}
  \begin{tabular}{l c c c c}
\hline
$\Htlep \geq$ &Irreducible	&Reducible	&Total Exp.	&Observed\\\hline
\multicolumn{5}{c}{\threeL, off-$Z$} \\
\hline
   0 \GeV	&54~$\pm 4$~$\pm 7$	&54~$\pm 6$~$\pm 23$	&107~$\pm 7$~$\pm 24$	&99\\
 100 \GeV	&32~$\pm 2$~$\pm 4$	&32~$\pm 4$~$\pm 16$	&65~$\pm 4$~$\pm 16$	&62\\
 150 \GeV	&22~$\pm 1$~$\pm 3$	&15~$\pm 2$~$\pm 8$	&37~$\pm 3$~$\pm 8$	&27\\
 200 \GeV	&9.7~$\pm 0.6$~$\pm 1.5$	&6~$\pm 2$~$\pm 4$	&16~$\pm 2$~$\pm 4$	&15\\
 300 \GeV	&3.6~$\pm 0.5$~$\pm 0.5$	&2.5~$\pm 1.2$~$\pm 1.8$	&6.2~$\pm 1.3$~$\pm 1.9$	&4\\
\hline
\multicolumn{5}{c}{\twoLoneT, off-$Z$} \\
\hline
   0 \GeV	&6.4~$\pm 0.4$~$\pm 1.0$	&214~$\pm 5$~$\pm 50$	&220~$\pm 5$~$\pm 50$	&226\\
 100 \GeV	&4.4~$\pm 0.3$~$\pm 0.6$	&109~$\pm 3$~$\pm 26$	&113~$\pm 3$~$\pm 26$	&113\\
 150 \GeV	&1.7~$\pm 0.2$~$\pm 0.3$	&46~$\pm 2$~$\pm 11$	&47~$\pm 2$~$\pm 11$	&42\\
 200 \GeV	&0.8~$\pm 0.1$~$\pm 0.1$	&17~$\pm 1$~$\pm 4$	&17~$\pm 1$~$\pm 4$	&15\\
 300 \GeV	&0.2~$\pm 0.1$~$\pm 0.0$	&2.5~$\pm 0.4$~$\pm 0.6$	&2.7~$\pm 0.4$~$\pm 0.6$	&1\\
\hline
\multicolumn{5}{c}{\threeL, on-$Z$} \\
\hline
   0 \GeV	&389~$\pm 5$~$\pm 50$	&120~$\pm 8$~$\pm 40$	&508~$\pm 10$~$\pm 70$	&588\\
 100 \GeV	&285~$\pm 4$~$\pm 40$	&71~$\pm 6$~$\pm 26$	&356~$\pm 7$~$\pm 50$	&422\\
 150 \GeV	&122~$\pm 2$~$\pm 17$	&14~$\pm 3$~$\pm 7$	&136~$\pm 4$~$\pm 18$	&151\\
 200 \GeV	&49~$\pm 1$~$\pm 7$	&5~$\pm 2$~$\pm 4$	&54~$\pm 2$~$\pm 8$	&60\\
 300 \GeV	&12.3~$\pm 0.7$~$\pm 1.6$	&0.5~$\pm 0.5$~$\pm 0.5$	&12.7~$\pm 0.9$~$\pm 1.7$	&18\\
\hline
\multicolumn{5}{c}{\twoLoneT, on-$Z$} \\
\hline
   0 \GeV	&13.2~$\pm 0.5$~$\pm 2.2$	&1050~$\pm 10$~$\pm 260$	&1060~$\pm 10$~$\pm 260$	&914\\
 100 \GeV	&11.1~$\pm 0.5$~$\pm 1.9$	&670~$\pm10$~$\pm 160$	&680~$\pm10$~$\pm 160$	&587\\
 150 \GeV	&4.5~$\pm 0.3$~$\pm 0.8$	&66~$\pm 2$~$\pm 16$	&71~$\pm 2$~$\pm 16$	&75\\
 200 \GeV	&1.8~$\pm 0.2$~$\pm 0.3$	&19~$\pm 1$~$\pm 5$	&21~$\pm 1$~$\pm 5$	&24\\
 300 \GeV	&0.5~$\pm 0.1$~$\pm 0.1$	&3.0~$\pm 0.5$~$\pm 0.8$	&3.5~$\pm 0.5$~$\pm 0.8$	&7\\
\hline
\end{tabular}
   \label{t:Result_All_HTLep}
\end{center}
\end{table*}

\begin{table*}[tbp]\footnotesize
 \begin{center}
   \caption{Results for the $\met$, \Ht$<$100 \GeV~signal regions.  Irreducible sources include all backgrounds estimated with MC simulation.  Results are presented in number of expected events as $N~\pm$ (statistical uncertainty) $\pm$ (systematic uncertainty).}
  \begin{tabular}{l c c c c}
\hline
$\met \geq$ &Irreducible	&Reducible	&Total Exp.	&Observed\\\hline
\multicolumn{5}{c}{\threeL, off-$Z$} \\
\hline
   0 \GeV	&46~$\pm 4$~$\pm 6$	&41~$\pm 5$~$\pm 16$	&86~$\pm 6$~$\pm 17$	&89\\
  20 \GeV	&28~$\pm 4$~$\pm 3$	&28~$\pm 4$~$\pm 12$	&56~$\pm 6$~$\pm 12$	&65\\
  50 \GeV	&7.5~$\pm 0.5$~$\pm 1.0$	&15~$\pm 2$~$\pm 7$	&22~$\pm 2$~$\pm 7$	&25\\
  75 \GeV	&3.0~$\pm 0.3$~$\pm 0.4$	&7~$\pm 2$~$\pm 4$	&10~$\pm 2$~$\pm 4$	&10\\
\hline
\multicolumn{5}{c}{\twoLoneT, off-$Z$} \\
\hline
   0 \GeV	&5.3~$\pm 0.4$~$\pm 0.9$	&184~$\pm 4$~$\pm 40$	&190~$\pm 4$~$\pm 40$	&202\\
  20 \GeV	&4.4~$\pm 0.3$~$\pm 0.7$	&93~$\pm 3$~$\pm 20$	&98~$\pm 3$~$\pm 20$	&91\\
  50 \GeV	&1.5~$\pm 0.2$~$\pm 0.2$	&17~$\pm 1$~$\pm 4$	&19~$\pm 1$~$\pm 4$	&20\\
  75 \GeV	&0.6~$\pm 0.1$~$\pm 0.1$	&8.0~$\pm 0.8$~$\pm 1.8$	&8.5~$\pm 0.8$~$\pm 1.8$	&10\\
\hline
\multicolumn{5}{c}{\threeL, on-$Z$} \\
\hline
  20 \GeV	&340~$\pm 5$~$\pm 50$	&100~$\pm 7$~$\pm 31$	&439~$\pm 9$~$\pm 60$	&509\\
  50 \GeV	&105~$\pm 2$~$\pm 14$	&14~$\pm 3$~$\pm 5$	&119~$\pm 3$~$\pm 14$	&144\\
  75 \GeV	&40~$\pm 1$~$\pm 5$	&5~$\pm 1$~$\pm 2$	&46~$\pm 2$~$\pm 6$	&57\\
\hline
\multicolumn{5}{c}{\twoLoneT, on-$Z$} \\
\hline
  20 \GeV	&11.3~$\pm 0.5$~$\pm 1.9$	&984~$\pm10$~$\pm 240$	&1000~$\pm10$~$\pm 240$	&862\\
  50 \GeV	&4.6~$\pm 0.3$~$\pm 0.7$	&43~$\pm 2$~$\pm 11$	&48~$\pm 2$~$\pm 11$	&33\\
  75 \GeV	&2.0~$\pm 0.2$~$\pm 0.3$	&4.1~$\pm 0.6$~$\pm 1.0$	&6.1~$\pm 0.6$~$\pm 1.0$	&4\\
\hline
\end{tabular}
   \label{t:Result_Weak_MET}
\end{center}
\end{table*}

\begin{table*}[tbp]\footnotesize
 \begin{center}
   \caption{Results for the $\met$, \Ht$\geq 100$\GeV~signal regions.  Irreducible sources include all backgrounds estimated with MC simulation.  Results are presented in number of expected events as $N~\pm$ (statistical uncertainty) $\pm$ (systematic uncertainty).}
  \begin{tabular}{l c c c c}
\hline
$\met \geq$ &Irreducible	&Reducible	&Total Exp.	&Observed\\\hline
\multicolumn{5}{c}{\threeL, off-$Z$} \\
\hline
   0 \GeV	&7.7~$\pm 0.8$~$\pm 1.2$	&13~$\pm 2$~$\pm 7$	&21~$\pm 2$~$\pm 7$	&10\\
  20 \GeV	&6.0~$\pm 0.6$~$\pm 0.9$	&12~$\pm 2$~$\pm 6$	&18~$\pm 2$~$\pm 6$	&8\\
  50 \GeV	&3.2~$\pm 0.3$~$\pm 0.5$	&8~$\pm 2$~$\pm 5$	&11~$\pm 2$~$\pm 5$	&5\\
  75 \GeV	&2.2~$\pm 0.2$~$\pm 0.3$	&7~$\pm 2$~$\pm 4$	&9~$\pm 2$~$\pm 4$	&5\\
\hline
\multicolumn{5}{c}{\twoLoneT, off-$Z$} \\
\hline
   0 \GeV	&1.1~$\pm 0.1$~$\pm 0.2$	&30~$\pm 2$~$\pm 7$	&31~$\pm 2$~$\pm 7$	&24\\
  20 \GeV	&1.1~$\pm 0.1$~$\pm 0.2$	&23~$\pm 1$~$\pm 6$	&25~$\pm 1$~$\pm 6$	&20\\
  50 \GeV	&0.7~$\pm 0.1$~$\pm 0.1$	&14.5~$\pm 1.1$~$\pm 3.4$	&15.2~$\pm 1.1$~$\pm 3.4$	&13\\
  75 \GeV	&0.5~$\pm 0.1$~$\pm 0.1$	&9.3~$\pm 0.8$~$\pm 2.2$	&9.8~$\pm 0.8$~$\pm 2.3$	&8\\
\hline
\multicolumn{5}{c}{\threeL, on-$Z$} \\
\hline
  20 \GeV	&49~$\pm 1$~$\pm 7$	&20~$\pm 4$~$\pm 10$	&69~$\pm 4$~$\pm 12$	&79\\
  50 \GeV	&29~$\pm 1$~$\pm 4$	&7~$\pm 2$~$\pm 3$	&36~$\pm 2$~$\pm 5$	&43\\
  75 \GeV	&17.4~$\pm 0.7$~$\pm 2.1$	&5~$\pm 1$~$\pm 2$	&22~$\pm 2$~$\pm 3$	&28\\
\hline
\multicolumn{5}{c}{\twoLoneT, on-$Z$} \\
\hline
  20 \GeV	&1.9~$\pm 0.2$~$\pm 0.4$	&61~$\pm 2$~$\pm 15$	&63~$\pm 2$~$\pm 15$	&52\\
  50 \GeV	&1.1~$\pm 0.1$~$\pm 0.2$	&7.8~$\pm 0.8$~$\pm 1.9$	&8.9~$\pm 0.8$~$\pm 1.9$	&11\\
  75 \GeV	&0.7~$\pm 0.1$~$\pm 0.1$	&2.7~$\pm 0.4$~$\pm 0.7$	&3.4~$\pm 0.5$~$\pm 0.7$	&1\\
\hline
\end{tabular}
   \label{t:Result_Strong_MET}
\end{center}
\end{table*}

\begin{table*}[tbp]\footnotesize
 \begin{center}
   \caption{Results for the $\St$~signal regions.  Irreducible sources include all backgrounds estimated with MC simulation.  Results are presented in number of expected events as $N~\pm$ (statistical uncertainty) $\pm$ (systematic uncertainty).}
  \begin{tabular}{l c c c c}
\hline
$\St \geq$ &Irreducible	&Reducible	&Total Exp.	&Observed\\\hline
\multicolumn{5}{c}{\threeL, off-$Z$} \\
\hline
   0 \GeV	&54~$\pm 4$~$\pm 7$	&54~$\pm 6$~$\pm 23$	&107~$\pm 7$~$\pm 24$	&99\\
 150 \GeV	&32~$\pm 2$~$\pm 4$	&43~$\pm 4$~$\pm 20$	&75~$\pm 4$~$\pm 20$	&64\\
 300 \GeV	&12.0~$\pm 0.9$~$\pm 1.6$	&16~$\pm 2$~$\pm 8$	&28~$\pm 3$~$\pm 8$	&15\\
 500 \GeV	&3.3~$\pm 0.2$~$\pm 0.5$	&3.2~$\pm 1.2$~$\pm 2.4$	&6.5~$\pm 1.2$~$\pm 2.5$	&5\\
\hline
\multicolumn{5}{c}{\twoLoneT, off-$Z$} \\
\hline
   0 \GeV	&6.4~$\pm 0.4$~$\pm 1.0$	&214~$\pm 5$~$\pm 50$	&220~$\pm 5$~$\pm 50$	&226\\
 150 \GeV	&4.4~$\pm 0.3$~$\pm 0.7$	&106~$\pm 3$~$\pm 24$	&111~$\pm 3$~$\pm 24$	&101\\
 300 \GeV	&1.3~$\pm 0.2$~$\pm 0.2$	&31~$\pm 2$~$\pm 7$	&32~$\pm 2$~$\pm 7$	&25\\
 500 \GeV	&0.4~$\pm 0.1$~$\pm 0.2$	&6.6~$\pm 0.7$~$\pm 1.6$	&7.0~$\pm 0.7$~$\pm 1.6$	&6\\
\hline
\multicolumn{5}{c}{\threeL, on-$Z$} \\
\hline
   0 \GeV	&390~$\pm 5$~$\pm 50$	&120~$\pm 8$~$\pm 40$	&510~$\pm 10$~$\pm 70$	&588\\
 150 \GeV	&270~$\pm 3$~$\pm 40$	&57~$\pm 6$~$\pm 22$	&330~$\pm 7$~$\pm 40$	&399\\
 300 \GeV	&73~$\pm 1$~$\pm 10$	&16~$\pm 3$~$\pm 8$	&89~$\pm 4$~$\pm 13$	&103\\
 500 \GeV	&22.2~$\pm 0.9$~$\pm 2.8$	&3~$\pm 1$~$\pm 1$	&25~$\pm 2$~$\pm 3$	&29\\
\hline
\multicolumn{5}{c}{\twoLoneT, on-$Z$} \\
\hline
   0 \GeV	&13.2~$\pm 0.5$~$\pm 2.2$	&1050~$\pm 10$~$\pm 260$	&1060~$\pm 10$~$\pm 260$	&914\\
 150 \GeV	&10.7~$\pm 0.5$~$\pm 1.8$	&360~$\pm 5$~$\pm 90$	&370~$\pm 5$~$\pm 90$	&309\\
 300 \GeV	&2.9~$\pm 0.3$~$\pm 0.4$	&47~$\pm 2$~$\pm 12$	&50~$\pm 2$~$\pm 12$	&42\\
 500 \GeV	&0.9~$\pm 0.2$~$\pm 0.1$	&7.7~$\pm 0.8$~$\pm 1.9$	&8.7~$\pm 0.8$~$\pm 2.0$	&5\\
\hline
\end{tabular}
   \label{t:Result_All_ST}
\end{center}
\end{table*}

\begin{table*}[tbp]\footnotesize
 \begin{center}
   \caption{Results for the $\St$, high-\met~signal regions.  Irreducible sources include all backgrounds estimated with MC simulation.  Results are presented in number of expected events as $N~\pm$ (statistical uncertainty) $\pm$ (systematic uncertainty).}
  \begin{tabular}{l c c c c}
\hline
$\St \geq$ &Irreducible	&Reducible	&Total Exp.	&Observed\\\hline
\multicolumn{5}{c}{\threeL, off-$Z$} \\
\hline
   0 \GeV	&5.1~$\pm 0.4$~$\pm 0.7$	&13~$\pm 2$~$\pm 8$	&18~$\pm 2$~$\pm 8$	&15\\
 150 \GeV	&5.1~$\pm 0.4$~$\pm 0.7$	&13~$\pm 2$~$\pm 8$	&18~$\pm 2$~$\pm 8$	&15\\
 300 \GeV	&3.7~$\pm 0.3$~$\pm 0.5$	&10~$\pm 2$~$\pm 6$	&13~$\pm 2$~$\pm 6$	&9\\
 500 \GeV	&1.7~$\pm 0.2$~$\pm 0.2$	&2.9~$\pm 1.1$~$\pm 2.3$	&4.5~$\pm 1.1$~$\pm 2.3$	&4\\
\hline
\multicolumn{5}{c}{\twoLoneT, off-$Z$} \\
\hline
   0 \GeV	&1.0~$\pm 0.2$~$\pm 0.1$	&17~$\pm 1$~$\pm 4$	&18~$\pm 1$~$\pm 4$	&18\\
 150 \GeV	&1.0~$\pm 0.2$~$\pm 0.1$	&17~$\pm 1$~$\pm 4$	&18~$\pm 1$~$\pm 4$	&18\\
 300 \GeV	&0.6~$\pm 0.1$~$\pm 0.1$	&11.9~$\pm 0.9$~$\pm 2.9$	&12.4~$\pm 0.9$~$\pm 2.9$	&11\\
 500 \GeV	&0.2~$\pm 0.1$~$\pm 0.1$	&3.2~$\pm 0.5$~$\pm 0.8$	&3.4~$\pm 0.5$~$\pm 0.8$	&2\\
\hline
\multicolumn{5}{c}{\threeL, on-$Z$} \\
\hline
   0 \GeV	&58~$\pm 1$~$\pm 7$	&10~$\pm 2$~$\pm 4$	&68~$\pm 2$~$\pm 8$	&85\\
 150 \GeV	&58~$\pm 1$~$\pm 7$	&10~$\pm 2$~$\pm 4$	&68~$\pm 2$~$\pm 8$	&85\\
 300 \GeV	&32~$\pm 1$~$\pm 4$	&6~$\pm 1$~$\pm 2$	&37~$\pm 2$~$\pm 4$	&47\\
 500 \GeV	&11.8~$\pm 0.6$~$\pm 1.4$	&2.2~$\pm 1.1$~$\pm 0.7$	&14.0~$\pm 1.3$~$\pm 1.6$	&18\\
\hline
\multicolumn{5}{c}{\twoLoneT, on-$Z$} \\
\hline
   0 \GeV	&2.7~$\pm 0.3$~$\pm 0.4$	&6.8~$\pm 0.7$~$\pm 1.6$	&9.5~$\pm 0.8$~$\pm 1.7$	&5\\
 150 \GeV	&2.7~$\pm 0.3$~$\pm 0.4$	&6.7~$\pm 0.7$~$\pm 1.6$	&9.4~$\pm 0.8$~$\pm 1.7$	&4\\
 300 \GeV	&1.6~$\pm 0.2$~$\pm 0.2$	&3.5~$\pm 0.5$~$\pm 0.9$	&5.0~$\pm 0.5$~$\pm 0.9$	&2\\
 500 \GeV	&0.6~$\pm 0.1$~$\pm 0.1$	&0.4~$\pm 0.1$~$\pm 0.1$	&1.0~$\pm 0.2$~$\pm 0.1$	&0\\
\hline
\end{tabular}
   \label{t:Result_HighMET_ST}
\end{center}
\end{table*}

\clearpage

\section{Tables of expected and observed limits}
\label{app:limits}
The expected and observed 95\% confidence level upper limits on the expected event yields from new phenomena for all signal regions under study 
are shown in Tables~\ref{t:lim_htlep}-\ref{t:lim_st_highmet}.

\begin{table}[bp]
\begin{center}
 \caption{Limits in the \htlep{} bins shown as the upper limit on the visible cross section ($\sigma^{\mathrm{vis}}_{95} = N_{95}/\intLdt$).} 
 \begin{tabular}{lcccc}
 \hline \hline \htlep{}& Observed& Expected& ${}^{+1\sigma}$ ${}_{-1\sigma}$& ${}^{+2\sigma}$ ${}_{-2\sigma}$ \\   
{[}\GeV]   & {[}fb]& {[}fb]& {[}fb]& {[}fb] \\
\hline\multicolumn{5}{c}{ \threeL\, off-$Z$} \\ \hline 
$ >0 $& 11 & 11 & ${}^{5}$  ${}_{2}$ & ${}^{9}$  ${}_{4}$ \\ 
$ >100 $& 8.7 & 8.5 & ${}^{2.9}$  ${}_{1.6}$ & ${}^{6.9}$  ${}_{2.6}$ \\ 
$ >150 $& 4.0 & 4.6 & ${}^{1.8}$  ${}_{1.2}$ & ${}^{5.1}$  ${}_{1.9}$ \\ 
$ >200 $& 4.4 & 3.6 & ${}^{1.7}$  ${}_{1.0}$ & ${}^{4.9}$  ${}_{1.3}$ \\ 
$ >300 $& 1.6 & 1.9 & ${}^{1.0}$  ${}_{0.4}$ & ${}^{2.4}$  ${}_{0.6}$ \\ 
\hline\multicolumn{5}{c}{\twoLtau\, off-$Z$} \\ \hline 
$ >0 $& 25 & 23 & ${}^{13}$  ${}_{5}$ & ${}^{29}$  ${}_{9}$ \\ 
$ >100 $& 14 & 14 & ${}^{6}$  ${}_{3}$ & ${}^{17}$  ${}_{5}$ \\ 
$ >150 $& 6.1 & 6.4 & ${}^{3.4}$  ${}_{1.3}$ & ${}^{8.3}$  ${}_{3.1}$ \\ 
$ >200 $& 3.3 & 3.6 & ${}^{1.9}$  ${}_{1.2}$ & ${}^{5.0}$  ${}_{1.7}$ \\ 
$ >300 $& 1.2 & 1.5 & ${}^{1.0}$  ${}_{0.5}$ & ${}^{2.4}$  ${}_{0.8}$ \\ 
\hline\multicolumn{5}{c}{ \threeL\, on-$Z$} \\ \hline 
$ >0 $& 48 & 33 & ${}^{15}$  ${}_{8}$ & ${}^{32}$  ${}_{14}$ \\ 
$ >100 $& 38 & 25 & ${}^{11}$  ${}_{7}$ & ${}^{23}$  ${}_{11}$ \\ 
$ >150 $& 14 & 12 & ${}^{4}$  ${}_{4}$ & ${}^{10}$  ${}_{5}$ \\ 
$ >200 $& 7.2 & 6.5 & ${}^{2.8}$  ${}_{1.9}$ & ${}^{6.2}$  ${}_{2.6}$ \\ 
$ >300 $& 4.5 & 3.1 & ${}^{1.4}$  ${}_{0.7}$ & ${}^{3.5}$  ${}_{1.1}$ \\ 
\hline\multicolumn{5}{c}{\twoLtau\, on-$Z$} \\ \hline 
$ >0 $& 85 & 94 & ${}^{41}$  ${}_{22}$ & ${}^{96}$  ${}_{30}$ \\ 
$ >100 $& 53 & 61 & ${}^{26}$  ${}_{16}$ & ${}^{64}$  ${}_{27}$ \\ 
$ >150 $& 11.0 & 9.9 & ${}^{4.3}$  ${}_{2.2}$ & ${}^{11.0}$  ${}_{3.7}$ \\ 
$ >200 $& 5.2 & 4.5 & ${}^{2.0}$  ${}_{1.3}$ & ${}^{5.3}$  ${}_{1.9}$ \\ 
$ >300 $& 3.0 & 1.9 & ${}^{1.0}$  ${}_{0.6}$ & ${}^{2.7}$  ${}_{0.8}$ \\ 
 \hline \hline 
 \end{tabular} 
 \label{t:lim_htlep} 
 \end{center} 
 \end{table}

\begin{table}
\begin{center}
 \caption{Limits in the \met{} bins with \Ht{} $\geq$ 100~\GeV{} requirement shown as the upper limit on the visible cross section ($\sigma^{\mathrm{vis}}_{95} = N_{95}/\intLdt$).} 
 \begin{tabular}{lcccc}
 \hline \hline \met{}& Observed& Expected& ${}^{+1\sigma}$ ${}_{-1\sigma}$& ${}^{+2\sigma}$ ${}_{-2\sigma}$ \\   
{[}\GeV]   & {[}fb]& {[}fb]& {[}fb]& {[}fb] \\
\hline\multicolumn{5}{c}{ \threeL\, off-$Z$} \\ \hline 
$ >0 $& 2.6 & 3.1 & ${}^{1.5}$  ${}_{0.7}$ & ${}^{3.4}$  ${}_{1.4}$ \\ 
$ >50 $& 2.1 & 2.4 & ${}^{1.0}$  ${}_{0.8}$ & ${}^{2.3}$  ${}_{1.2}$ \\ 
$ >75 $& 2.1 & 2.3 & ${}^{1.1}$  ${}_{0.4}$ & ${}^{1.9}$  ${}_{0.9}$ \\ 
\hline\multicolumn{5}{c}{\twoLtau\, off-$Z$} \\ \hline 
$ >0 $& 4.2 & 4.8 & ${}^{2.5}$  ${}_{1.5}$ & ${}^{6.1}$  ${}_{2.1}$ \\ 
$ >50 $& 3.1 & 3.3 & ${}^{1.8}$  ${}_{1.2}$ & ${}^{4.4}$  ${}_{1.6}$ \\ 
$ >75 $& 2.6 & 2.1 & ${}^{0.8}$  ${}_{0.6}$ & ${}^{1.9}$  ${}_{1.0}$ \\ 
\hline\multicolumn{5}{c}{ \threeL\, on-$Z$} \\ \hline 
$ >20 $& 11.0 & 8.7 & ${}^{2.5}$  ${}_{3.2}$ & ${}^{7.0}$  ${}_{4.1}$ \\ 
$ >50 $& 6.4 & 4.9 & ${}^{2.3}$  ${}_{1.2}$ & ${}^{5.4}$  ${}_{1.8}$ \\ 
$ >75 $& 5.1 & 3.8 & ${}^{1.6}$  ${}_{1.0}$ & ${}^{3.8}$  ${}_{1.4}$ \\ 
\hline\multicolumn{5}{c}{\twoLtau\, on-$Z$} \\ \hline 
$ >20 $& 5.9 & 7.3 & ${}^{2.9}$  ${}_{1.4}$ & ${}^{6.8}$  ${}_{3.5}$ \\ 
$ >50 $& 3.4 & 2.8 & ${}^{1.2}$  ${}_{0.9}$ & ${}^{3.8}$  ${}_{1.2}$ \\ 
$ >75 $& 1.2 & 1.5 & ${}^{0.4}$  ${}_{0.4}$ & ${}^{1.0}$  ${}_{0.6}$ \\ 
 \hline 
 \end{tabular} 
 \label{t:lim_metstrong} 
 \end{center} 
 \end{table}

\begin{table}
\begin{center}
 \caption{Limits in the \met{} bins with \Ht{} $\leq$ 100~\GeV{} requirement shown as the upper limit on the visible cross section ($\sigma^{\mathrm{vis}}_{95} = N_{95}/\intLdt$).} 
 \begin{tabular}{lcccc}
 \hline \hline \met{}& Observed& Expected& ${}^{+1\sigma}$ ${}_{-1\sigma}$& ${}^{+2\sigma}$ ${}_{-2\sigma}$ \\   
{[}\GeV]   & {[}fb]& {[}fb]& {[}fb]& {[}fb] \\
\hline\multicolumn{5}{c}{ \threeL\, off-$Z$} \\ \hline 
$ >0 $& 11 & 10 & ${}^{4}$  ${}_{2}$ & ${}^{8}$  ${}_{4}$ \\ 
$ >50 $& 5.3 & 4.7 & ${}^{1.9}$  ${}_{1.0}$ & ${}^{4.5}$  ${}_{1.6}$ \\ 
$ >75 $& 3.1 & 3.0 & ${}^{1.0}$  ${}_{0.6}$ & ${}^{1.8}$  ${}_{1.0}$ \\ 
\hline\multicolumn{5}{c}{\twoLtau\, off-$Z$} \\ \hline 
$ >0 $& 23 & 21 & ${}^{9}$  ${}_{6}$ & ${}^{23}$  ${}_{9}$ \\ 
$ >50 $& 4.3 & 4.0 & ${}^{2.3}$  ${}_{1.2}$ & ${}^{5.0}$  ${}_{1.7}$ \\ 
$ >75 $& 3.1 & 2.6 & ${}^{1.1}$  ${}_{0.7}$ & ${}^{3.1}$  ${}_{1.1}$ \\ 
\hline\multicolumn{5}{c}{ \threeL\, on-$Z$} \\ \hline 
$ >20 $& 41 & 30 & ${}^{10}$  ${}_{9}$ & ${}^{20}$  ${}_{14}$ \\ 
$ >50 $& 16 & 10 & ${}^{4}$  ${}_{3}$ & ${}^{11}$  ${}_{5}$ \\ 
$ >75 $& 8.0 & 5.4 & ${}^{2.6}$  ${}_{1.3}$ & ${}^{6.2}$  ${}_{1.9}$ \\ 
\hline\multicolumn{5}{c}{\twoLtau\, on-$Z$} \\ \hline 
$ >20 $& 80 & 88 & ${}^{39}$  ${}_{23}$ & ${}^{94}$  ${}_{43}$ \\ 
$ >50 $& 4.4 & 5.5 & ${}^{3.2}$  ${}_{1.4}$ & ${}^{7.6}$  ${}_{2.1}$ \\ 
$ >75 $& 1.8 & 2.2 & ${}^{0.4}$  ${}_{0.7}$ & ${}^{1.0}$  ${}_{1.0}$ \\ 
 \hline 
 \end{tabular} 
 \label{t:lim_metweak} 
 \end{center} 
 \end{table}

\begin{table}
\begin{center}
 \caption{Limits in the \st{} bins shown as the upper limit on the visible cross section ($\sigma^{\mathrm{vis}}_{95} = N_{95}/\intLdt$).} 
 \begin{tabular}{lcccc}
 \hline \st{}& Observed& Expected& ${}^{+1\sigma}$ ${}_{-1\sigma}$& ${}^{+2\sigma}$ ${}_{-2\sigma}$ \\   
{[}\GeV]   & {[}fb]& {[}fb]& {[}fb]& {[}fb] \\
\hline\multicolumn{5}{c}{ \threeL\, off-$Z$} \\ \hline 
$ >0 $& 11 & 11 & ${}^{5}$  ${}_{2}$ & ${}^{9}$  ${}_{4}$ \\ 
$ >150 $& 8.1 & 8.8 & ${}^{3.0}$  ${}_{2.2}$ & ${}^{7.2}$  ${}_{3.9}$ \\ 
$ >300 $& 3.1 & 3.7 & ${}^{1.7}$  ${}_{0.7}$ & ${}^{3.8}$  ${}_{1.6}$ \\ 
$ >500 $& 2.1 & 2.1 & ${}^{1.1}$  ${}_{0.6}$ & ${}^{2.3}$  ${}_{0.9}$ \\ 
\hline\multicolumn{5}{c}{\twoLtau\, off-$Z$} \\ \hline 
$ >0 $& 25 & 23 & ${}^{13}$  ${}_{5}$ & ${}^{29}$  ${}_{9}$ \\ 
$ >150 $& 12 & 13 & ${}^{6}$  ${}_{4}$ & ${}^{14}$  ${}_{5}$ \\ 
$ >300 $& 3.9 & 4.9 & ${}^{2.5}$  ${}_{1.5}$ & ${}^{6.4}$  ${}_{2.3}$ \\ 
$ >500 $& 2.2 & 2.4 & ${}^{1.3}$  ${}_{0.5}$ & ${}^{3.4}$  ${}_{1.2}$ \\ 
\hline\multicolumn{5}{c}{ \threeL\, on-$Z$} \\ \hline 
$ >0 $& 48 & 33 & ${}^{15}$  ${}_{8}$ & ${}^{32}$  ${}_{14}$ \\ 
$ >150 $& 37 & 25 & ${}^{9}$  ${}_{7}$ & ${}^{21}$  ${}_{11}$ \\ 
$ >300 $& 11 & 9 & ${}^{4}$  ${}_{2}$ & ${}^{9}$  ${}_{3}$ \\ 
$ >500 $& 4.8 & 3.9 & ${}^{1.7}$  ${}_{1.0}$ & ${}^{4.3}$  ${}_{1.1}$ \\ 
\hline\multicolumn{5}{c}{\twoLtau\, on-$Z$} \\ \hline 
$ >0 $& 85 & 94 & ${}^{41}$  ${}_{22}$ & ${}^{96}$  ${}_{30}$ \\ 
$ >150 $& 28 & 35 & ${}^{13}$  ${}_{11}$ & ${}^{34}$  ${}_{15}$ \\ 
$ >300 $& 5.9 & 6.8 & ${}^{2.8}$  ${}_{1.8}$ & ${}^{8.1}$  ${}_{2.4}$ \\ 
$ >500 $& 1.9 & 2.5 & ${}^{1.4}$  ${}_{1.0}$ & ${}^{3.5}$  ${}_{1.2}$ \\ 
 \hline 
 \end{tabular} 
 \label{t:lim_st} 
 \end{center} 
 \end{table}

\begin{table}
\begin{center}
 \caption{Limits in the \st{} bins with \met{} $\geq$ 75~\GeV{} requirement shown as the upper limit on the visible cross section ($\sigma^{\mathrm{vis}}_{95} = N_{95}/\intLdt$). } 
 \begin{tabular}{lcccc}
 \hline \st{}& Observed& Expected& ${}^{+1\sigma}$ ${}_{-1\sigma}$& ${}^{+2\sigma}$ ${}_{-2\sigma}$ \\   
{[}\GeV]   & {[}fb]& {[}fb]& {[}fb]& {[}fb] \\
\hline\multicolumn{5}{c}{ \threeL\, off-$Z$} \\ \hline 
$ >0 $& 3.8 & 3.9 & ${}^{1.5}$  ${}_{0.7}$ & ${}^{3.4}$  ${}_{1.3}$ \\ 
$ >150 $& 3.8 & 3.9 & ${}^{1.5}$  ${}_{1.0}$ & ${}^{3.6}$  ${}_{1.3}$ \\ 
$ >300 $& 2.8 & 3.0 & ${}^{1.2}$  ${}_{0.7}$ & ${}^{3.2}$  ${}_{1.1}$ \\ 
$ >500 $& 2.1 & 2.0 & ${}^{0.8}$  ${}_{0.4}$ & ${}^{2.2}$  ${}_{0.8}$ \\ 
\hline\multicolumn{5}{c}{\twoLtau\, off-$Z$} \\ \hline 
$ >0 $& 3.9 & 3.8 & ${}^{2.1}$  ${}_{1.2}$ & ${}^{5.2}$  ${}_{1.8}$ \\ 
$ >150 $& 4.0 & 3.9 & ${}^{2.0}$  ${}_{1.3}$ & ${}^{4.8}$  ${}_{1.7}$ \\ 
$ >300 $& 2.9 & 3.1 & ${}^{1.6}$  ${}_{1.0}$ & ${}^{4.1}$  ${}_{1.3}$ \\ 
$ >500 $& 1.5 & 1.6 & ${}^{0.7}$  ${}_{0.5}$ & ${}^{2.1}$  ${}_{0.7}$ \\ 
\hline\multicolumn{5}{c}{ \threeL\, on-$Z$} \\ \hline 
$ >0 $& 10.0 & 6.9 & ${}^{3.0}$  ${}_{1.3}$ & ${}^{7.4}$  ${}_{2.5}$ \\ 
$ >150 $& 10.0 & 7.1 & ${}^{2.8}$  ${}_{2.2}$ & ${}^{7.0}$  ${}_{2.5}$ \\ 
$ >300 $& 6.8 & 4.9 & ${}^{2.1}$  ${}_{1.0}$ & ${}^{5.1}$  ${}_{2.1}$ \\ 
$ >500 $& 3.9 & 3.0 & ${}^{1.2}$  ${}_{0.7}$ & ${}^{3.4}$  ${}_{1.3}$ \\ 
\hline\multicolumn{5}{c}{\twoLtau\, on-$Z$} \\ \hline 
$ >0 $& 1.6 & 2.4 & ${}^{1.4}$  ${}_{0.9}$ & ${}^{3.8}$  ${}_{1.3}$ \\ 
$ >150 $& 1.4 & 2.5 & ${}^{1.5}$  ${}_{0.8}$ & ${}^{3.8}$  ${}_{1.5}$ \\ 
$ >300 $& 1.5 & 2.0 & ${}^{1.1}$  ${}_{0.8}$ & ${}^{2.7}$  ${}_{1.1}$ \\ 
$ >500 $& 0.9 & 1.1 & ${}^{0.8}$  ${}_{0.4}$ & ${}^{2.0}$  ${}_{0.4}$ \\ 
 \hline 
 \end{tabular} 
 \label{t:lim_st_highmet} 
 \end{center} 
 \end{table}

\onecolumngrid
\clearpage
\begin{flushleft}
{\Large The ATLAS Collaboration}

\bigskip

G.~Aad$^{\rm 48}$,
T.~Abajyan$^{\rm 21}$,
B.~Abbott$^{\rm 111}$,
J.~Abdallah$^{\rm 12}$,
S.~Abdel~Khalek$^{\rm 115}$,
A.A.~Abdelalim$^{\rm 49}$,
O.~Abdinov$^{\rm 11}$,
R.~Aben$^{\rm 105}$,
B.~Abi$^{\rm 112}$,
M.~Abolins$^{\rm 88}$
% ...
\bigskip

% ...
$^{11}$ Institute of Physics, Azerbaijan Academy of Sciences, Baku, Azerbaijan\\
$^{12}$ Institut de F\'{i}sica d'Altes Energies and Departament de F\'{i}sica de la Universitat Aut\`{o}noma de Barcelona and ICREA, Barcelona, Spain\\
% ...
$^{21}$ Physikalisches Institut, University of Bonn, Bonn, Germany\\
% ...
$^{48}$ Fakult\"{a}t f\"{u}r Mathematik und Physik, Albert-Ludwigs-Universit\"{a}t, Freiburg, Germany\\
$^{49}$ Section de Physique, Universit\'{e} de Gen\`{e}ve, Geneva, Switzerland\\
% ...
$^{88}$ Department of Physics and Astronomy, Michigan State University, East Lansing MI, United States of America\\
% ...
$^{105}$ Nikhef National Institute for Subatomic Physics and University of Amsterdam, Amsterdam, Netherlands\\
% ...
$^{111}$ Homer L. Dodge Department of Physics and Astronomy, University of Oklahoma, Norman OK, United States of America\\
$^{112}$ Department of Physics, Oklahoma State University, Stillwater OK, United States of America\\
% ...
$^{115}$ LAL, Universit\'{e} Paris-Sud and CNRS/IN2P3, Orsay, France\\
% ...
\end{flushleft}


\label{app:ATLASColl}

\end{document}